\documentclass[11pt,american]{article}
\usepackage[T1]{fontenc}
\usepackage[latin9]{inputenc}
\usepackage{geometry}
\usepackage{color}
\usepackage{babel}
\usepackage{booktabs}
\usepackage{units}
\usepackage{mathrsfs}
\usepackage{mathtools}
\usepackage{amsmath}
\usepackage{amssymb}
\usepackage{stmaryrd}
\usepackage{graphicx}
\usepackage[unicode=true]
 {hyperref}

\makeatletter

\newcommand*\LyXThinSpace{\,\hspace{0pt}}
\providecommand{\tabularnewline}{\\}

\newcommand{\lyxaddress}[1]{
\par {\raggedright #1
\vspace{1.4em}
\noindent\par}
}

\@ifundefined{date}{}{\date{}}
\usepackage{cite}
\usepackage{upgreek}

\makeatother

\begin{document}

\title{\textbf{Analog Programmable-Photonic Information}}

\author{Andr\'es Macho-Ortiz,\textsuperscript{a,$\ast$} Ra\'ul L\'opez-March,\textsuperscript{a}
Pablo Mart\'inez-Carrasco,\textsuperscript{a} \\
F. Javier Fraile-Pel\'aez,\textsuperscript{b} and Jos\'e Capmany\textsuperscript{a,c,$\ast$} }
\maketitle

\lyxaddress{\begin{center}
\textsuperscript{a}{\small{}ITEAM Research Institute, Universitat
Polit\`ecnica de Val\`encia, Valencia 46022, Spain}\\
\textsuperscript{b}{\small{}Dept. Teor}\'i{\small{}a de la Señal
y Comunicaciones, Universidad de Vigo E.I. Telecomunicaci}\'o{\small{}n,
Campus Universitario, E-36202 Vigo (Pontevedra), Spain}\\
\textsuperscript{c}{\small{}iPronics, Programmable Photonics, S.L,
Camino de Vera s/n, Valencia 46022, Spain}
\par\end{center}}

\noindent $\ast$corresponding author: \href{mailto:amachor@iteam.upv.es}{amachor@iteam.upv.es},
\href{mailto:jcapmany@iteam.upv.es}{jcapmany@iteam.upv.es}

\vspace{1cm}

\section*{Abstract}

The limitations of digital electronics in handling real-time matrix
operations for emerging computational tasks \textendash{} such as
artificial intelligence, drug design, and medical imaging \textendash{}
have prompted renewed interest in analog computing. Programmable Integrated
Photonics (PIP) has emerged as a promising technology for scalable,
low-power, and high-bandwidth analog computation. While prior work
has explored PIP implementations of quantum and neuromorphic computing,
both approaches face significant limitations due to misalignments
between their mathematical models and the native capabilities of photonic
hardware. Building on the recently proposed Analog Programmable-Photonic
Computation (APC) \textendash{} a computation theory explicitly matched
to the technological features of PIP \textendash{} we introduce its
critical missing component: an information theory. We present Analog
Programmable-Photonic Information (API), a mathematical framework
that addresses fundamental concepts beyond APC by examining the amount
of information that can be generated, computed and recovered in a
PIP platform. API also demonstrates the robustness of APC against
errors arising from system noise and hardware imperfections, enabling
scalable computation without the extensive error-correction overhead
required in quantum computing. Together, APC and API provide a unified
foundation for on-chip photonic computing, offering a complementary
alternative to digital, quantum and neuromorphic paradigms, and positioning
PIP as a cornerstone technology for next-generation information processing.\\

\noindent \textbf{Keywords:} integrated optics, programmable integrated
photonics, optical computing 
\noindent \begin{center}
\newpage{}
\par\end{center}

\section*{1 Introduction}

\noindent Computational science is the discipline dedicated to the
study and development of systems capable of processing information
autonomously \cite{key-P1}. Any computational system is constructed
by combining three essential pieces \cite{key-P1,key-P2,key-P3,key-P4,key-P5,key-P6,key-P7}:
an information theory, which provides a mathematical framework for
describing, generating and recovering user information; a computation
theory, which defines the mathematical transformations required to
address diverse problems and applications; and a technology that realizes
these theories through physical devices and systems.

The earliest known computational system is the Antikythera mechanism,
invented in Greece between 150 and 100 BC, which processed information
using an analog approach based on mechanical technology (gears and
wheels modeling the position of the Moon and the Sun in their orbits)
\cite{key-P8}. Nowadays, the cornerstone of our information society
is the computational landscape of digital electronics, which has emerged
thanks to the collective development of digital computation and information
theories, alongside electronic technology \cite{key-P2,key-P3,key-P4}. 

Over the past 50 years, digital electronics has exponentially scaled
its information processing capacity, driven by continuous advances
in integrated electronic circuits. Electronic microprocessors have
been able to duplicate the density of transistors, power efficiency,
and clock frequency every 18-24 months, as predicted by Moore's and
Dennard\textquoteright s laws \cite{key-P9,key-P10}. Nevertheless,
fundamental physical limits of electronic transistors are currently
leading to the eventual demise of these computing laws \cite{key-P11,key-P12,key-P13}.
Consequently, a wide range of ground-breaking applications that require
real-time matrix information processing cannot be efficiently conducted
with the digital electronic paradigm. These include, for example,
artificial intelligence \cite{key-P14}, drug design \cite{key-P15},
quantum simulation and optimization \cite{key-P16,key-P17}, robotic
control \cite{key-P18}, computational fluid dynamics \cite{key-P19},
financial modeling and risk analysis \cite{key-P20}, medical diagnostic
imaging \cite{key-P21}, and genomic analysis \cite{key-P22}, to
mention a few. 

As a result, recent years have witnessed a strong renaissance of non-electronic
analog computing systems \cite{key-P8,key-P23,key-P24,key-P25}. Hot-topic
research focuses on analog computational models based on matrix algebra,
implemented with CMOS-compatible and scalable technologies offering
complementary hardware requirements to those of electronics in energy
consumption, reconfigurability, bandwidth, or parallelism \cite{key-P24,key-P25,key-P26,key-P27,key-P28,key-P29}. 

In this scenario, a novel system-on-chip technology has recently emerged
fulfilling all the aforementioned requirements: \emph{programmable
integrated photonics} (PIP) \cite{key-P30,key-P31}. PIP is a technological
platform that leverages the natural ability of integrated photonic
circuits to carry out high-dimensional matrix signal processing through
the optical interference of multiple input waves. This is achieved
using meshes of 2$\times$2 optical systems constructed from phase
shifters, resonators, beam splitters, and beam combiners \cite{key-P31}. 

Outstandingly, the manufacturing of PIP circuitry is rapidly changing
to a robust landscape that could reach economies of scale comparable
to those of the microelectronic industry within the next 10-20 years
\cite{key-P30}. In addition, silicon PIP platforms can be seamlessly
integrated with electronic chips by taking advantage of their CMOS
compatibility while providing features that complement integrated
electronics, such as low power consumption, high reconfigurability,
high bandwidth, and massive parallelism \cite{key-P32}. This combination
of characteristics makes PIP a promising technological candidate to
perform analog computing tasks that complement digital electronics
in applications demanding real-time matrix operations \cite{key-P33,key-P34}.
\newpage{}

To date, the main analog computation theories based on matrix algebra
that have been explored with PIP hardware are quantum and neuromorphic
computation \cite{key-P24,key-P25,key-P26,key-P27}. However, these
mathematical models were not originally conceived to be realized with
PIP circuitry, giving rise to computational limitations inherited
from the complexity of their implementation using basic integrated
photonic devices \cite{key-P25,key-P35}. In photonic quantum computing,
the main challenges include ensuring low optical losses and high fidelity
in the units of information (the quantum bits or qubits), as errors
introduced during photon generation and interference, as well as indirect
effects of environmental noise, can significantly degrade the computational
performance \cite{key-P25}. Furthermore, scaling quantum PIP systems
to multiple qubits is hindered by the large number of qubits required
for quantum error correction \cite{key-P36}. In photonic neuromorphic
computing, limitations arise from the difficulty of implementing scalable
non-linear optical components that mimic neuronal activation functions
along with achieving efficient, high-density connectivity between
nodes to emulate neural network architectures \cite{key-P37,key-P38}. 

Recently, a new computation theory has been presented, termed \emph{Analog
Programmable-Photonic Computation} (APC), which has been explicitly
conceived to be matched with the technological features of PIP \cite{key-P39}.
To this end, the unit of information \textendash{} the \emph{analog
bit} or \emph{anbit} (a two-dimensional (2D) vector, not a \textquotedblleft unit\textquotedblright{}
in the strict physical sense but rather an abstract \textquotedblleft container\textquotedblright{}
of information, akin to the qubit in quantum computation, capable
of encoding a variable amount of information) \textendash{} and the
basic computational operations \textendash{} the single-anbit gates
(2$\times$2 matrices) \textendash{} have been defined, endowing them
with mathematical properties that mirror the inherent ability of PIP
to perform vector-by-matrix multiplications \cite{key-P30}. As a
direct consequence, this enables circumventing some of the fundamental
limitations of photonic quantum and neuromorphic computing, which
arise from the mismatch between the mathematical properties of these
computation theories and the technological features of PIP hardware. 

Remarkably, the computational performance of APC systems is not constrained
by optical losses or non-linear operations. Optical losses, in fact,
play a crucial role in executing non-unitary anbit gates, while non-linear
operations are unnecessary for most of the primary computational problems
addressed by APC (although they can be implemented with Mach-Zehnder
interferometers and ring resonators) \cite{key-P39}. Moreover, APC
is expected to exhibit greater tolerance to errors caused by system
noise compared to quantum computing, as the classical wave superposition
inherent to anbits cannot be annihilated during its generation, interference,
or reception. This concurrently would reduce the necessity of using
a large number of anbits for error correction, thereby simplifying
the scalability of APC architectures.

Nonetheless, these unique properties of APC, in combination with other
significant characteristics identified in this work, could not be
previously assessed in ref.\,\cite{key-P39} due to the absence of
a critical complementary tool: an \emph{information theory}. In the
same vein as digital and quantum computing are underpinned by specific
information-theoretical frameworks \cite{key-P1,key-P2,key-P3,key-P4,key-P5},
APC necessitates the development of a new information theory, referred
to as \emph{Analog Programmable-Photonic Information} (API). 

Here, we establish the foundations of API by addressing fundamental
concepts that extend beyond the scope of APC but are crucial to unlocking
the full computational potential of PIP. Specifically, in the Results
section, we examine: \textbf{1)} the average amount of information
that can be generated (the entropy of the transmitter), computed (the
channel capacity), and recovered (the accessible information) in a
PIP platform; \textbf{2)} the principles to design anbit codes and
modulation formats, which define the mapping between user information,
anbits, and the optical waves within the circuits; \textbf{3)} the
mathematical formalism for modeling noise and non-ideal device operation
in APC systems; \textbf{4)} the tolerance of APC to computational
errors induced by system noise and the non-ideal behavior of PIP devices;
and \textbf{5)} strategies for mitigating such errors. These information-theoretical
principles are experimentally validated in the Materials and Methods
section. Finally, in the Discussion section, we present a qualitative
comparison among the main properties of API \emph{vs} digital information
(DI), and quantum information (QI), positioning both APC and API theories
as independent but complementary research fields essential for realizing
the complete potential of this new information-processing paradigm. 

\section*{2 Results}

\noindent Any computational platform or communication system (e.g.,
an optical fiber network) can be broadly conceptualized as an information-processing
system encompassing a transmitter, a channel, and a receiver, through
which information is sequentially generated, propagated, and ultimately
recovered.

The primary distinction between a communication system and a computational
circuit lies in the channel. In the former, information must be propagated
without modification between the transmitter and receiver. In contrast,
in the latter, information is not only propagated but also modified
\textendash{} in this case using PIP circuits implementing computational
operations \textendash{} to solve a specific mathematical problem
of interest. Hence, as illustrated in Fig.\,1, API theory has a more
general perspective than APC for analyzing computational PIP architectures.
In particular, APC is limited to modeling the computational operations
and algorithms executed within the channel, whereas API encompasses
and unifies the entire information-processing system. 

The channel, composed of anbit gates, can compute either a single
or multiple anbits simultaneously \cite{key-P39}. Accordingly, we
should respectively distinguish between simple and composite systems
in API. Here, for the sake of simplicity in introducing the basic
principles of API, let us focus our attention on \emph{simple systems},
where the anbits generated by the transmitter are sequentially processed
by single-anbit gates in the channel. Composite systems, associated
with multi-anbit gates, are briefly addressed in the Discussion section.
In the following, we outline the design principles governing the transmitter,
channel, and receiver in simple API systems.

\subsection*{2.1 Anbit transmitter}

\noindent The transmitter of an API system consists of three independent
blocks (Fig. 1): (\emph{i}) an \emph{originator source}, which generates
information according to a specific computational problem that must
be solved by an APC architecture; (\emph{ii}) an \emph{encoder}, which
maps this information onto a set of anbits, subsequently transformed
by the computational gates of the channel; and (\emph{iii}) a \emph{modulator},
which physically implements the anbits with optical waves for propagation
through the channel. While the originator source and encoder are integrated
electronic systems, the modulator is an integrated photonic circuit
that carries out an electro-optic conversion of information prior
to its transmission through the channel. 

\subsubsection*{2.1.1 Originator source}

\noindent In this block, we tackle two significant goals. Firstly,
the mathematical description of an originator source in API, tailored
to the primary applications demanded by APC. Secondly, the quantification
of the average amount of information that can be generated by the
originator source, that is, the \emph{pre-codification entropy}.

As commented above, APC will execute computational problems requiring
matrix operations, which are inefficiently handled by digital electronics
\cite{key-P39}. This implies that digital computing and APC are expected
to coexist on the same system-on-chip platform, necessitating compatibility
between the originator sources of DI and API. This compatibility is
achieved by conceptualizing user information in API in the same way
as in DI: a sequence of random events $\zeta_{i}$ belonging to a
\emph{discrete} sample space $\mathcal{S}=\left\{ \zeta_{i};i=1,\ldots,M\right\} $,
consistent with the discrete nature of the state space of the digital
bit \cite{key-P1,key-P2,key-P3,key-P4}. The number of random events
is determined by the level of accuracy required to solve a specific
computational problem in APC. Next, given that the channel of simple
API systems consists of 2$\times$2 PIP circuits implementing basic
vector-by-matrix multiplications, it might initially seem reasonable
to describe the sample space using a 2D discrete random vector, in
line with the mathematical definition of the anbit \cite{key-P39}.
However, a discrete random vector can be equivalently represented
by a discrete real random variable $X=\left\{ \zeta_{i}\in\mathcal{S}/X\left(\zeta_{i}\right)=x_{i}\in\mathbb{R}\right\} $,
provided that their probability mass functions (pmf) are identical
(see Supplementary Note 1 for a detailed discussion of this equivalence).
In this scenario, the use of a discrete random variable allows us
to simplify the mathematical framework of API without any loss of
generality in the description of the originator source. Finally, keeping
in mind the classical nature of information, it is straightforward
to quantify the entropy of the source (the pre-codification entropy)
by using \emph{Shannon\textquoteright s entropy} $H\left(X\right)=-\sum_{i}p_{i}\log_{2}p_{i}$
(bits), where $p_{i}=p\left(x_{i}\right)$ is the pmf of $X$ \cite{key-P1,key-P4}. 

\subsubsection*{2.1.2 Encoder}

\noindent As sketched in Fig. 1, the encoder maps the symbols $x_{i}$
of the originator source $X$ onto a set of anbits $\left|\psi_{i}\right\rangle $:
\begin{equation}
\bigr|\psi_{i}\bigr\rangle=r_{i}\left(\cos\frac{\theta_{i}}{2}\bigl|0\bigr\rangle+e^{\mathrm{j}\varphi_{i}}\sin\frac{\theta_{i}}{2}\bigl|1\bigr\rangle\right),\ \ \ i=1,\ldots,M\label{eq:1}
\end{equation}
which can be geometrically represented as a collection of different
points on the generalized Bloch sphere (GBS), each with a radius $r_{i}>0$,
where $r_{i}^{2}$ corresponds to the optical power required to physically
implement the anbit $\bigr|\psi_{i}\bigr\rangle$ at the modulator
(see below). Each anbit is located in the GBS through a position vector
(or Bloch vector) given by $\mathbf{r}_{i}=r_{i}\left(\sin\theta_{i}\cos\varphi_{i}\hat{\mathbf{x}}+\sin\theta_{i}\sin\varphi_{i}\hat{\mathbf{y}}+\cos\theta_{i}\hat{\mathbf{z}}\right)$.
Here, the parameters $r_{i}$ (radius), $\theta_{i}$ (elevation angle),
and $\varphi_{i}$ (azimuthal angle) are referred to as the \emph{effective
degrees of freedom} (EDFs) of the anbit $\bigr|\psi_{i}\bigr\rangle$
\cite{key-P39}. The random behavior of the encoder is characterized
by the average anbit $\bigl|\psi_{X}\bigr\rangle=\sum_{i}p_{i}\bigl|\psi_{i}\bigr\rangle$,
which inherits the probabilistic distribution of the originator source
$X$ and plays a role analogous to that of a mixed state in QI. Nevertheless,
unlike a mixed quantum state, which must be described by a density
operator \cite{key-P5}, API does not require an operator to handle
mixed states. Supplementary Note 2 provides a detailed discussion
on the key differences between pure and mixed classical states, as
well as the applicability of the density operator formalism in API.\newpage{}

Notably, the correspondence between symbols and anbits in the GBS
leads to the concept of \emph{analog constellation}. An analog constellation
should be designed to safeguard information against the main physical
impairments of the system \textendash{} noise and non-ideal behavior
of PIP devices \textendash{} which induce computational errors by
deviating anbits from their ideal location in the GBS. Accordingly,
an optimal analog constellation can mitigate such errors, thereby
improving accuracy in solving computational problems in APC. To achieve
this, we establish: (\emph{a}) state-comparative parameters that quantify
the proximity of two anbits within the GBS, and (\emph{b}) general
design criteria for analog constellations based on these parameters.

The similarity between two anbits can be quantified through various
parameters. A straightforward approach could be the extrapolation
of the fidelity and trace distance from QI \cite{key-P5}. Unfortunately,
these parameters have limited utility and lack geometric intuitiveness
within the API framework, as the radius of the GBS may differ from
unity (see Supplementary Note 3 for further discussions about the
suitability of the fidelity and trace distance in API). Contrariwise,
in API, we define the GBS distance:
\begin{equation}
D_{\mathrm{GBS}}\left(\bigl|\psi_{i}\bigr\rangle,\bigl|\psi_{k}\bigr\rangle\right)\coloneqq\frac{1}{2}\left\Vert \mathbf{r}_{i}-\mathbf{r}_{k}\right\Vert ,\label{eq:2}
\end{equation}
which is conceptually simpler than the fidelity or trace distance,
as it is a metric that quantifies the (true) Euclidean distance between
anbits (the factor 1/2 is introduced to account for the geometric
scaling inherent to the construction of the GBS \cite{key-P39}).
The GBS distance reaches its minimum value $D_{\mathrm{GBS}}=0$ when
the states are the same, and its maximum value $D_{\mathrm{GBS}}=\left(r_{i}+r_{k}\right)/2$
when the states are orthogonal (located at opposite points on the
GBS). Unlike the trace distance, this metric is not normalized, reflecting
the arbitrary radius of the GBS. Likewise, it is noteworthy that $D_{\mathrm{GBS}}$
can also be applied to mixed classical states, enabling a comparison
between different encoders. Further properties of this metric, along
with alternative state-comparative parameters for API, are detailed
in Supplementary Note\,3.

Remarkably, state-comparative parameters such as the GBS distance
pave the way for designing analog constellations. As mentioned above,
an optimal arrangement of the analog constellation is of paramount
importance to reduce computational errors, as it minimizes the error
probability at the receiver and maximizes the channel capacity (see
Subsections 2.2 and 2.3). Figure 2 shows different classes of constellations
that can be conceived by varying the EDFs of the anbits. Constellations
with a constant radius are particularly attractive, as they simplify
the optimization problem to two dimensions and enable most computational
problems in APC to be solved using unitary PIP circuitry, which performs
energy-efficient matrix operations by inducing rotations on a constant-radius
GBS \cite{key-P39,key-P40}. A general (although not exclusive) criterion
for optimizing such constellations involves maximizing the GBS distance
among the most probable anbits, while engineering the optical power
consumption required to implement the constellation. This optimization
criterion may be fulfilled through the following two-step procedure.
Firstly, select the radius $r$ of the GBS, which provides information
about the optical power $\mathcal{P}$ demanded to realize a constellation
of M anbits ($\mathcal{P}=Mr^{2}$). The optical power of each anbit
($r^{2}$) must remain below the threshold separating the linear and
non-linear regimes of integrated waveguides to prevent undesired non-linear
effects in the channel (in silicon PIP waveguides the threshold power
is around $\sim$ 17 dBm \cite{key-P41}). Secondly, the anbits should
be positioned on the surface of the GBS to maximize the average GBS
distance $\overline{D}_{\mathrm{GBS}}=\sum_{i,k}p\left(x_{i},x_{k}\right)D_{\mathrm{GBS}}\left(\bigl|\psi_{i}\bigr\rangle,\bigl|\psi_{k}\bigr\rangle\right)$.
This optimization problem can be tackled using, e.g., the K-means
method (or Lloyd\textquoteright s algorithm) \cite{key-P42}, the
gradient method \cite{key-P43}, or genetic algorithms \cite{key-P44}.
Additional design criteria to optimize analog constellations are suggested
in the Discussion section.

On the other hand, before delving into the theory of the modulator,
we should explore how to quantify the \emph{post-codification entropy},
that is, the average amount of information stored in the analog constellation.
Bearing in mind the mathematical similarity between anbits and qubits
\cite{key-P39}, one could ask whether the post-codification entropy
in API should be quantified similarly to QI, that is, using von Neumann\textquoteright s
entropy. In QI, von Neumann\textquoteright s entropy is essential
to measure the post-codification entropy, as it reflects fundamental
limitations imposed by the postulates of quantum mechanics on the
ability to unambiguously distinguish symbols $x_{i}$ encoded in non-orthogonal
quantum states \cite{key-P5}. Nonetheless, as experimentally demonstrated
in Materials and Methods, classical information can be retrieved without
errors from non-orthogonal anbits. In addition, it should be noted
that the probability distribution of the originator source is preserved
by the encoder in the mixed state $\bigl|\psi_{X}\bigr\rangle=\sum_{i}p_{i}\bigl|\psi_{i}\bigr\rangle$,
provided there exists a one-to-one correspondence between symbols
$x_{i}$ and anbits $\bigl|\psi_{i}\bigr\rangle$, which is the desired
scenario. This suffices to conclude that the post-codification entropy
must be quantified through Shannon\textquoteright s entropy, thus
coinciding with the pre-codification entropy $H\left(X\right)$. 

Finally, we introduce a figure of merit to measure the efficiency
of the encoder: the ratio of the post-codification entropy to the
number of anbits of the constellation, termed the \emph{bit-anbit
ratio} (BAR), and given by the expression $\mathrm{BAR}_{X}\coloneqq H\left(X\right)/M$
(bits/anbit). As inferred from this equation, the encoder efficiency
is thus maximized by optimizing $H\left(X\right)$ via the pmf of
$X$. The BAR parameter will later prove useful for quantifying the
channel capacity in anbits.

\subsubsection*{2.1.3 Modulator}

\noindent The modulation block physically implements the analog constellation
with optical waves using PIP technology. Since each anbit in the constellation
is defined through a 2D complex vector $\bigl|\psi_{i}\bigr\rangle\equiv\left(\psi_{i,0},\psi_{i,1}\right)$,
where $\psi_{i,0}=r_{i}\cos\theta_{i}/2$ and $\psi_{i,0}=r_{i}e^{\mathrm{j}\varphi_{i}}\sin\theta_{i}/2$
are the components (or anbit amplitudes), then two classical wave
packets (or complex envelopes) with a phase delay $\varphi_{i}$ suffice
to implement an anbit with an optical field $\mathbf{E}_{\mathrm{in}}$
(Fig.\,1). These wave packets are multiplexed in one of the degrees
of freedom of light (space, mode, frequency, time or polarization),
thus defining a \emph{modulation format} \cite{key-P39}. 

In PIP, the basic building block is a 2$\times$2 circuit that transforms
2D signals propagating through a pair of spatially separated waveguides
\cite{key-P30}. Hence, the \emph{space-anbit modulation} (SAM) is
the natural approach to implement an anbit constellation. Figure 3
depicts the PIP hardware required to realize the SAM. A laser diode,
combined with a Mach-Zehnder modulator (MZM) and a reconfigurable
beam splitter implemented using a tunable basic unit of PIP {[}32{]},
generates two complex envelopes $\psi_{i,0}$ and $\psi_{i,1}$ with
complementary moduli satisfying the condition $\left|\psi_{i,0}\right|^{2}+\left|\psi_{i,1}\right|^{2}=r_{i}^{2}$.
These envelopes may be either continuous or pulsed waves; in the latter
case, their shapes are tailored by the MZM. While pulsed waves are
essential for sequential information processing in real-time computing
applications, continuous waves simplify experimental proof-of-concept
implementations of API systems (see Materials and Methods). Finally,
the desired anbit $\bigl|\psi_{i}\bigr\rangle$ of the constellation
is obtained by adjusting the phase delay $\varphi_{i}$ between the
envelopes using two phase shifters.\newpage{}

Outstandingly, the physical implementation of the anbit amplitudes
via complex envelopes facilitates the geometric interpretation of
analog constellations, as there is a one-to-one correspondence between
the modulus $r_{i}$ of the anbit $\bigl|\psi_{i}\bigr\rangle$ and
the optical power\linebreak{} $\mathcal{P}_{i}=\left|\psi_{i,0}\right|^{2}+\left|\psi_{i,1}\right|^{2}=r_{i}^{2}$
required to generate that anbit at the modulator. Therefore, the optical
power consumption associated with an M-anbit constellation can be
directly calculated as $\mathcal{P}=\sum_{i=1}^{M}r_{i}^{2}$.

\subsection*{2.2 Anbit receiver}

\noindent The optical waves generated by the modulator propagate through
the channel (the PIP circuits that implement the computational gates).
Nevertheless, before introducing the fundamentals of the channel,
we first examine the anbit receiver, as some basic concepts presented
here are crucial for the subsequent analysis of the channel.

The receiver of an API system is composed of the counterpart blocks
of the transmitter (Fig. 1): (\emph{i}) a \emph{demodulator}, which
transforms the optical waves at the channel output into a specific
anbit $\bigl|\phi_{j}\bigr\rangle$ ($j=1,\ldots,N$); (\emph{ii})
a \emph{decoder}, which maps each anbit $\bigl|\phi_{j}\bigr\rangle$
onto a random symbol $y_{j}$; and (\emph{iii}) a \emph{recipient
source}, which retrieves the user information (i.e. the solution of
the computational problem) from the set of symbols $\left\{ y_{j}\right\} _{j=1}^{N}$,
described by a discrete real random variable $Y$. The set and number
of symbols $\left\{ y_{j}\right\} _{j=1}^{N}$ and anbits $\left\{ \bigl|\phi_{j}\bigr\rangle\right\} _{j=1}^{N}$
allowed at the receiver is determined by the accuracy required to
solve a specific computational problem in APC (mirroring the situation
at the transmitter, where the set and number of emitted symbols and
anbits are likewise selected). Whilst the demodulation block is implemented
using integrated photonic technology to perform an opto-electrical
(O/E) conversion of information, the decoder and recipient source
are integrated electronic circuits that operate under principles analogous
to those governing the encoder and originator source at the transmitter.
Accordingly, we now focus exclusively on the demodulator, whose underlying
principles and hardware differ fundamentally from those of the modulator. 

The transformation of the optical waves at the channel output into
an (electrical) anbit $\bigl|\phi_{j}\bigr\rangle$ is realized at
the demodulator by executing two tasks: (\emph{i}) an \emph{O/E conversion}
and (\emph{ii}) a \emph{signal filtering}. The O/E conversion, based
on direct or coherent detection of light, generates electrical currents
that encode the ideal anbits $\bigl|\phi_{j=1,\ldots,N}\bigr\rangle$
(the states allowed at the receiver) along with system noise and additional
perturbations arising from the non-ideal behavior of PIP devices (Subsection
2.2.1). The noise sources in a PIP platform include laser noise, thermal
noise from phase shifters, amplified spontaneous emission noise from
optical amplifiers, and both shot and thermal noise from photodiodes
\cite{key-P32}. The non-ideal behavior of PIP components emerges
from manufacturing imperfections \cite{key-P31}. The combination
of both physical impairments \textendash{} noise and non-ideal device
operation \textendash{} induces random perturbations in the EDFs of
the ideal anbits, giving rise to \emph{noisy anbits} (denoted $\bigl|\phi\bigr\rangle$),
whose possible locations within the GBS are visualized as distinct
three-dimensional (3D) regions. Such 3D regions define the analog
constellation at the output of the O/E converter (Fig. 1). The successive
signal-filtering task recovers the ideal anbits $\bigl|\phi_{j}\bigr\rangle$
from the noisy anbits $\bigl|\phi\bigr\rangle$ using one of the following
mutually exclusive strategies: \emph{anbit estimation} or \emph{anbit
measurement} (Subsection 2.2.2). Whilst anbit estimation calculates
the ideal anbits from the noisy anbits by using estimation theory
\cite{key-P45}, anbit measurement mimics a digital measurement by
defining decision regions within the received constellation to identify
the ideal anbits \textendash{} similar to the decision regions employed
in digital constellations \cite{key-P46}. The level of noise present
in the constellation determines the optimal signal-filtering strategy.
In the next subsections, we describe in detail the demodulation process
used to recover the ideal anbits $\bigl|\phi_{j}\bigr\rangle$ from
the received optical field $\mathbf{E}_{\mathrm{out}}$ at the channel
output. 

\subsubsection*{2.2.1 Opto-electrical conversion}

\noindent Firstly, the O/E converter transforms the field $\mathbf{E}_{\mathrm{out}}$,
which propagates two complex envelopes $\phi_{0}$ and $\phi_{1}$,
into a noisy (electrical) anbit $\bigl|\phi\bigr\rangle=\phi_{0}\bigl|0\bigr\rangle+\phi_{1}\bigl|1\bigr\rangle$
(Fig. 4). Consistent with the terminology defined in ref.\,\cite{key-P39},
this task is termed \emph{coherent} or \emph{differential} O/E conversion,
depending on whether $\mathbf{E}_{\mathrm{out}}$ is coherently or
directly detected. The former, whose circuitry and functionality are
detailed in ref.\,\cite{key-P39}, can recover the individual moduli
and phases of $\phi_{0}$ and $\phi_{1}$, leading to an anbit $\bigl|\phi\bigr\rangle$
with 4 EDFs, which are indispensable for computing complex matrices.
The latter provides information about $\left|\phi_{0}\right|$, $\left|\phi_{1}\right|$,
and the differential phase $\varphi$ between $\phi_{0}$ and $\phi_{1}$,
resulting in an anbit $\bigl|\phi\bigr\rangle=\left|\phi_{0}\right|\bigl|0\bigr\rangle+e^{\mathrm{j}\varphi}\left|\phi_{1}\right|\bigl|1\bigr\rangle$
with 3 EDFs, sufficient to solve computational problems based on real
matrices. In this work, we take a closer look at the differential
O/E conversion since it is the most energy-efficient option in a PIP
platform.

In particular, we present two basic hardware designs for implementing
differential O/E conversion: an \emph{unbalanced} architecture {[}Fig.
4(a){]} and a \emph{quadrature} architecture {[}Fig. 4(b){]}. Their
functionality is simple. The field $\mathbf{E}_{\mathrm{out}}$ is
composed of two complex envelopes $\phi_{0}$ and $\phi_{1}$, with
a phase shift $\varphi$ between them. In both schemes, the moduli
$\left|\phi_{0}\right|$ and $\left|\phi_{1}\right|$ are recovered
via the photocurrents $I_{0}\propto\left|\phi_{0}\right|^{2}$ and
$I_{1}\propto\left|\phi_{1}\right|^{2}$. Moreover, the differential
phase $\varphi$ is retrieved from an interference between $\phi_{0}$
and $\phi_{1}$. In the unbalanced design, the interference is induced
with a 50:50 beam combiner (a multi-mode interferometer) and converted
into the electrical domain using an unbalanced PIN photodiode that
generates the photocurrent $I_{\varphi}\propto\sin\varphi$. This
design suffices to experimentally demonstrate the principles of API
in Materials and Methods utilizing a minimum number of hardware components,
but only provides access to half of the GBS since the \textquotedblleft sine\textquotedblright{}
and \textquotedblleft cosine\textquotedblright{} components of the
differential phase are not simultaneously recovered. Nonetheless,
the complete GBS must be reconstructed to solve computational problems
in APC. The quadrature architecture circumvents this limitation by
inducing the interference between $\phi_{0}$ and $\phi_{1}$ with
a $90^{\circ}$ optical hybrid and recovering the differential phase
from the photocurrents $I_{\varphi,\mathrm{I}}\propto\cos\varphi$
and $I_{\varphi,\mathrm{Q}}\propto\sin\varphi$. A more detailed analysis
of both differential O/E converters is provided in Supplementary Note
4. 

\subsubsection*{2.2.2 Signal filtering: anbit estimation \emph{vs} anbit measurement}

\noindent Secondly, a signal-filtering task is required to extract
the ideal anbit $\bigl|\phi_{j}\bigr\rangle$ from the noisy anbit
$\bigl|\phi\bigr\rangle$ obtained at the output of the O/E converter.
As mentioned above, this can be accomplished through anbit estimation,
based on estimation theory \cite{key-P45}, or anbit measurement,
grounded in decision theory \cite{key-P46}. In scenarios where the
noisy anbits do not overlap in the GBS (i.e., when the 3D regions
defining their possible locations in the received constellation are
disjoint), anbit estimation becomes the optimal signal-filtering strategy
due to its simplicity. Contrariwise, in presence of overlapping among
the noisy anbits in the GBS, anbit measurement provides the most accurate
signal-filtering strategy {[}Fig. 5(a){]}. We discuss these two approaches
in detail. 

\newpage{}

\emph{Anbit estimation} approximates the ideal anbit $\bigl|\phi_{j}\bigr\rangle$
by the expectation vector $\bigl|\overline{\phi}_{j}\bigr\rangle$
of the noisy anbit $\bigl|\phi\bigr\rangle$, which can be regarded
as a 3D random vector composed of the continuous real random variables
$r$, $\theta$, and $\varphi$ (the EDFs of $\bigl|\phi\bigr\rangle$).
In this vein, the EDFs of $\bigl|\overline{\phi}_{j}\bigr\rangle$
are estimated as $\overline{r}_{j}=\widehat{\mathrm{E}}\left(r\right)$,
$\overline{\theta}_{j}=\widehat{\mathrm{E}}\left(\theta\right)$,
and $\overline{\varphi}_{j}=\widehat{\mathrm{E}}\left(\varphi\right)$,
where $\widehat{\mathrm{E}}$ is the expectation operator \cite{key-P45}.
This procedure requires prior evaluation of the probability density
function (pdf) of each random variable, which can be inferred from
the photocurrents at the output of the O/E converter (see Supplementary
Note 4). The error associated with an anbit estimation is characterized
by the deviation between $\bigl|\overline{\phi}_{j}\bigr\rangle$
and $\bigl|\phi_{j}\bigr\rangle$ using the GBS distance (or any state-comparative
parameter defined in Supplementary Note 3). 

\emph{Anbit measurement} selects the ideal anbit $\bigl|\phi_{j}\bigr\rangle$
from the discrete set $\left\{ \bigl|\phi_{1}\bigr\rangle,\ldots,\bigl|\phi_{N}\bigr\rangle\right\} $
by performing a decision on the noisy anbit $\bigl|\phi\bigr\rangle$
that should minimize the error probability of recovering an incorrect
symbol $y_{j}$ at the receiver. To achieve this, we should first
derive an expression to calculate such error probability or \emph{Symbol
Error Rate} (SER). In most computational problems of APC, which require
matrix inversion operations, the channel is composed of reversible
gates, described by non-singular matrices \cite{key-P39}. This scenario
is termed \emph{bijective channels}, as there is a one-to-one correspondence
between the emitted anbit $\bigl|\psi_{i}\bigr\rangle$ and the ideal
anbit $\bigl|\phi_{j}\bigr\rangle$ that should be received, which
can be modeled by using the same subindex for simplicity $\bigl|\psi_{i}\bigr\rangle\overset{1:1}{\leftrightarrow}\bigl|\phi_{i}\bigr\rangle$
(note that an equal number of anbits is transmitted and received,
$M=N$). Therefore, the SER can be calculated as the complementary
probability of error-free transmission $\textrm{SER}=1-\sum_{i}p\left(x_{i},y_{i}\right)$
(non-bijective channels integrating non-reversible gates are discussed
in Supplementary Note 5). In this landscape, an anbit measurement
should minimize the SER by optimizing a set of decision regions $D_{i}$
($i=1,\ldots,N$) in the received analog constellation using the \emph{maximum
a posteriori probability} (MAP) criterion \cite{key-P46}, as detailed
below. 

However, the MAP decision rule cannot be directly applied in the GBS,
as this geometric representation does not preserve the linear perturbation
induced by an additive noise $\left|n\right\rangle $ on the anbits
$\bigl|\phi_{i}\bigr\rangle$. In particular, the position vector
of a noisy anbit of the form $\bigl|\phi\bigr\rangle=\bigl|\phi_{i}\bigr\rangle+\left|n\right\rangle $
cannot be calculated by summing the position vectors of $\bigl|\phi_{i}\bigr\rangle$
and $\left|n\right\rangle $ (this can be verified through a basic
example, for instance, the Bloch vector of $\left|0\right\rangle +\left|1\right\rangle $
is not equal to the sum of the Bloch vectors of $\left|0\right\rangle $
and $\left|1\right\rangle $). This entails a significant limitation
of the GBS in the context of API, particularly given that the dominant
noise sources in PIP systems are additive in nature (Supplementary
Note 6). Notably, this issue is addressed by representing the anbits
in a vector space $\mathbb{S}$ that must fulfill the following condition:
the position vector $\mathbf{r}$ of a noisy anbit $\bigl|\phi\bigr\rangle=\bigl|\phi_{i}\bigr\rangle+\left|n\right\rangle $
must satisfy that $\mathbf{r}=\mathbf{r}_{i}^{\prime}+\mathbf{n}$,
where $\mathbf{r}_{i}^{\prime}$ and $\mathbf{n}$ are respectively
the position vectors of $\bigl|\phi_{i}\bigr\rangle$ and $\left|n\right\rangle $
in $\mathbb{S}$. In general, a suitable (but not unique) vector space
is $\mathbb{S}\equiv\mathbb{R}^{3}$, where the GBS representation
corresponds to the \emph{half-angle GBS}, see Fig. 5(b) and Supplementary
Note 6 for more details. The half-angle GBS is a fundamental tool
in API since it concurrently simplifies the optimization of the anbit
measurement and the channel capacity in most practical scenarios. 

Specifically, the decision regions that minimize the SER using MAP
decision should be defined as (Supplementary Note 5):
\begin{equation}
D_{i}\coloneqq\left\{ \mathbf{r}\in\mathbb{S}/\ p_{i}f\left(\mathbf{r}|\mathbf{r}_{i}^{\prime}\right)>p_{j}f\left(\mathbf{r}|\mathbf{r}_{j}^{\prime}\right),\ \forall j\in\left\{ 1,\ldots,N\right\} /j\neq i\right\} .\label{eq:3}
\end{equation}
If the position vector $\mathbf{r}$ of the noisy anbit $\bigl|\phi\bigr\rangle$
belongs to $D_{i}$, then the outcome of the measurement is the anbit
$\bigl|\phi_{i}\bigr\rangle$, and the SER is given by the expression:

\newpage{}

\begin{equation}
\textrm{SER}=1-\sum_{i=1}^{N}p_{i}\int_{D_{i}}f\left(\mathbf{r}|\mathbf{r}_{i}^{\prime}\right)\mathrm{d}^{3}r,\label{eq:4}
\end{equation}
where $p_{i}=p\left(x_{i}\right)$ is the pmf of the originator source
and $f\left(\mathbf{r}|\mathbf{r}_{i}^{\prime}\right)$ is the \emph{conditional
pdf} accounting for the probability distribution of the noisy anbit
$\bigl|\phi\bigr\rangle$ in the $\mathbb{S}$-space when $x_{i}$
is the symbol emitted by the transmitter and $\bigl|\phi_{i}\bigr\rangle$
(or $\mathbf{r}_{i}^{\prime}$) is the ideal anbit that should be
measured. In this context, $f\left(\mathbf{r}|\mathbf{r}_{i}^{\prime}\right)$
is determined by the statistical properties of the system\textquoteright s
physical impairments \textendash{} noise and the non-ideal behavior
of PIP circuits. As discussed in Supplementary Note 6, the main noise
sources induce zero-mean random fluctuations in the EDFs of $\bigl|\phi_{i}\bigr\rangle$,
while hardware non-idealities introduce a constant perturbation in
$\bigl|\phi_{i}\bigr\rangle$. Hence, the combined effect of the dominant
noise sources and the non-ideal operation of PIP devices can be modeled
as an additive noise state $\left|n\right\rangle $ inducing random
perturbations on the moduli and differential phase of $\bigl|\phi_{i}\bigr\rangle$,
with the expectation value of $\left|n\right\rangle $ accounting
for the hardware imperfections. Under this assumption, the noisy anbit
can be expressed as $\bigl|\phi\bigr\rangle=\bigl|\phi_{i}\bigr\rangle+\left|n\right\rangle $
and the conditional pdf becomes $f\left(\mathbf{r}|\mathbf{r}_{i}^{\prime}\right)=f_{\mathbf{N}}\left(\mathbf{n}=\mathbf{r}-\mathbf{r}_{i}^{\prime}\right)$,
where $f_{\mathbf{N}}\left(\mathbf{n}\right)$ denotes the random
distribution of $\left|n\right\rangle $ in the $\mathbb{S}$-space.
Alternative procedures exist for describing $\bigl|\phi\bigr\rangle$
\textendash{} particularly in cases where information is encoded in
a single EDF \textendash{} but yield the same expression for the conditional
pdf (see Supplementary Note 6). 

The result established in Eq. (4) leads us to formulate the \emph{anbit
measurement theorem}: ``\emph{If the constellation at the output
of the O/E conversion is composed of non-overlapping noisy anbits,
then there exists a set of decision regions} $\left\{ D_{1},\ldots,D_{N}\right\} $
\emph{that ensures a zero SER}''. The proof of this theorem is straightforward.
If the 3D regions defined by the noisy anbits in the received constellation
are not overlapped, then the conditional pdfs are disjoint. Consequently,
it is always possible to define a set of decision regions $\left\{ D_{1},\ldots,D_{N}\right\} $
fulfilling the condition $\int_{D_{i}}f\bigl(\mathbf{r}|\mathbf{r}_{i}^{\prime}\bigr)\mathrm{d}^{3}r=1$
(for all $i=1,\ldots,N$), which inserted into Eq. (4) gives rise
to the sought result ($\textrm{SER}=0$). Remarkably, the anbit measurement
theorem also implies that we can perfectly distinguish between non-orthogonal
anbits $\bigl|\phi_{i}\bigr\rangle$ and $\bigl|\phi_{j}\bigr\rangle$,
provided that $D_{\mathrm{GBS}}\left(\bigl|\phi_{i}\bigr\rangle,\bigl|\phi_{j}\bigr\rangle\right)\neq0$,
since classical superposition is not annihilated in an anbit measurement.
This represents a crucial difference with QI, where orthogonality
is a necessary and sufficient condition for error-free discrimination
between two quantum states \cite{key-P5}. Furthermore, note that
an anbit measurement involves a vector-space optimization problem,
whereas a quantum measurement requires solving an intricate matrix-based
optimization problem \cite{key-P5}. 

In order to gain insight into the theory of anbit measurement, we
now examine a basic example. Consider an originator source $X$ that
emits two equiprobable symbols $x_{1}$ and $x_{2}$, which are encoded
into the anbits:
\begin{equation}
\bigl|\psi_{1}\bigr\rangle=\cos\frac{\theta}{2}\left|0\right\rangle +\sin\frac{\theta}{2}\left|1\right\rangle ,\ \ \ \bigl|\psi_{2}\bigr\rangle=\cos\frac{\theta}{2}\left|0\right\rangle -\sin\frac{\theta}{2}\left|1\right\rangle ,\label{eq:5}
\end{equation}
with $0<\theta\leq\pi/2$. The states are non-orthogonal for all values
of the EDF $\theta$, except when $\theta=\pi/2$, see Fig.\,5(c).
These anbits are transmitted through a channel that does not execute
any computational operation, that is, the ideal anbits to be measured
are $\bigl|\phi_{i}\bigr\rangle=\bigl|\psi_{i}\bigr\rangle$\linebreak{} ($i=1,2$).
Nevertheless, the system introduces an additive noise modeled by a
ket of the form $\left|n\right\rangle =n_{0}\left|0\right\rangle +n_{1}\left|1\right\rangle $,
where $n_{0}$ and $n_{1}$ are independent and identically distributed
Gaussian random variables with zero mean and variance $\sigma^{2}$.
Such considerations about the system noise are consistent with the
dominant noise sources identified in passive linear PIP circuits (see
Supplementary Note 6). As a result, the noisy anbits at the output
of the O/E converter take the form $\bigl|\phi\bigr\rangle=\bigl|\phi_{i}\bigr\rangle+\left|n\right\rangle $,
which define the received analog constellation, represented in the
half-angle GBS, as shown in Fig. 5(d).

In this example, the anbit measurement is optimized by designing decision
regions ($D_{1}$ and $D_{2}$) in the half-angle GBS using the MAP
decision rule {[}Eq. (3){]}. In this geometric representation, the
anbits $\bigl|\phi\bigr\rangle$, $\bigl|\phi_{1}\bigr\rangle$, $\bigl|\phi_{2}\bigr\rangle$,
and the noise ket $\left|n\right\rangle $ are respectively described
by the position vectors (we use Cartesian coordinates): $\mathbf{r}=\left(x,y,z\right)$,
$\mathbf{r}_{1}^{\prime}=\left(\sin\theta/2,0,\cos\theta/2\right)$,
$\mathbf{r}_{2}^{\prime}=\left(-\sin\theta/2,0,\cos\theta/2\right)$,
and $\mathbf{n}=\left(n_{1},0,n_{0}\right)$. Next, using the noise
distribution $f_{\mathbf{N}}\left(\mathbf{n}\right)$, obtained as
the product of the marginal pdfs of $n_{0}$ and $n_{1}$:
\begin{equation}
f_{\mathbf{N}}\left(\mathbf{n}\right)=\frac{1}{2\pi\sigma^{2}}\exp\left(-\frac{n_{0}^{2}+n_{1}^{2}}{2\sigma^{2}}\right),\label{eq:6}
\end{equation}
we derive the conditional pdfs $f\left(\mathbf{r}|\mathbf{r}_{i}^{\prime}\right)=f_{\mathbf{N}}\left(\mathbf{n}=\mathbf{r}-\mathbf{r}_{i}^{\prime}\right)$
required to determine the optimal decision regions. In this case,
the regions that minimize the SER are $D_{1}=\left\{ \mathbf{r}\in\mathbb{R}^{3}/x>0\right\} $
and $D_{2}=\left\{ \mathbf{r}\in\mathbb{R}^{3}/x<0\right\} $, see
Fig. 5(e). Substituting into Eq. (4), the SER reduces to the closed-form
expression:
\begin{equation}
\textrm{SER}=\frac{1}{2}\left[1-\mathrm{erf}\left(\frac{1}{\sqrt{2}\sigma}\sin\frac{\theta}{2}\right)\right]=\frac{1}{2}\mathrm{erfc}\left(\frac{1}{\sqrt{2}\sigma}\sin\frac{\theta}{2}\right),\label{eq:7}
\end{equation}
where erf is the error function, defined as $\mathrm{erf}\left(z\right)\coloneqq\left(2/\sqrt{\pi}\right)\int_{0}^{z}e^{-w^{2}}\mathrm{d}w$,
and erfc is the complementary error function, $\mathrm{erfc}\left(z\right)\coloneqq1-\mathrm{erf}\left(z\right)$
\cite{key-P47}.

Figure 5(f) shows the SER as a function of $\theta$, for two cases:
$\sigma=0$ (noiseless channel) and $\sigma=0\textrm{.}22$ (a noisy
channel with a noise standard deviation intentionally set much higher
than the experimental value, $\sigma\sim10^{-3}$ as detailed in Materials
and Methods, deliberately chosen for didactic purposes to highlight
the effect of noise). In the absence of noise, the SER is zero for
$\theta>0$, as there is no overlap between states in the constellation.
In contrast, under noisy conditions, the maximum SER occurs as $\theta\rightarrow0$,
since the overlap between noisy anbits increases as their separation
decreases. Conversely, the SER drops to zero as $\theta\rightarrow\pi/2$,
because the noisy anbits become fully distinguishable in the received
constellation {[}Fig. 5(d){]}, in agreement with the anbit measurement
theorem. Therefore, consistent with the encoder design principles
introduced in Subsection 2.1, an optimal analog constellation mitigates
computational errors by minimizing the error probability at the receiver.

For completeness, we also compare these results in API with the SER
obtained in QI when considering a noiseless quantum channel where
the emitted quantum states are the same as the classical states of
this example {[}see Fig. 5(f); theory detailed in Supplementary Note
8{]}. It is worth highlighting that, in a noiseless channel, the SER
in API is substantially lower than the SER in QI when emitting non-orthogonal
states. This difference stems from the fact that, in API, state superposition
is preserved after measurement, whereas in QI it is annihilated, precluding
the possibility of distinguishing with certainty between non-orthogonal
qubits. Likewise, the SER in API with noisy conditions ($\sigma=0\textrm{.}22$
) can also be found lower than the SER in QI with noiseless conditions.
This result constitutes a first theoretical proof that APC exhibits
greater tolerance to noise-induced errors than the quantum computing
paradigm, thereby reducing the constraint of introducing a large number
of redundant units of information for error correction when scaling
computational systems. 

\subsection*{2.3 Anbit channel}

\noindent In simple API systems, the channel is composed of single-anbit
gates, which propagate and compute a sequence of individual anbits
between the transmitter and receiver using PIP circuitry (Fig. 1).
In this subsection, we discuss how to quantify the \emph{channel capacity},
that is, the maximum average amount of information that can be propagated
and computed without errors by a single-anbit gate. 

Bearing in mind that API only deals with classical information, the
definition of the channel capacity ($C$) is provided by Shannon\textquoteright s
theory \cite{key-P1,key-P4}: the maximum mutual information (in bits)
between the originator and recipient sources $X$ and $Y$, optimized
over all possible pmfs of $X$. Consequently, within the API framework,
Shannon\textquoteright s channel capacity retains its general properties
\textendash{} positivity, continuity, and uniqueness \textendash{}
and the channel-coding theorem likewise remains valid, identifying
$C$ as the limit on the maximum amount of information generated by
$X$ that can be reliably transmitted over the channel \cite{key-P1}.
See Supplementary Note 7 for further discussion of the general properties
of $C$ and the channel-coding theorem in the context of API. 

However, the question of how to calculate the channel capacity in
simple API systems remains open. Specifically, in API, the evaluation
of the channel capacity requires different approaches depending on
whether anbit estimation or anbit measurement is employed at the receiver,
as information recovery relies on distinct signal-filtering strategies,
thereby affecting the mutual information between $X$ and $Y$.

API systems based on \emph{anbit estimation} may be described via
a relation between the originator and recipient sources of the form
$Y=g\left(X\right)+\mathcal{N}$, with the $g$ function accounting
for the computational operation of the channel and the random variable
$\mathcal{N}$ modeling an additive Gaussian noise, which aligns with
our discussions about noise in Subsection 2.2 and Supplementary Note
6. Under these conditions, as demonstrated in Supplementary Note 7,
the channel capacity is governed by the \emph{Shannon-Hartley theorem}
\cite{key-P1}: $C\left(\textrm{bits}\right)\leq0\textrm{.}5\log_{2}\left(1+\sigma_{X}^{2}/\sigma_{\mathcal{N}}^{2}\right)$,
where $\sigma_{X}^{2}$ and $\sigma_{\mathcal{N}}^{2}$ are the variances
of $X$ and $\mathcal{N}$, respectively. The equality holds for bijective
channels (reversible gates), while the inequality applies to non-bijective
channels (irreversible gates).

In API systems based on \emph{anbit measurement}, although the channel
capacity may also be examined through the Shannon-Hartley theorem,
it does not capture the influence of the decision regions utilized
to minimize the SER. In order to include the impact of the anbit measurement
procedure on the channel capacity, the mutual information between
$X$ and $Y$ should be calculated incorporating the decision regions.
This gives rise to an expression of the channel capacity in bijective
channels of the form (see Supplementary Note 7, where non-bijective
channels are also discussed):
\begin{equation}
C=\max_{p_{i},\mathbf{r}_{i}^{\prime},D_{j}}\left\{ \sum_{i,j=1}^{N}p_{i}\int_{D_{j}}f\left(\mathbf{r}|\mathbf{r}_{i}^{\prime}\right)\mathrm{d}^{s}r\log_{2}\frac{\int_{D_{j}}f\left(\mathbf{r}|\mathbf{r}_{i}^{\prime}\right)\mathrm{d}^{s}r}{\int_{D_{j}}f\left(\mathbf{r}\right)\mathrm{d}^{s}r}\right\} \ \ \ \textrm{(bits)},\label{eq:8}
\end{equation}
with $f\left(\mathbf{r}\right)=\sum_{k}p_{k}f\left(\mathbf{r}|\mathbf{r}_{k}^{\prime}\right)$.
Multiplying the above expression by the BAR parameter, the channel
capacity can equivalently be expressed in anbits. Furthermore, incorporating
the Nyquist sampling rate \cite{key-P1,key-P46}, $C$ may be quantified
in bits per second or in anbits per second. \newpage{}

As detailed in Eq. (8), maximizing the mutual information in anbit-based
measurement systems requires optimizing not only the pmf $\left\{ p_{1},\ldots,p_{N}\right\} $
of $X$, but also the received constellation $\left\{ \mathbf{r}_{1}^{\prime},\ldots,\mathbf{r}_{N}^{\prime}\right\} $
and the decision regions $\left\{ D_{1},\ldots,D_{N}\right\} $. In
most practical scenarios, channel capacity is achieved by utilizing
a uniform pmf, selecting a received constellation that minimizes overlap
among the noisy anbits at the output of the O/E converter, and defining
decision regions that minimize the SER. Along these lines, note that
the received constellation can be optimized either directly, through
appropriate selection of the ideal anbits $\bigl|\phi_{i}\bigr\rangle$
(or $\mathbf{r}_{i}^{\prime}$) allowed in the GBS (or half-angle
GBS) at the receiver, or indirectly, by optimizing the transmitted
constellation to minimize anbit overlap at the output of the O/E converter. 

On the other hand, by treating the pmf of $X$ and the received constellation
as fixed parameters in Eq. (8), the mutual information is optimized
exclusively through the measurement process, by determining the decision
regions that minimize the SER. This provides insights into the maximum
average amount of information that can be recovered at the receiver,
a concept referred to as \emph{accessible information} by analogy
with the terminology employed in QI \cite{key-P48}. 

In API, the upper bound of accessible information may be easily explored
via a noiseless channel. In such conditions, the received anbits do
not overlap (regardless of the constellation employed) and, therefore,
one can always define a set of decision regions satisfying that $\int_{D_{j}}f\bigl(\mathbf{r}|\mathbf{r}_{i}^{\prime}\bigr)\mathrm{d}^{3}r=\delta_{ij}$
(the Kronecker delta) by virtue of the anbit measurement theorem (Subsection
2.2). As a result, Eq. (8) reduces to $C=\max\left\{ H\left(X\right)\right\} =\log_{2}N$
(bits), which represents the upper bound of accessible information
in API. Outstandingly, the same bound emerges even for noisy channels,
provided that there is no overlap in the received constellation, as
the anbit measurement theorem ensures the condition $\int_{D_{j}}f\bigl(\mathbf{r}|\mathbf{r}_{i}^{\prime}\bigr)\mathrm{d}^{3}r=\delta_{ij}$
still holds. 

As seen, the upper bound of accessible information in API is determined
by the logarithm of the number of received anbits or symbols (the
Shannon bound). In contrast, in QI, this limit is established by the
Holevo bound ($\chi$), which is given by the logarithm of the dimension
of the Hilbert space when encoding into uniformly distributed, mutually
orthogonal pure quantum states \cite{key-P5,key-P48}. This implies
that the upper bound of accessible information in single-anbit systems
with $N>2$ exceeds the Holevo bound in single-qubit systems ($\chi=\log_{2}2=1$
bit), as verified experimentally in Materials and Methods. This finding
emerges from the ability to distinguish non-orthogonal states with
certainty in API. 

As a basic example of channel capacity, we revisit the scenario proposed
in Eq. (5), where an anbit measurement was optimized in an API system
emitting two equiprobable anbits. Now, we aim to calculate the channel
capacity of this system using Eq. (8) along with the decision regions
depicted in Fig. 5(e) and the conditional pdfs $f\bigl(\mathbf{r}|\mathbf{r}_{i}^{\prime}\bigr)$
provided below Eq.\,(6). After some algebraic manipulation, we find
that (see Supplementary Note 8):
\begin{equation}
C=\frac{1}{2}\sum_{k=1}^{2}\mathrm{erfc}\left(\frac{\left(-1\right)^{k}}{\sqrt{2}\sigma}\sin\frac{\theta}{2}\right)\log_{2}\left[\mathrm{erfc}\left(\frac{\left(-1\right)^{k}}{\sqrt{2}\sigma}\sin\frac{\theta}{2}\right)\right]\ \ \ \textrm{(bits)}.\label{eq:9}
\end{equation}
Figure 6 shows the channel capacity given by the above equation considering
the same cases analyzed when computing the SER in Fig. 5(f): $\sigma=0$
(noiseless channel) and $\sigma=0\textrm{.}22$ (noisy channel). In
the \emph{noiseless} case, the channel capacity is $C=\log_{2}2=1$
bit for $\theta>0$, since there is no overlap between anbits ($\textrm{SER}=0$).
This is consistent with the upper bound of the accessible information
(the Shannon bound). In the \emph{noisy} case, the minimum channel
capacity ($C\rightarrow0$ bits) takes place when the error probability
maximizes ($\textrm{SER}\rightarrow1/2$), since the anbits become
completely indistinguishable when the distance between them drops
to zero ($\theta\rightarrow0$). Contrariwise, the channel capacity
maximizes to $C=1$ bit when $\textrm{SER}=0$. This corresponds to
the optimal arrangement of the analog constellation ($\theta=\pi/2$),
which reduces computational errors by concurrently minimizing the
SER and maximizing $C$, as postulated in the encoder design principles
(Subsection 2.1).

In addition, we compare these results of channel capacity in API with
those obtained in QI when emitting the same states {[}Eq. (5){]} through
a noiseless quantum channel (Supplementary Note 8). As depicted in
Fig. 6, in noiseless conditions, the channel capacity in API is significantly
higher than in QI when emitting non-orthogonal states, as state superposition
is preserved after measurement only in API. This is also the underlying
reason that explains why the channel capacity in API with noisy conditions
($\sigma=0\textrm{.}22$) is higher than the channel capacity in QI
with noiseless conditions when information is encoded into the same
states of the Bloch sphere. These results emphasize the robustness
of APC against system noise compared to the quantum computing paradigm.

\section*{3 Materials and Methods }

\noindent In this section, we experimentally demonstrate the fundamental
principles of API. We validate three key results: (1) the generation
and reception of different classes of analog constellations in the
GBS, (2) the characterization of perturbations induced by system noise
and non-ideal behavior of PIP devices on the anbits, and (3) the empirical
verification of the mathematical framework developed to calculate
the SER and channel capacity in a transmission of multiple anbits.

The experiments are conducted using the laboratory setup shown in
Fig. 7(a). A tunable continuous-wave external cavity laser (TUNICS
T100S-HP), operating at 1550 nm with a linewidth of 400 kHz, is connected
to a PIP circuit comprising a SAM hardware (Fig. 3), a universal single-anbit
U-gate \cite{key-P39}, and an unbalanced differential O/E converter
{[}Fig. 4(a){]}. For simplicity, and without loss of generality in
demonstrating the API principles, the gate is programmed as the identity
matrix, representing a bijective channel. Photocurrents generated
by the O/E converter are captured by a signal processing module, which
subsequently performs either anbit estimation or anbit measurement.
The PIP circuit was fabricated by Advanced Micro Foundry using a silicon-on-insulator
platform. A micrograph of the fabricated chip is shown in Fig. 7(b)
(see Supplementary Note 9 for further details of the manufacturing
process). 

Figure 7(c) illustrates diverse classes of analog constellations that
have been generated by varying a single EDF of the anbits. These constellations
are observed at the output of the O/E converter. As discussed in Subsection
2.2, the received anbits (blue points) are perturbed by system noise
\textendash{} inducing zero-mean random fluctuations in the EDFs (red
points) \textendash{} and by non-idealities of PIP devices \textendash{}
which introduce a constant perturbation in the EDFs. In these constellations,
the impact of hardware imperfections is negligible, as the mean value
of the received anbits closely matches the ideal anbits that are expected
to be received.

An effective means of revealing the non-ideal behavior of PIP devices
is to generate the anbits corresponding to the poles of the GBS and
perform an estimation of the received states. As shown in Fig.\LyXThinSpace 7(d),
the mean value of the anbits obtained at the output of the O/E converter
(blue points) deviate from their ideal locations $\left|0\right\rangle $
and $\left|1\right\rangle $. This deviation is primarily caused by
the tunable basic unit in the SAM hardware (Fig. 3), whose finite
extinction ratio prevents light from being fully confined to a single
waveguide \textendash{} an ideal scenario corresponding to the anbit
$\left|0\right\rangle $ (upper waveguide) or $\left|1\right\rangle $
(lower waveguide). Additionally, it is worth noting that the error
in the mean value of the elevation angle $\theta$ differs between
the standard anbits $\left|0\right\rangle $ and $\left|1\right\rangle $,
since the tunable basic unit exhibits different extinction ratios
(ER) in its bar and cross configurations ($\mathrm{ER}_{\mathrm{bar}}=-36$
dB, $\mathrm{ER}_{\mathrm{cross}}=-38$ dB). This finding suggests
that the standard anbits may not constitute the most suitable vector
basis for solving certain computational problems in APC.

Next, we evaluate the SER and channel capacity (accessible information)
in a transmission of $M$ equiprobable anbits located on the equator
of the GBS with a differential phase ranging from 0.78 rad to 0.99
rad. Specifically, the anbits $\left|\psi_{i}\right\rangle $ transmitted
through the channel \textendash{} and thus the ideal anbits $\left|\phi_{i}\right\rangle $
that should be measured \textendash{} are:
\begin{equation}
\bigl|\psi_{i}\bigr\rangle=\frac{1}{40}\left(\left|0\right\rangle +e^{\mathrm{j}\varphi_{i}}\left|1\right\rangle \right)\equiv\bigl|\phi_{i}\bigr\rangle,\ \ \ \left(i=1,\ldots,M\right)\label{eq:10}
\end{equation}
with $\varphi_{i}=\varphi_{1}+\left(i-1\right)\Delta\varphi/\left(M-1\right)$,
$\varphi_{1}=0\textrm{.}78$ rad, $\varphi_{M}=0\textrm{.}99$ rad,
and $\Delta\varphi=\varphi_{M}-\varphi_{1}=0\textrm{.}21$ rad. The
optical power required to implement each anbit is $\mathcal{P}=\left(1/40\right)^{2}\ \textrm{W}\equiv-2$
dBm. As the number of transmitted anbits ($M$) increases, the overlap
between the states at the output of the O/E converter ($\left|\phi\right\rangle $)
grows due to system noise {[}Fig.\LyXThinSpace 7(e){]}. Since these
noisy anbits $\left|\phi\right\rangle $ overlap, anbit measurement
is the most accurate signal-filtering strategy for the receiver. Here,
the goal is to optimize the anbit measurement and the resulting channel
capacity.

Following the general procedure detailed in Subsection 2.2, the anbit
measurement problem may be formulated by describing the system\textquoteright s
physical impairments via the ket $\left|n\right\rangle =\left|\phi\right\rangle -\left|\phi_{i}\right\rangle $.\linebreak{}
This approach leads to a 3D optimization problem. Nonetheless, in
this particular case, information is encoded onto a single EDF (the
differential phase). Consequently, the optimization problem can be
reduced to one dimension using an alternative procedure to describe
$\left|\phi\right\rangle $ (Supplementary Note 6). As inferred from
Figs.\LyXThinSpace 7(e, f), the noisy anbits may be represented with
an arbitrary state of the form $\left|\phi\right\rangle =\left(1/40\right)\left(\left|0\right\rangle +e^{\mathrm{j}\varphi}\left|1\right\rangle \right)$,
where $\varphi=\varphi_{i}+\eta_{i}$ and $\eta_{i}$ is a random
variable accounting for the system\textquoteright s physical impairments,
whose pdf $f_{\mathcal{N}_{i}}\left(\eta_{i}\right)$ can be approximated
by a Gaussian distribution with mean $\mu_{i}\simeq0$ and variance
$\sigma_{i}^{2}\sim10^{-5}$. Along this line, the following noteworthy
observations are in order: (\emph{i}) a null mean $\mu_{i}$ indicates
that the non-idealities of PIP devices can be neglected (consistent
with the theory in Supplementary Note 6); (\emph{ii}) the variance
$\sigma_{i}^{2}$ is of the same order of magnitude for all noisy
anbits, which simplifies subsequent theoretical calculation of SER
and $C$; (\emph{iii}) a suitable vector space $\mathbb{S}$ to optimize
the measurement is $\mathbb{S}\equiv\mathbb{R}$, where the vectors
involved in the measurement process are $\mathbf{r}_{i}^{\prime}=\varphi_{i}\hat{\mathbf{x}}$,
$\mathbf{r}=\varphi\hat{\mathbf{x}}$, and $\mathbf{n}_{i}=\eta_{i}\hat{\mathbf{x}}$;
(\emph{iv}) the position vector $\mathbf{r}$ of $\left|\phi\right\rangle $
satisfies the condition $\mathbf{r}=\mathbf{r}_{i}^{\prime}+\mathbf{n}_{i}$;
(\emph{v}) the conditional pdfs are given by the expression $f\left(\mathbf{r}|\mathbf{r}_{i}^{\prime}\right)=f_{\mathcal{N}_{i}}\left(\eta_{i}=\varphi-\varphi_{i}\right)$,
denoted $f_{i}\left(\varphi\right)$ for short in Fig.\LyXThinSpace 7(f). 

As shown in Fig.\LyXThinSpace 7(f), the optimal decision regions $D_{i}$
that minimize the SER are defined by the intersection points $\chi_{i}$
of the conditional pdfs $f_{i}\left(\varphi\right)$, such that $D_{i}=\left\{ \chi_{i-1}<\varphi<\chi_{i}\right\} $.
By substituting the conditional pdfs and their corresponding decision
regions into Eqs. (4) and (8), we obtain the experimental SER and
channel capacity of this API system, as depicted in Fig.\LyXThinSpace 7(g).
In line with the anbit measurement theorem, the SER remains near zero
for a small number of transmitted anbits ($M\leq7$), as there is
negligible overlap between conditional pdfs (azimuthal separation
between adjacent anbits $\varphi_{i+1}-\varphi_{i}\geq2^{\circ}$).
As $M$ increases and the azimuthal separation reduces beyond this
threshold, the overlap grows, leading to an approximately linear rise
in the SER. Regarding channel capacity, when state overlap is minimal
($M\leq7$, $\varphi_{i+1}-\varphi_{i}\geq2^{\circ}$), $C$ closely
approaches the source entropy, $C\simeq H\left(X\right)=\log_{2}M$
(bits). This supports a key theoretical prediction of API, introduced
in Subsection 2.3: the channel capacity of a noisy API system with
negligible state overlap converges to that of a noiseless API system,
$C\simeq\log_{2}M$, even when the states are non-orthogonal. As a
result, a noisy single-anbit system can readily exceed the channel
capacity (accessible information) of a noiseless single-qubit system,
which is fundamentally limited to 1 bit by the Holevo bound \cite{key-P5,key-P48}.
In contrast, as state overlap increases ($M>7$, $\varphi_{i+1}-\varphi_{i}<2^{\circ}$),
the channel capacity saturates, reaching a maximum of approximately
$C_{\mathrm{max}}=3$ bits.

Alternatively, both the SER and channel capacity can be theoretically
predicted by approximating the random variables $\eta_{i}$ as independent
and identically distributed Gaussian variables with zero mean and
variance $\sigma^{2}$. The zero-mean assumption is justified by the
fact that the mean of $\varphi$ closely matches the target value
$\varphi_{i}$. The assumption of identical variance is supported
by the observation that all conditional pdfs exhibit variances on
the same order, $\sigma^{2}\sim10^{-5}$. Under these assumptions,
the SER and channel capacity can be derived analytically, following
the procedure detailed in Supplementary Note 9:
\begin{align}
\textrm{SER}_{\textrm{theoretical}} & \simeq\textrm{erfc}\left(\frac{\Delta\varphi}{2\sqrt{2}\sigma\left(M-1\right)}\right),\label{eq:11}\\
C_{\textrm{theoretical}} & \simeq\frac{1}{2M}\sum_{i,j=1}^{M}\mathrm{ERF}\left(j,i\right)\log_{2}\frac{M\cdot\mathrm{ERF}\left(j,i\right)}{\sum_{k=1}^{M}\mathrm{ERF}\left(j,k\right)}\ \ \ \textrm{(bits)},\label{eq:12}
\end{align}
where:
\begin{equation}
\mathrm{ERF}\left(j,i\right)=\mathrm{erf}\left(\frac{\Delta\varphi\left(j-i+\nicefrac{1}{2}\right)}{\sqrt{2}\sigma\left(M-1\right)}\right)-\mathrm{erf}\left(\frac{\Delta\varphi\left(j-i-\nicefrac{1}{2}\right)}{\sqrt{2}\sigma\left(M-1\right)}\right).\label{eq:13}
\end{equation}
As shown in Fig.\LyXThinSpace 7(g), these theoretical approximations
closely match experimental results for $\sigma^{2}\equiv4\textrm{.}5\cdot10^{-5}$.

Finally, as inferred from Fig.\LyXThinSpace 7(g), an azimuthal separation
greater than $2^{\circ}$ between adjacent anbits is sufficient to
ensure negligible state overlap in an analog constellation of this
API system. Leveraging this result, we generate 900 non-overlapping
anbits in a GBS with constant radius by varying both the azimuthal
and elevation angles, applying a $6^{\circ}$ separation in each angle
between adjacent states. At the output of the O/E conversion, we perform
anbit estimation on the received constellation, depicted in Fig. 7(h).
This result demonstrates the potential of API, which can readily achieve
a channel capacity of $C=\log_{2}900\simeq10$ bits in a single-anbit
system. 

\section*{4 Discussion}

\noindent This work lays the theoretical foundations of API, a new
information theory conceived to demonstrate the combined potential
of APC and PIP technology for enabling on-chip photonic computing
with exceptional tolerance to errors induced by system noise and imperfections
of optical devices. Our results suggest that extensive error-correction
overhead is not required in APC architectures, thus simplifying their
scalability in the near- and mid-term. Furthermore, the principles
of API demonstrate that APC systems can easily surpass the average
amount of information that may be computed and recovered in basic
quantum computing systems, even in the presence of noise in the PIP
circuits.\newpage{}

Remarkably, the \emph{mitigation of errors} by engineering the units
of information \textendash{} in our case by optimizing \emph{discrete
analog constellations} of anbits in the GBS and their associated signal-filtering
strategies \textendash{} is a central feature of API. This should
be further analyzed by designing constellations tailored to the computational
problems of APC \cite{key-P39}, aiming to simultaneously minimize
computational errors and optical power consumption in the PIP platform
\textendash{} for instance, by maximizing the average GBS distance
in constellations that exploit variations in the three EDFs of the
anbits, or by placing higher-probability anbits on spheres of smaller
radius, in analogy with probabilistic constellation shaping techniques
used in digital coherent communication systems \cite{key-P49}. In
parallel, the GBS and the state-comparative parameters introduced
in this work serve as a technological testbed for characterizing non-ideal
behavior of basic PIP components, such as tunable basic units (lower-error
pole generation correlates with tunable basic units exhibiting higher
extinction rations in both cross and bar configurations).

Compared to existing information theories, API shares both similarities
and differences with DI and QI (Fig. 8). Shannon\textquoteright s
theory, a \emph{universal} classical framework, describes any system
that processes classical information \cite{key-P1}. This suggests
that DI and API may be interpreted as subclasses \textendash{} or
distinct realizations \textendash{} of Shannon\textquoteright s theory,
each operating under specific strategies for the encoder, modulator,
channel, demodulator, and decoder. The primary similarities between
DI and API arise at the originator and recipient sources, as well
as in the measurement process, which relies on decision regions. On
the other hand, QI is a theoretical framework that models information-processing
systems propagating classical or quantum information through a quantum
channel \cite{key-P5}. Hence, Shannon\textquoteright s theory and
QI must coexist when classical transmitters and receivers are connected
via a quantum channel \cite{key-P50}. In this scenario, API might
be regarded as a conceptual link between Shannon\textquoteright s
theory and QI, enabled by similar (but not identical) strategies at
the encoder and modulator, both of which exploit vector superposition
within a Hilbert space. This perspective positions APC as a valuable
didactic toolbox for illustrating the subtle, yet fundamental, distinctions
between classical and quantum computational systems. Nevertheless,
the key differences between API and QI lie in quantum measurement
and quantum entanglement, physical phenomena with no classical counterpart
\cite{key-P5,key-P48,key-P50}.

Interestingly, API theory is not only restricted to PIP-computing
applications but also generalizes this technology to complement fiber-based
\emph{communications} by leveraging an underlying compatibility between
API and DI. Since the signal generated by a digital modulator is a
1D complex analog wave \cite{key-P46}, the output of two digital
modulators (namely $\psi_{0}\left(t\right)$ and $\psi_{1}\left(t\right)$)
may be described by a continuous-time anbit $\left|\psi\left(t\right)\right\rangle =\psi_{0}\left(t\right)\left|0\right\rangle +\psi_{1}\left(t\right)\left|1\right\rangle $
\cite{key-P39}. In this vein, advanced multidimensional digital modulation
schemes significantly enhancing spectral efficiency and data throughput
(e.g., a 2D quadrature amplitude modulation) may be explored by implementing
$\left|\psi\left(t\right)\right\rangle $, for instance, via the polarization-anbit
modulation \cite{key-P39} in a standard single-mode fiber or using
the SAM in a multi-core fiber (among other options). In this context,
combining anbits with optical fiber media could facilitate digital
signal processing, enabling in-fiber operations such as multiplexing
while providing scalable and energy-efficient solutions for data centers
as well as metropolitan and backbone networks. 

Although the present work introduces the foundations of API, substantial
research is still required to complete this information theory. In
future contributions, we will focus on three main directions: 1) analyzing
the distribution of diverse noise sources in the GBS and characterizing
the dominant hardware imperfections; 2) designing advanced O/E converters
capable of real-time correction of these physical impairments; and
3) extending API theory to composite systems, which model channels
composed of multi-anbit gates \cite{key-P39}. In particular, understanding
\emph{composite API systems} will be essential for scaling computational
architectures in APC. This endeavor involves completing the theoretical
principles of the encoder (by generalizing state-comparative parameters
to multiple anbits and incorporating mixed classical states to represent
multiple encoders within a single computational system), the channel
(by developing a mathematical formalism to calculate the channel capacity),
and the decoder (by designing multi-anbit O/E converters and conceiving
the theory of multi-anbit estimation and measurement). Both API and
APC theories must be developed in tandem to unlock the full potential
of PIP technology in tackling advanced computational challenges, blazing
a trail for a paradigm shift in our information society.

\subsection*{Author contributions}

\noindent Andr\'es Macho Ortiz conceived the idea of the new information
theory and developed its mathematical framework. Ra\'ul L\'opez
March and Pablo Mart\'inez Carrasco conducted the experimental work.
Andr\'es Macho Ortiz and Ra\'ul L\'opez March carried out the numerical
simulations. Francisco Javier Fraile-Pel\'aez and Jos\'e Capmany
supervised the work. All authors participated in refining the theory
and preparing the manuscript. 

\subsection*{Disclosures}

\noindent The authors have declared no conflict of interest.

\subsection*{Code and Data Availability}

\noindent The code and data that support the findings of this study
are available from the corresponding authors upon reasonable request. 

\subsection*{Acknowledgements}

\noindent This work was supported by ERC-ADG-2022-101097092 ANBIT,
ERC-POC-2023-101138302 NU MESH, GVA PROMETEO 2021/015 research excellency
award, Ministerio de Ciencia y Universidades Plan Complementario de
Comunicaci\'on Cu\'antica projects QUANTUMABLE-1 and QUANTUMABLE-2,
and HUB de Comunicaciones Cu\'anticas.

\newpage{}

\noindent \begin{center}
\newpage{}
\par\end{center}

\noindent \begin{center}
\includegraphics[width=15.5cm,height=22cm,keepaspectratio]{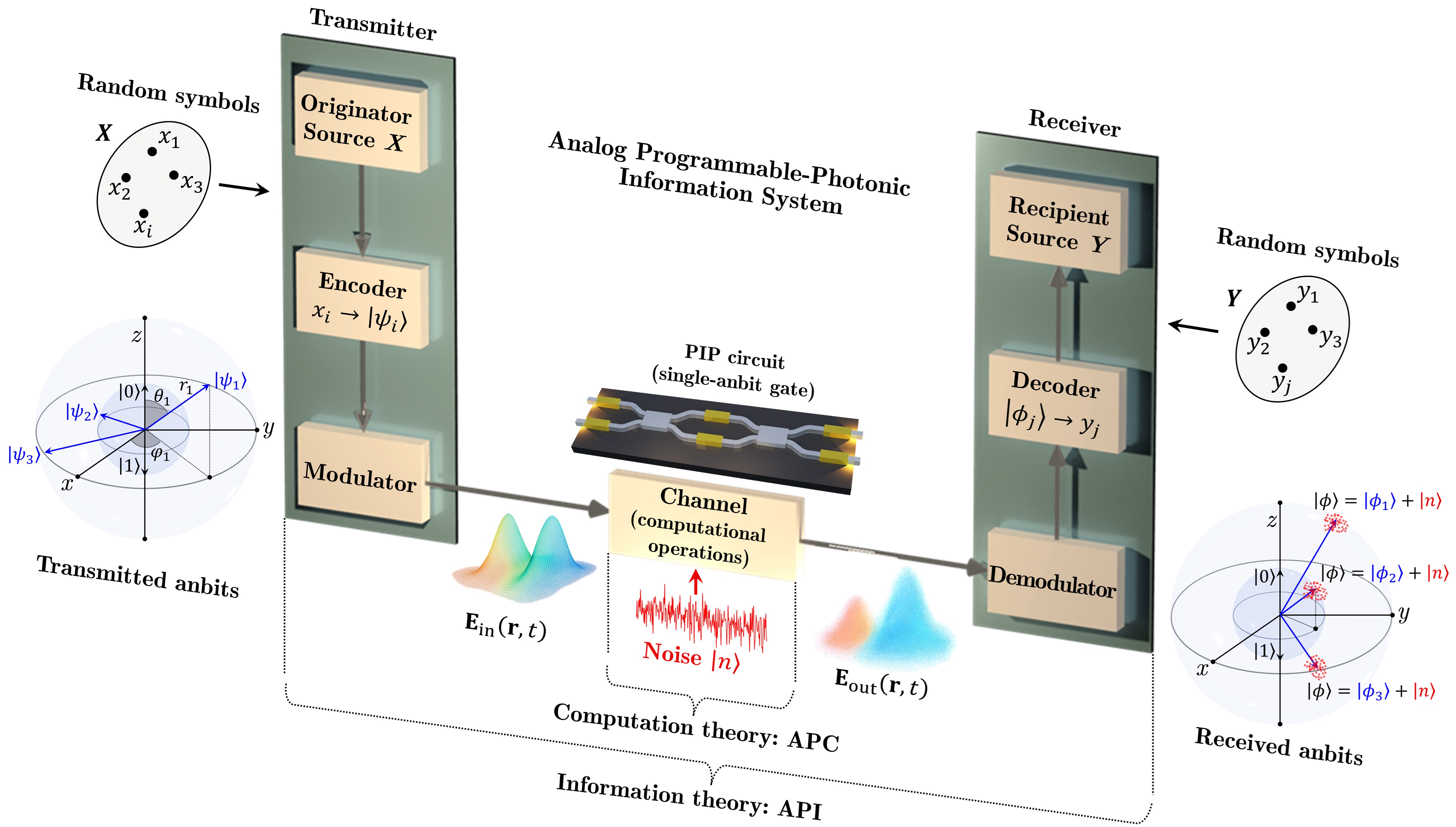}
\par\end{center}

\noindent \textbf{\small{}Fig.\,1 }{\small{}Analog programmable-photonic
information system. The information theory introduced in this work,
Analog Programmable-Photonic Information (API), analyzes any PIP computational
architecture based on the Analog Programmable-Photonic Computation
(APC) model \cite{key-P39} as an information-processing system composed
of a transmitter, a channel, and a receiver through which information
is sequentially generated, computed, and recovered. The channel not
only propagates information but also transforms it using PIP circuitry
to solve a specific mathematical problem. This perspective unifies
the entire information-processing system and positions API as a foundational
research field, addressing fundamental questions that extend beyond
the scope of APC. }{\small \par}
\noindent \begin{center}
\newpage{}
\par\end{center}

\noindent \begin{center}
\includegraphics[width=8.5cm,height=20cm,keepaspectratio]{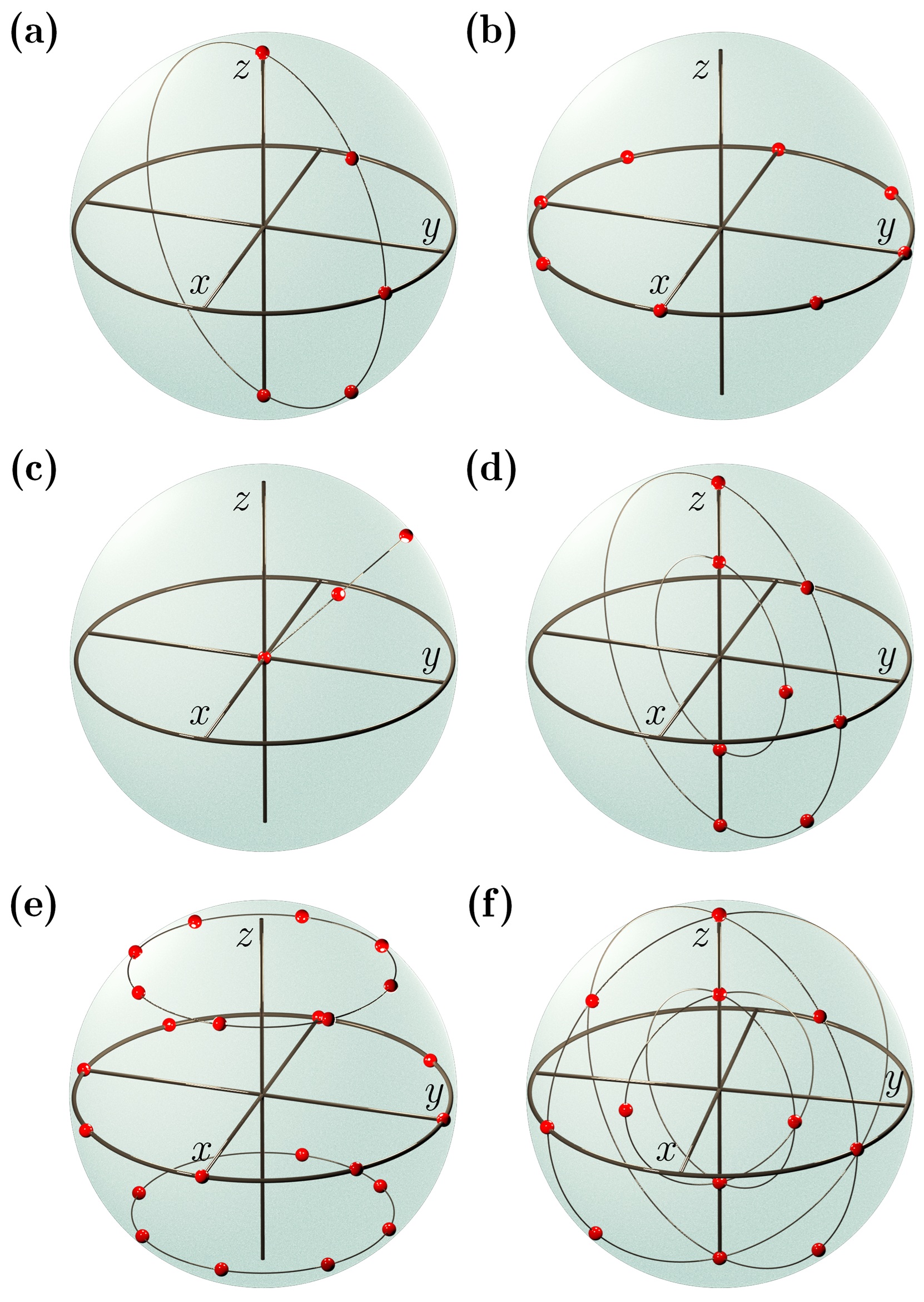}
\par\end{center}

\noindent \textbf{\small{}Fig.\,2 }{\small{}Examples of diverse classes
of analog constellations designed by varying the effective degrees
of freedom (EDF) of the anbits. (a) Single-EDF constellation based
on the elevation angle. (b) Single-EDF constellation based on the
azimuthal angle. (c) Single-EDF constellation based on the radius.
(d) Two-EDF constellation varying the elevation angle and the radius.
(e) Two-EDF constellation varying the azimuthal and elevation angles.
(f) Three-EDF constellation varying the elevation angle, the azimuthal
angle, and the radius. }{\small \par}
\noindent \begin{center}
\newpage{}
\par\end{center}

\noindent \begin{center}
\includegraphics[width=11cm,height=10cm,keepaspectratio]{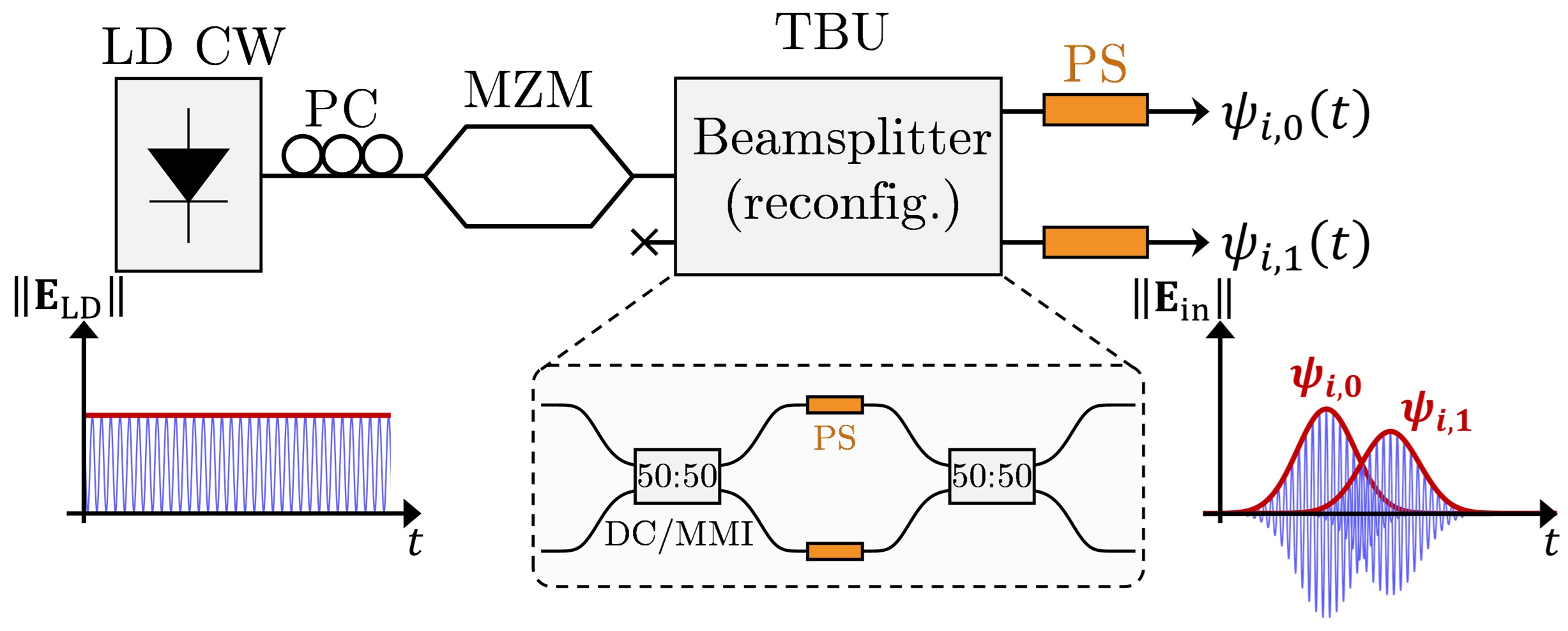}
\par\end{center}

\noindent \textbf{\small{}Fig.\,3 }{\small{}Hardware implementation
of the space-anbit modulator (SAM). Two wave packets (or complex envelopes),
$\psi_{i,0}$ and $\psi_{i,1}$, with independently controllable moduli
are generated using a continuous-wave (CW) laser diode (LD), a polarization
controller (PC), a Mach-Zehnder modulator (MZM), and a reconfigurable
beam splitter implemented via a tunable basic unit (TBU) of PIP. The
phase of each wave packet \textendash{} and hence the phase delay
between them \textendash{} is controlled by two phase shifters (PSs).}{\small \par}
\noindent \begin{center}
\newpage{}
\par\end{center}

\noindent \begin{center}
\includegraphics[width=10cm,height=20cm,keepaspectratio]{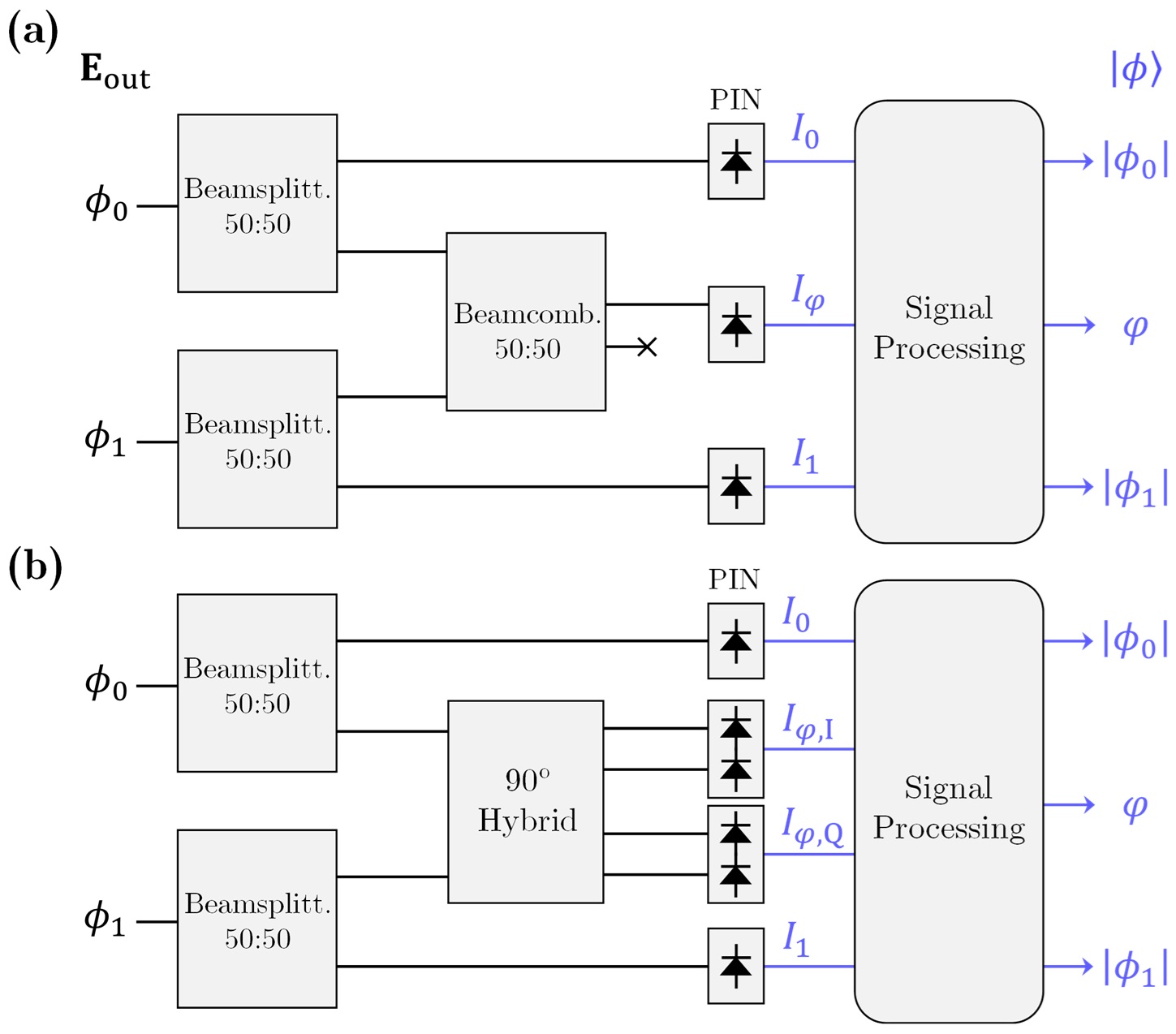}
\par\end{center}

\noindent \textbf{\small{}Fig.\,4 }{\small{}Differential O/E converters.
(a) Unbalanced architecture. (b) Quadrature architecture. The 50:50
beam splitters are realized using Y-junctions, whereas the 50:50 beam
combiner and the $90^{\circ}$ optical hybrid can be implemented using
multi-mode interferometers \cite{key-P32}.}{\small \par}
\noindent \begin{center}
\newpage{}
\par\end{center}

\noindent \begin{center}
\includegraphics[width=9.5cm,height=17cm,keepaspectratio]{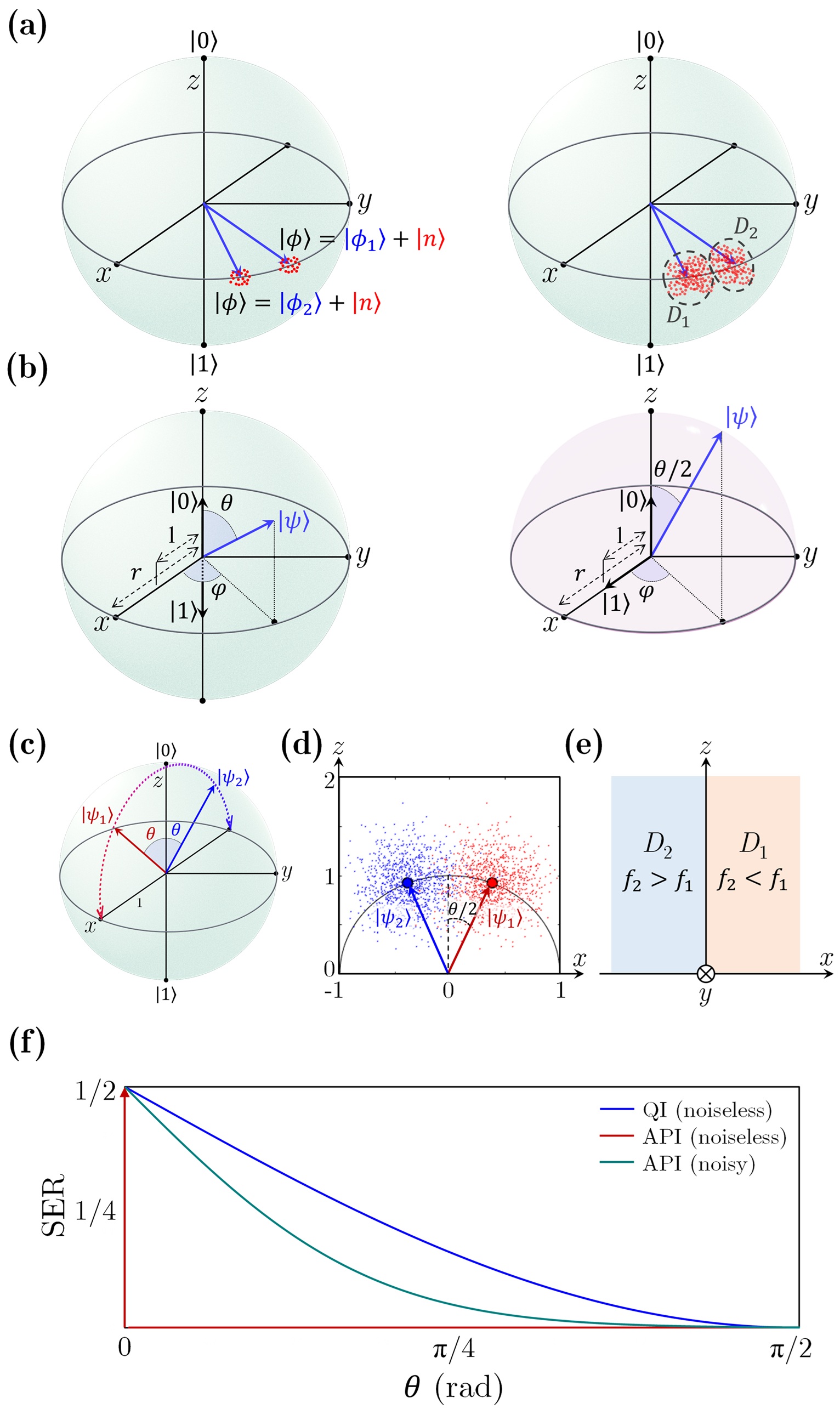}
\par\end{center}

\noindent \textbf{\small{}Fig.\,5 }{\small{}Principles of anbit receiver.
(a) non-overlapping }\emph{\small{}vs}{\small{} overlapping analog
constellations. In the non-overlapping case, the ideal anbits $\left|\phi_{1,2}\right\rangle $
are estimated by averaging the noisy anbits $\left|\phi\right\rangle $.
In the overlapping case, the ideal anbits are recovered from the noisy
anbits $\left|\phi\right\rangle $ using decision regions $D_{1,2}$.
(b) GBS representation }\emph{\small{}vs}{\small{} half-angle GBS
representation. (c) Analog constellation described by Eq.\,(5). User
information is encoded in the elevation angle $0<\theta\leq\pi/2$.
(d) Received analog constellation, represented in the half-angle GBS,
along with the system noise described by the ket $\left|n\right\rangle =n_{0}\left|0\right\rangle +n_{1}\left|1\right\rangle $,
where $n_{0}$ and $n_{1}$ are independent zero-mean Gaussian random
variables with standard deviation $\sigma=0\textrm{.}22$. (e) Optimal
decision regions in the half-angle GBS. (f) Symbol error rate (SER)
in API under varying noise conditions and comparison with the SER
found in QI when the same pair of states defined in Eq. (5) (qubits)
is transmitted through a noiseless quantum channel. API exhibits a
lower SER than QI even under noisy conditions, highlighting the robustness
of APC to system noise.}{\small \par}
\noindent \begin{center}
\newpage{}
\par\end{center}

\noindent \begin{center}
\includegraphics[width=10cm,height=10cm,keepaspectratio]{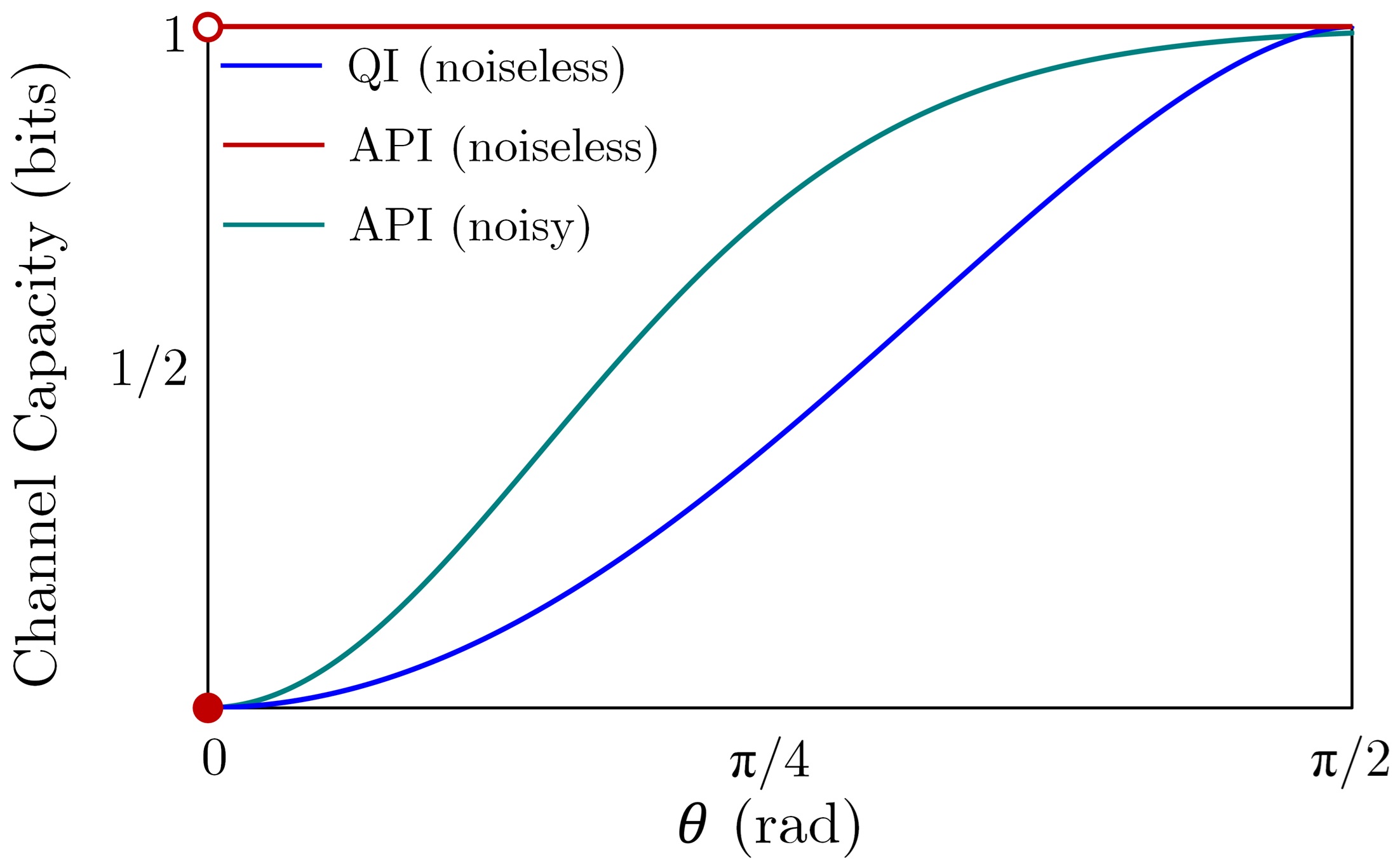}
\par\end{center}

\noindent \textbf{\small{}Fig.\,6 }{\small{}Channel capacity in API
}\emph{\small{}vs}{\small{} QI. Classical and quantum systems emit
the same pair of states (anbits in API and qubits in QI), illustrated
in Fig.\,5(c) as a function of the elevation angle ($\theta$). The
states are propagated through a noiseless quantum channel (blue line),
a noiseless classical channel (red line), and a noisy classical channel
(green line). The noisy classical channel introduces additive Gaussian
noise (with zero mean and a standard deviation $\sigma=0\textrm{.}22$)
in the amplitudes of the emitted anbits. Notably, API maintains a
higher channel capacity than QI even under noisy conditions, reflecting
the robustness of APC against system noise.}{\small \par}
\noindent \begin{center}
\newpage{}
\par\end{center}

\noindent \begin{center}
\includegraphics[width=15.5cm,height=22cm,keepaspectratio]{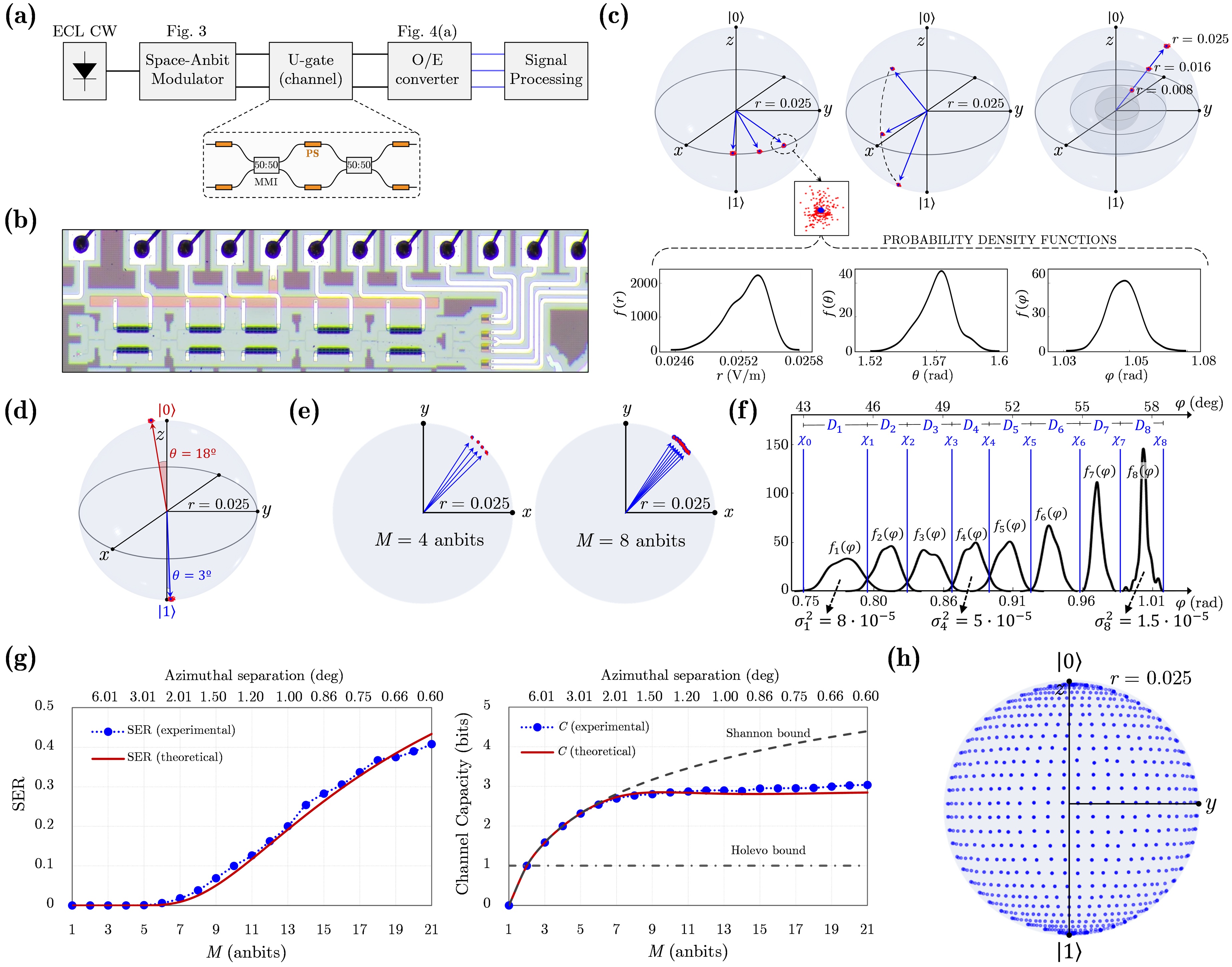}
\par\end{center}

\noindent \textbf{\small{}Fig.\,7 }{\small{}Experimental validation
of API principles. (a) Laboratory setup comprising a continuous-wave
(CW) external cavity laser (ECL), a space-anbit modulator (SAM, see
Fig.\,3), a universal single-anbit U-gate \cite{key-P39}, and an
unbalanced differential O/E converter {[}see Fig.\,4(a){]}. (b) Micrograph
of the fabricated PIP chip. (c) Diverse analog constellations generated
by the SAM and retrieved at the output of the O/E converter. The received
anbits (blue points) are perturbed by system noise (red points), which
induces random perturbations in the EDFs, characterized by the corresponding
probability density functions (pdfs). (d) Hardware imperfections inducing
a constant deviation in the poles of the GBS (blue points). (e) Analog
constellation located on the equator of the GBS {[}Eq.\,(10){]} with
different number of anbits ($M$). (f) Conditional pdfs used to optimize
the anbit measurement and associated decision regions $D_{i}$. (g)
Experimental and theoretical {[}Eqs.\,(11) and (12){]} symbol error
rate (SER) and channel capacity for the $M$-anbit analog constellation
shown in panel e). (h) Analog constellation generated by the SAM and
estimated at the demodulator, comprising 900 non-overlapping anbits,
corresponding to a channel capacity of $C=\log_{2}900\simeq10$ bits. }{\small \par}
\noindent \begin{center}
\newpage{}
\par\end{center}

\noindent \begin{center}
\includegraphics[width=8cm,height=8cm,keepaspectratio]{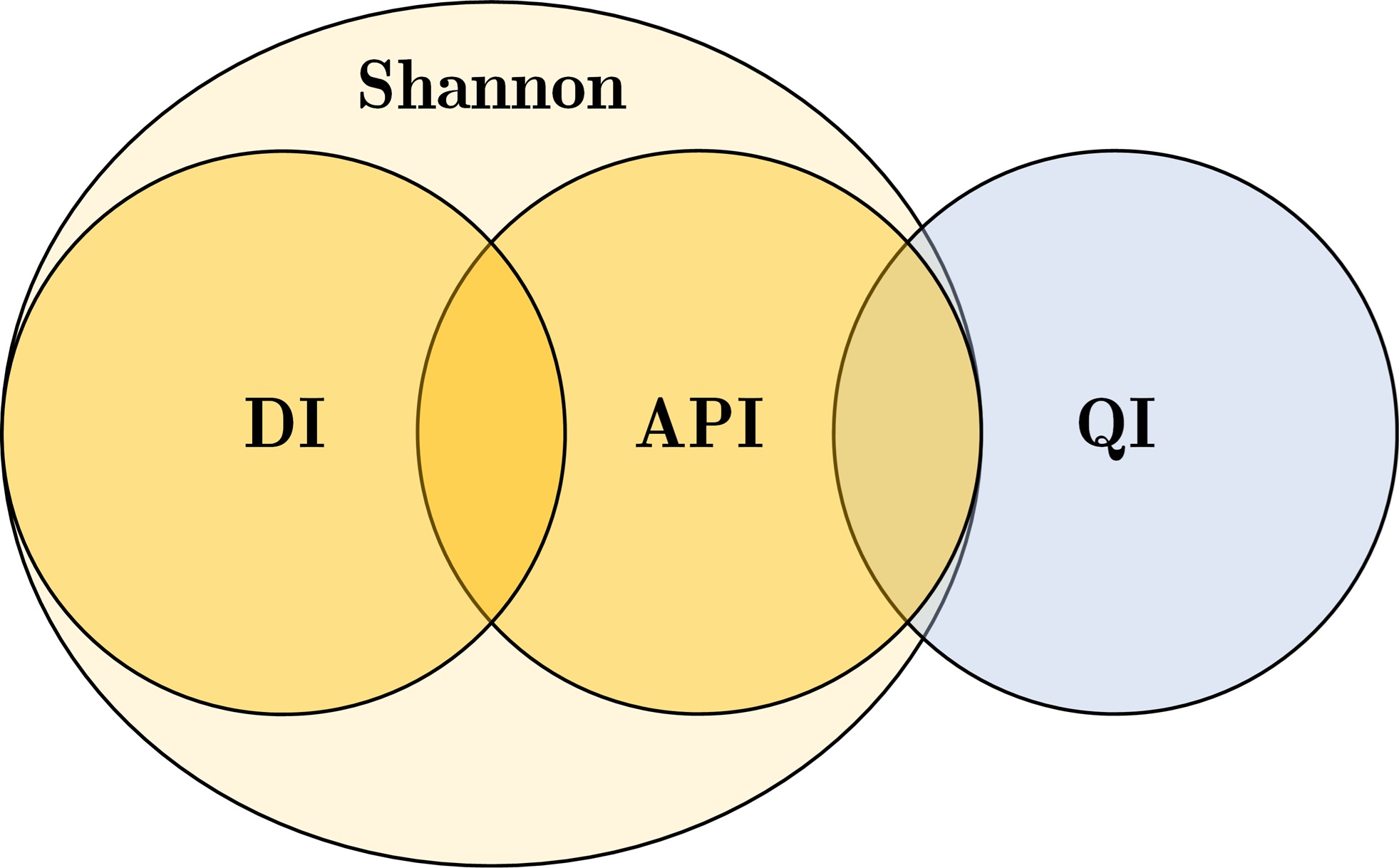}
\par\end{center}

\noindent \textbf{\small{}Fig.\,8 }{\small{}Conceptual relationship
between information theories. Analog Programmable-Photonic Information
(API) intersects with both Digital Information (DI) and Quantum Information
(QI), while remaining grounded within Shannon\textquoteright s classical
theory. This position highlights API as a potential link between classical
and quantum paradigms in photonic computing. }{\small \par}
\noindent \begin{center}
\newpage{}
\par\end{center}

\noindent \begin{center}
\textbf{\LARGE{}Supplementary Information}
\par\end{center}{\LARGE \par}

\vspace{1cm}

\tableofcontents{}

\newpage{}

\section*{Supplementary Note 1: originator source\label{sec:1}}

\addcontentsline{toc}{section}{Supplementary Note 1: originator source}

\noindent As commented in the main text, an Analog Programmable-Photonic
Computation (APC) system deals with mathematical problems requiring
\emph{matrix} operations using Programmable Integrated Photonic (PIP)
circuits. In PIP meshes, the \emph{system input }can be regarded as
a multivariate random variable or random \emph{vector} $\mathbf{A}=\left(A_{1},\ldots,A_{n}\right)$,
that is, the information generated by the originator source (the sample
space or alphabet) is mapped onto $n$-tuplas $\left(a_{1},\ldots,a_{n}\right)$
belonging to a multi-dimensional vector space (the state space or
range of $\mathbf{A}$). 

In both Digital Information (DI) and Analog Programmable-Photonic
Information (API), the sample space and the state space are \emph{discretized}
to ensure compatibility between the originator sources of both information
paradigms. This implies that the state space consists of a finite
set of $M$ distinct $n$-tuples. In such a situation, a one-to-one
mapping can always be established between each $n$-tupla and a specific
real number $x_{i}$:
\begin{equation}
\bigl\{\bigl(a_{1}^{\left(i\right)},\ldots,a_{n}^{\left(i\right)}\bigr)\bigr\}_{i=1,\ldots,M}\overset{1:1}{\longleftrightarrow}\left\{ x_{i}\right\} _{i=1,\ldots,M}\tag{S1.1}\label{eq:S1.1}
\end{equation}
The set of real numbers $\left\{ x_{i}\right\} _{i=1,\ldots,M}$ defines
the range of a discrete real random variable $X$. The bijective correspondence
established by Eq.\,(\ref{eq:S1.1}) ensures that the joint probability
mass function (pmf) of $\mathbf{A}$ and the pmf of $X$ are identical:
\begin{equation}
p\bigl(\mathbf{A}=\bigl(a_{1}^{\left(i\right)},\ldots,a_{n}^{\left(i\right)}\bigr)\bigr)\equiv p\left(X=x_{i}\right).\tag{S1.2}\label{eq:S1.2}
\end{equation}
Consequently, the sample space of the originator source in API can
be equivalently described either by the random vector $\mathbf{A}$
or by the random variable $X$. 

As a didactical example, let us consider the case of rolling two dice.
The outcomes of this experiment may be described by $M=36$ different
2-tuples $\left\{ \left(1,1\right),\left(1,2\right),\ldots,\left(6,5\right),\left(6,6\right)\right\} $,
which define the range of a random vector $\mathbf{A}$. Each 2-tuple
can be mapped onto a different real number, e.g., belonging to the
set $\left\{ 1,\ldots,36\right\} $, thus defining the range of a
discrete real random variable $X$ (Fig.\,S1). Accordingly, the pmfs
of $\mathbf{A}$ and $X$ are found to be identical, that is, $p\left(\mathbf{A}=\left(1,1\right)\right)\equiv p\left(X=1\right)$,
$p\left(\mathbf{A}=\left(1,2\right)\right)\equiv p\left(X=2\right)$,
..., $p\left(\mathbf{A}=\left(6,6\right)\right)\equiv p\left(X=36\right)$.
As seen, this scenario can be equivalently modeled either by the random
vector $\mathbf{A}$ or by the random variable $X$.
\noindent \begin{center}
\includegraphics[width=12cm,height=8cm,keepaspectratio]{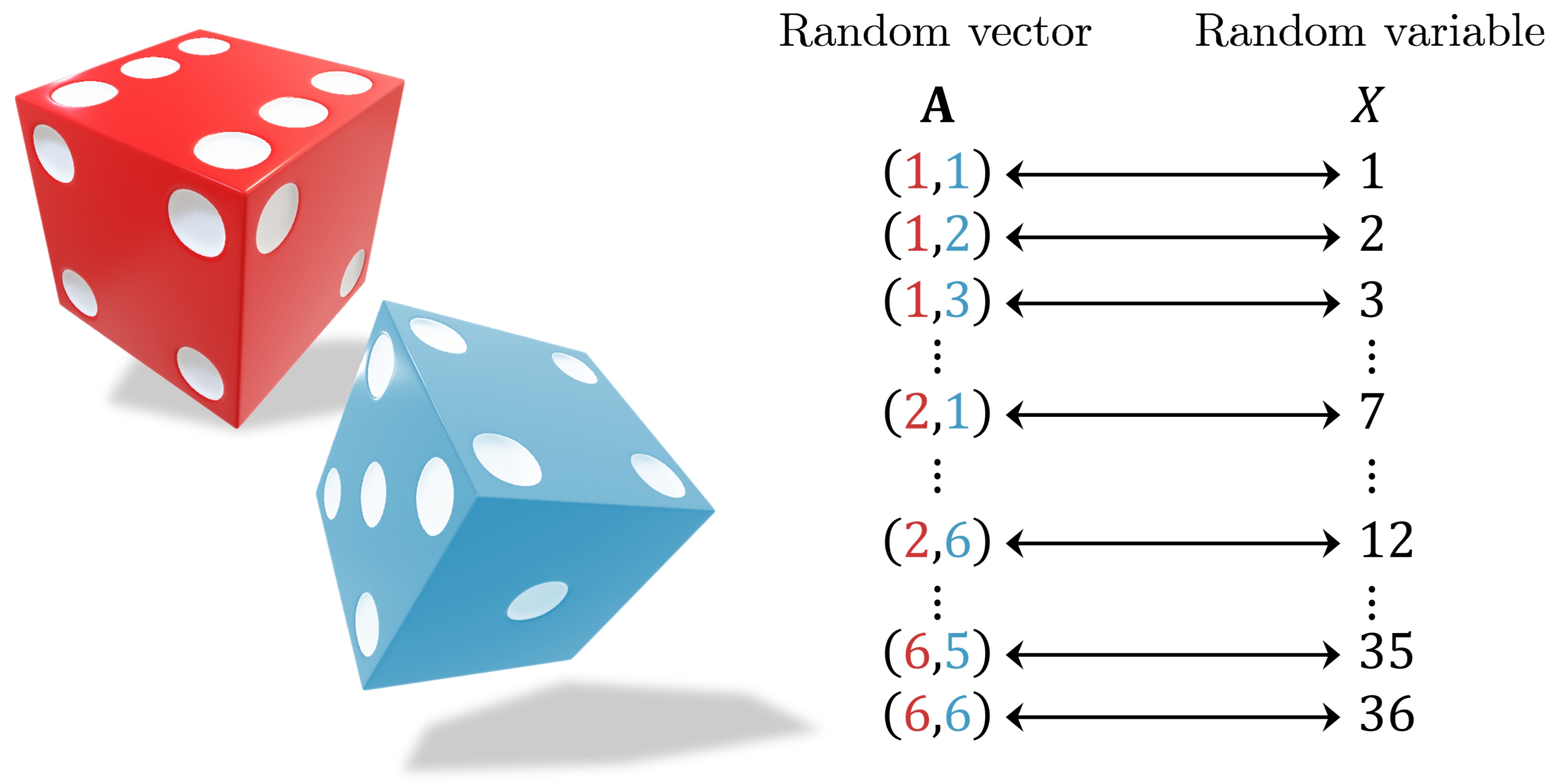}
\par\end{center}

\noindent {\small{}Figure S1. The outcome of rolling two dice can
be represented either as a random vector or as a single random variable.}{\small \par}

\newpage{}

\section*{Supplementary Note 2: pure \emph{vs} mixed classical states\label{sec:2}}

\addcontentsline{toc}{section}{Supplementary Note 2: pure \emph{vs} mixed classical states}

\noindent In this section, we further investigate the conceptual distinction
between pure and mixed states in the context of API. In addition,
inspired by Quantum Information (QI), we extrapolate the concept of
the quantum density operator to the classical world within the API
framework and assess its practical relevance.

\subsection*{2.1 Preliminary concepts: pure \emph{vs} mixed quantum states }

\addcontentsline{toc}{subsection}{2.1 Preliminary concepts: pure \emph{vs} mixed quantum states}

\noindent To support non-expert readers, this subsection reviews the
basic difference between pure and mixed quantum states, which can
be easily visualized through the following didactical example. Consider
a QI system with an originator source $X$ that emits a single symbol
$x_{1}$ with unit probability, $p_{1}=1$. This symbol is encoded
into a single particle, described by the wave function $\psi_{1}\left(x\right)=\bigl\langle x\bigr|\psi_{1}\bigr\rangle$,
or equivalently, by the ket $\bigr|\psi_{1}\bigr\rangle$. The particle
propagates through the channel with probability $p_{1}=1$. In this
case, the sample space of the originator source and the encoder \emph{is
not partitioned}; only one symbol and one state are involved. Therefore,
the quantum system is described by a \emph{pure} \emph{state}. The
probability density function (pdf) of the quantum system is given
by:
\begin{equation}
f_{1}\left(x\right)=\bigl|\psi_{1}\left(x\right)\bigr|^{2}=\bigl|\bigl\langle x\bigr|\psi_{1}\bigr\rangle\bigr|^{2}=\bigl\langle x\bigr|\psi_{1}\bigr\rangle\bigl\langle\psi_{1}\bigr|x\bigr\rangle\equiv\bigl\langle x\bigr|\widehat{\rho}_{1}\bigl|x\bigr\rangle,\tag{S2.1}\label{eq:S2.1}
\end{equation}
where $\widehat{\rho}_{1}\coloneqq\bigr|\psi_{1}\bigr\rangle\bigl\langle\psi_{1}\bigr|$
is the \emph{density operator associated with the pure state}. Hence,
from the one-to-one correspondence between $\psi_{1}\left(x\right)$,
$\bigr|\psi_{1}\bigr\rangle$ and $\widehat{\rho}_{1}$, it follows
that the statistical properties of a quantum system in a pure state
can be equivalently described using any of these three mathematical
tools.

Now, consider the same QI system, but with the originator source $X$
emitting two different symbols $x_{1}$ and $x_{2}$, with probabilities
$p_{1}$ and $p_{2}$, respectively. Each symbol is encoded into a
distinct particle, described by the wave functions $\psi_{1}\left(x\right)=\bigl\langle x\bigr|\psi_{1}\bigr\rangle$
and $\psi_{2}\left(x\right)=\bigl\langle x\bigr|\psi_{2}\bigr\rangle$,
or equivalently, by the kets $\bigr|\psi_{1}\bigr\rangle$ and $\bigr|\psi_{2}\bigr\rangle$,
or by the density operators $\widehat{\rho}_{1}=\bigr|\psi_{1}\bigr\rangle\bigl\langle\psi_{1}\bigr|$
and $\widehat{\rho}_{2}=\bigr|\psi_{2}\bigr\rangle\bigl\langle\psi_{2}\bigr|$.
The first (second) particle propagates through the channel with probability
$p_{1}$ ($p_{2}$). In this case, the sample space of the originator
source and the encoder \emph{is partitioned} in two distinct random
events; two symbols and two quantum states are involved. Hence, the
quantum system is described by a \emph{mixed} \emph{state}. The pdf
$f\left(x\right)$ of the quantum system reflects a \emph{statistical
mixture} of the pdfs ($i=1,2$):
\begin{align}
f_{i}\left(x\right) & =\bigl|\psi_{i}\left(x\right)\bigr|^{2}=\bigl|\bigl\langle x\bigr|\psi_{i}\bigr\rangle\bigr|^{2}=\bigl\langle x\bigr|\psi_{i}\bigr\rangle\bigl\langle\psi_{i}\bigr|x\bigr\rangle\equiv\bigl\langle x\bigr|\widehat{\rho}_{i}\bigl|x\bigr\rangle.\tag{S2.2}\label{eq:S2.2}
\end{align}

\noindent Consequently, the total pdf $f\left(x\right)$ must be calculated
using the law of total probability \cite{key-S1}:
\begin{equation}
f\left(x\right)=\sum_{i}p_{i}f_{i}\left(x\right)=\sum_{i}p_{i}\bigl\langle x\bigr|\widehat{\rho}_{i}\bigl|x\bigr\rangle=\bigl\langle x\bigr|\sum_{i}p_{i}\widehat{\rho}_{i}\bigl|x\bigr\rangle\equiv\bigl\langle x\bigr|\widehat{\rho}\bigl|x\bigr\rangle,\tag{S2.3}\label{eq:S2.3}
\end{equation}
with $\sum_{i}p_{i}=1$ and:
\begin{equation}
\widehat{\rho}\coloneqq\sum_{i}p_{i}\widehat{\rho}_{i}=\sum_{i}p_{i}\bigr|\psi_{i}\bigr\rangle\bigl\langle\psi_{i}\bigr|,\tag{S2.4}\label{eq:S2.4}
\end{equation}
being the \emph{density operator associated with the mixed state}.
As shown in Eq.\,(\ref{eq:S2.3}), $f\left(x\right)$ is in one-to-one
correspondence only with $\widehat{\rho}$, and not with any specific
wave function or ket. This highlights that the statistical properties
of a quantum system in a mixed state cannot be captured by a single
ket alone; instead, they require the use of the density operator to
fully characterize the mixture, ensuring that the law of total probability
is fulfilled.

\subsection*{2.2 Pure \emph{vs} mixed classical states and classical density operator}

\addcontentsline{toc}{subsection}{2.2 Pure \emph{vs} mixed classical states and classical density operator}

\noindent To explore the interpretation of pure and mixed states in
the context of API, we replicate the examples introduced in the preceding
subsection. Consider an API system with an originator source $X$
that emits a single symbol $x_{1}$ with unit probability, $p_{1}=1$.
This symbol is encoded into a single anbit, described by the ket $\bigr|\psi_{1}\bigr\rangle$:
\begin{equation}
\bigr|\psi_{1}\bigr\rangle=r_{1}\left(\cos\frac{\theta_{1}}{2}\bigl|0\bigr\rangle+e^{\mathrm{j}\varphi_{1}}\sin\frac{\theta_{1}}{2}\bigl|1\bigr\rangle\right).\tag{S2.5}\label{eq:S2.5}
\end{equation}
The anbit propagates through the channel with probability $p_{1}=1$.
In this case, the sample space of the originator source and the encoder
\emph{is not partitioned}; only one symbol and one state are involved.
Therefore, the API system is described by a \emph{pure} \emph{classical
state}. Here, in contrast to QI, the statistical properties of the
API system are only given by the pmf of the originator source, since
the system is governed by classical deterministic physical laws (Maxwell's
equations). This implies that the ``virtual'' wave function $\psi_{1}\left(x\right)=\bigl\langle x\bigr|\psi_{1}\bigr\rangle$
does not convey statistical information about the system. In fact,
while $\bigl|\psi_{i}\left(x\right)\bigr|^{2}$ is associated to a
pdf in QI yielding insight into the particle\textquoteright s spatial
distribution, in API $\bigl|\psi_{i}\left(x\right)\bigr|^{2}$ may
be interpreted instead as a ``power density function'', revealing
information about the optical power ($\mathcal{P}_{1}$) required
to physically implement the anbit $\bigr|\psi_{1}\bigr\rangle$ at
the modulator:
\begin{equation}
\mathcal{P}_{1}=r_{1}^{2}=\bigl\langle\psi_{1}\bigr|\psi_{1}\bigr\rangle=\int_{\infty}\bigl|\psi_{1}\left(x\right)\bigr|^{2}\mathrm{d}x=\int_{\infty}\bigl\langle x\bigr|\psi_{1}\bigr\rangle\bigl\langle\psi_{1}\bigr|x\bigr\rangle\mathrm{d}x=\int_{\infty}\bigl\langle x\bigr|\widehat{\rho}_{1}\bigr|x\bigr\rangle\mathrm{d}x.\tag{S2.6}\label{eq:S2.6}
\end{equation}
This equation allows us to introduce the operator $\widehat{\rho}_{1}\coloneqq\bigr|\psi_{1}\bigr\rangle\bigl\langle\psi_{1}\bigr|$,
which can be referred to as \emph{the} \emph{density operator of the
pure classical state} $\bigr|\psi_{1}\bigr\rangle$.

Now, consider the same API system, but with the originator source
$X$ emitting two different symbols $x_{1}$ and $x_{2}$, with probabilities
$p_{1}$ and $p_{2}$, respectively. Each symbol is encoded into a
distinct anbit, described by the kets $\bigr|\psi_{1}\bigr\rangle$
and $\bigr|\psi_{2}\bigr\rangle$ given by the expression:
\begin{equation}
\bigr|\psi_{i}\bigr\rangle=r_{i}\left(\cos\frac{\theta_{i}}{2}\bigl|0\bigr\rangle+e^{\mathrm{j}\varphi_{i}}\sin\frac{\theta_{i}}{2}\bigl|1\bigr\rangle\right),\tag{S2.7}\label{eq:S2.7}
\end{equation}
for all $i=1,2$. The first (second) anbit propagates through the
channel with probability $p_{1}$ ($p_{2}$). In this case, the sample
space of the originator source and the encoder \emph{is partitioned}
in two distinct random events; two symbols and two states are involved.
Hence, the API system is described by a \emph{mixed} \emph{classical
state}. Here, as in the pure-state case, the statistical properties
of the API system are determined solely by the pmf of the originator
source. This implies that the average ket:
\begin{equation}
\bigl|\psi_{X}\bigr\rangle\coloneqq\widehat{\mathrm{E}}\bigl(\bigl|\psi_{i}\bigr\rangle\bigr)=\sum_{i}p_{i}\bigl|\psi_{i}\bigr\rangle,\tag{S2.8}\label{eq:S2.8}
\end{equation}
defined via the expectation operator $\widehat{\mathrm{E}}$ \cite{key-S1},
suffices to describe the statistics of the system at the output of
the encoder. As commented in the main text, the average anbit $\bigl|\psi_{X}\bigr\rangle$
plays a role analogous to that of a mixed state in QI. Interestingly,
the average optical power ($\mathcal{P}_{X}$)\linebreak{} required to physically
implement the analog constellation $\bigl\{\bigl|\psi_{i}\bigr\rangle\bigr\}_{i}$
can be calculated using Eq.\,(\ref{eq:S2.6}) as:

\newpage{}

\begin{align}
\mathcal{P}_{X} & =\widehat{\mathrm{E}}\bigl(\mathcal{P}_{i}\bigr)=\sum_{i}p_{i}\mathcal{P}_{i}\nonumber \\
 & =\sum_{i}p_{i}\int_{\infty}\bigl\langle x\bigr|\widehat{\rho}_{i}\bigr|x\bigr\rangle\mathrm{d}x=\int_{\infty}\bigl\langle x\bigr|\sum_{i}p_{i}\widehat{\rho}_{i}\bigr|x\bigr\rangle\mathrm{d}x\equiv\int_{\infty}\bigl\langle x\bigr|\widehat{\rho}_{X}\bigr|x\bigr\rangle\mathrm{d}x,\tag{S2.9}\label{eq:S2.9}
\end{align}
where:
\begin{equation}
\widehat{\rho}_{X}\coloneqq\sum_{i}p_{i}\widehat{\rho}_{i}=\sum_{i}p_{i}\bigr|\psi_{i}\bigr\rangle\bigl\langle\psi_{i}\bigr|,\tag{S2.10}\label{eq:S2.10}
\end{equation}
is defined as the \emph{the} \emph{density operator of the mixed classical
state} $\bigl|\psi_{X}\bigr\rangle$. Defined analogously to its quantum
counterpart, it is consequently a Hermitian and positive operator.
However, in contrast to QI, the classical density operator is not
required to analyze the post-codification entropy, which is provided
by Shannon's entropy in API, as mentioned in the paper. Instead, within
the API framework, the density operator offers insight into the \emph{average
optical power} needed to physically implement an analog constellation
at the modulator.

In forthcoming contributions to the API paradigm, we will explore
in greater detail the main properties of the classical density operator
and its associated density matrix. This will enable the analysis of
various electromagnetic characteristics at the modulation block, not
only the average optical power, but also optical interference between
anbit amplitudes, which can be examined through the coherences of
the density matrix, i.e., its off-diagonal elements.

Finally, to further illustrate the potential of the classical density
operator formalism for future works, we briefly comment on its utility
in the analysis of composite API systems. As an example, consider
a two-anbit computational system with an input state defined via the
Cartesian product $\bigl|\Psi_{AB}\bigr\rangle=\bigl|\psi_{A}\bigr\rangle\times\bigl|\psi_{B}\bigr\rangle$
\cite{key-S2}. The optical power $\mathcal{P}_{AB}$ required by
the modulator to implement $\bigl|\Psi_{AB}\bigr\rangle$ is:
\begin{equation}
\mathcal{P}_{AB}=\bigl\langle\Psi_{AB}|\Psi_{AB}\bigr\rangle=\bigl\langle\psi_{A}|\psi_{A}\bigr\rangle+\bigl\langle\psi_{B}|\psi_{B}\bigr\rangle\equiv\mathcal{P}_{A}+\mathcal{P}_{B},\tag{S2.11}\label{eq:S2.11}
\end{equation}
where $\mathcal{P}_{A}$ and $\mathcal{P}_{B}$ are the optical powers
needed to implement the anbits $\bigl|\psi_{A}\bigr\rangle$ and $\bigl|\psi_{B}\bigr\rangle$,
respectively. In such a scenario, $\mathcal{P}_{AB}$ can alternatively
be calculated from the density operator of the composite system $\widehat{\rho}_{AB}=\widehat{\rho}_{A}+\widehat{\rho}_{B}$,
as deduced from the equation:
\begin{align}
\mathcal{P}_{AB}=\mathcal{P}_{A}+\mathcal{P}_{B} & =\int_{\infty}\bigl\langle x\bigr|\widehat{\rho}_{A}\bigr|x\bigr\rangle\mathrm{d}x+\int_{\infty}\bigl\langle x\bigr|\widehat{\rho}_{B}\bigr|x\bigr\rangle\mathrm{d}x\nonumber \\
 & =\int_{\infty}\left(\bigl\langle x\bigr|\widehat{\rho}_{A}\bigr|x\bigr\rangle+\bigl\langle x\bigr|\widehat{\rho}_{B}\bigr|x\bigr\rangle\right)\mathrm{d}x\nonumber \\
 & =\int_{\infty}\bigl\langle x\bigr|\left(\widehat{\rho}_{A}+\widehat{\rho}_{B}\right)\bigr|x\bigr\rangle\mathrm{d}x\nonumber \\
 & \equiv\int_{\infty}\bigl\langle x\bigr|\widehat{\rho}_{AB}\bigr|x\bigr\rangle\mathrm{d}x.\tag{S2.12}\label{eq:S2.12}
\end{align}
Remarkably, the construction of composite (i.e., multi-anbit) systems
via the summation of density operators associated with simple (i.e.,
single-anbit) systems reveals mathematical principles in API that
diverge from those of QI theory. The classical density operator framework
may thus provide a pathway to uncovering fundamental physical differences
between these two information paradigms. 

\newpage{}

\section*{Supplementary Note 3: state-comparative parameters\label{sec:3}}

\addcontentsline{toc}{section}{Supplementary Note 3: state-comparative parameters}

\noindent Here, we introduce and analyze a comprehensive set of state-comparative
parameters within the API framework to quantify the closeness between
two classical states. Noting that there exit some mathematical similarities
between QI and API, we first explore the possibility of extrapolating
the quantum fidelity and quantum trace distance into the API context.
We then propose and develop specific state-comparative parameters
for the API paradigm, tailored to its distinctive structure.

\subsection*{3.1 Fidelity }

\addcontentsline{toc}{subsection}{3.1 Fidelity}

\noindent Consider that we are interested in generating the ideal
anbit: 
\begin{equation}
\bigl|\psi\bigr\rangle=r\left(\cos\frac{\theta}{2}\bigl|0\bigr\rangle+e^{\mathrm{j}\varphi}\sin\frac{\theta}{2}\bigl|1\bigr\rangle\right),\tag{S3.1}\label{eq:S3.1}
\end{equation}
but the real anbit that is generated is $\bigl|\varphi\bigr\rangle$,
for example, due to hardware imperfections of the modulator. This
error can be quantified by using the quantum fidelity, which reduces
for pure quantum states (described by kets) to the expression \cite{key-S3}:
\begin{equation}
F\left(\bigl|\psi\bigr\rangle,\bigl|\varphi\bigr\rangle\right)\coloneqq\left|\bigl\langle\psi\bigr|\varphi\bigr\rangle\right|\geq0.\tag{S3.2}\label{eq:S3.2}
\end{equation}
In API, this definition applies to both pure and mixed classical states,
which are represented by kets (see Supplementary Note 2). In particular,
\emph{within the API framework}, fidelity satisfies the following
properties:
\begin{enumerate}
\item \emph{Extremal values}. Fidelity reaches its minimum value $F=0$
if and only if the anbits are orthogonal, i.e., located at opposite
points on the generalized Bloch sphere (GBS). Conversely, fidelity
reaches its maximum value $F=r^{2}$ if and only if the anbits are
identical. As observed, the maximum value may differ from unity, indicating
that fidelity lacks geometric intuitiveness in API.
\item \emph{Symmetry}. It is direct to verify that $F\left(\bigl|\psi\bigr\rangle,\bigl|\varphi\bigr\rangle\right)=F\left(\bigl|\varphi\bigr\rangle,\bigl|\psi\bigr\rangle\right)$.
\item \emph{Not a metric}. Since $F\neq0$ when the anbits are identical,
it follows that the triangle inequality cannot be fulfilled, a basic
property for any metric. 
\item \emph{Base independent}. This property directly emerges from the inner
product, which is a base-independent application.
\item \emph{Unitary invariance}. Fidelity is invariant under unitary operations
(U-gates \cite{key-S2}):
\begin{equation}
F\bigl(\widehat{\mathrm{U}}\bigl|\psi\bigr\rangle,\widehat{\mathrm{U}}\bigl|\varphi\bigr\rangle\bigr)=F\bigl(\bigl|\psi\bigr\rangle,\bigl|\varphi\bigr\rangle\bigr).\tag{S3.3}\label{eq:S3.3}
\end{equation}
The proof is straightforward, as any unitary operator preserves the
inner product \cite{key-S4}.
\item \emph{Monotonicity}. In QI, an operation cannot reduce fidelity. However,
in API, a computational operation may reduce, preserve, or increase
fidelity. While a U-gate preserves fidelity, a G-gate can increase
(e.g., $G=2I$) or reduce (e.g., $G=(1/2)I$) fidelity. In API, fidelity
does not satisfy a specific monotonicity criterion.
\item \emph{Composite systems}. In multi-anbit computational systems composed
by using the tensor product \cite{key-S2}, fidelity fulfills the
same multiplicativity condition as in QI \cite{key-S5}:
\begin{equation}
F\bigl(\bigl|\psi_{X}\bigr\rangle\otimes\bigl|\psi_{Y}\bigr\rangle,\bigl|\varphi_{X}\bigr\rangle\otimes\bigl|\varphi_{Y}\bigr\rangle\bigr)=F\bigl(\bigl|\psi_{X}\bigr\rangle,\bigl|\varphi_{X}\bigr\rangle\bigr)\cdot F\bigl(\bigl|\psi_{Y}\bigr\rangle,\bigl|\varphi_{Y}\bigr\rangle\bigr).\tag{S3.4}\label{eq:S3.4}
\end{equation}
In contrast, in multi-anbit computational systems composed via the
Cartesian product \cite{key-S2}, fidelity satisfies the triangle
inequality:
\begin{equation}
F\bigl(\bigl|\psi_{X}\bigr\rangle\times\bigl|\psi_{Y}\bigr\rangle,\bigl|\varphi_{X}\bigr\rangle\times\bigl|\varphi_{Y}\bigr\rangle\bigr)\leq F\bigl(\bigl|\psi_{X}\bigr\rangle,\bigl|\varphi_{X}\bigr\rangle\bigr)+F\bigl(\bigl|\psi_{Y}\bigr\rangle,\bigl|\varphi_{Y}\bigr\rangle\bigr).\tag{S3.5}\label{eq:S3.5}
\end{equation}
The proof of these properties is direct by using the definition of
the inner product in the tensor product space and in the Cartesian
product space.\footnote{Consider $\bigl|\Psi_{XY}\bigr\rangle=\bigl|\psi_{X}\bigr\rangle\otimes\bigl|\psi_{Y}\bigr\rangle$
and $\bigl|\Phi_{XY}\bigr\rangle=\bigl|\varphi_{X}\bigr\rangle\otimes\bigl|\varphi_{Y}\bigr\rangle$.
We find that:
\[
\left|\bigl\langle\Psi_{XY}\bigr|\Phi_{XY}\bigr\rangle\right|=\left|\bigl\langle\psi_{X}\bigr|\varphi_{X}\bigr\rangle\cdot\bigl\langle\psi_{Y}\bigr|\varphi_{Y}\bigr\rangle\right|=\left|\bigl\langle\psi_{X}\bigr|\varphi_{X}\bigr\rangle\right|\cdot\left|\bigl\langle\psi_{Y}\bigr|\varphi_{Y}\bigr\rangle\right|,
\]
which demonstrates Eq.\,(\ref{eq:S3.4}). Now, taking $\bigl|\Psi_{XY}\bigr\rangle=\bigl|\psi_{X}\bigr\rangle\times\bigl|\psi_{Y}\bigr\rangle$
and $\bigl|\Phi_{XY}\bigr\rangle=\bigl|\varphi_{X}\bigr\rangle\times\bigl|\varphi_{Y}\bigr\rangle$,
we note that:
\[
\left|\bigl\langle\Psi_{XY}\bigr|\Phi_{XY}\bigr\rangle\right|=\left|\bigl\langle\psi_{X}\bigr|\varphi_{X}\bigr\rangle+\bigl\langle\psi_{Y}\bigr|\varphi_{Y}\bigr\rangle\right|\leq\left|\bigl\langle\psi_{X}\bigr|\varphi_{X}\bigr\rangle\right|+\left|\bigl\langle\psi_{Y}\bigr|\varphi_{Y}\bigr\rangle\right|,
\]
which leads to Eq.\,(\ref{eq:S3.5}).}
\item \emph{Curvature}. In API, fidelity is a convex function in the first
entry:
\begin{equation}
F\left(\sum_{i}p_{i}\bigl|\psi_{i}\bigr\rangle,\bigl|\varphi\bigr\rangle\right)\leq\sum_{i}p_{i}F\left(\bigl|\psi_{i}\bigr\rangle,\bigl|\varphi\bigr\rangle\right),\tag{S3.6}\label{eq:S3.6}
\end{equation}
given that $\left|\sum_{i}p_{i}\bigl\langle\psi_{i}\bigr|\varphi\bigr\rangle\right|\leq\sum_{i}p_{i}\left|\bigl\langle\psi_{i}\bigr|\varphi\bigr\rangle\right|$,
which demonstrates this property. By symmetry, fidelity is also convex
in the second entry.
\end{enumerate}

\subsection*{3.2 Normalized fidelity}

\addcontentsline{toc}{subsection}{3.2 Normalized fidelity}

\noindent Keeping in mind that, unlike in QI, fidelity lacks geometric
intuitiveness in API, primarily because its maximum value differs
from unity; a natural approach is to normalize this parameter as follows:
\begin{equation}
F_{\mathrm{N}}\left(\bigl|\psi\bigr\rangle,\bigl|\varphi\bigr\rangle\right)\coloneqq\frac{F\left(\bigl|\psi\bigr\rangle,\bigl|\varphi\bigr\rangle\right)}{F\left(\bigl|\psi\bigr\rangle,\bigl|\psi\bigr\rangle\right)}=\left|\frac{\bigl\langle\psi\bigr|\varphi\bigr\rangle}{\bigl\langle\psi\bigr|\psi\bigr\rangle}\right|.\tag{S3.7}\label{eq:S3.7}
\end{equation}
Specifically, $F_{\mathrm{N}}$ quantifies the normalized projection
of $\bigl|\varphi\bigr\rangle$ onto $\left|\psi\right\rangle $.
In this way, the extremal values of $F_{\mathrm{N}}$ range from 0
(when $\bigl|\varphi\bigr\rangle\bot\bigl|\psi\bigr\rangle$) to 1
(when $\bigl|\varphi\bigr\rangle=\bigl|\psi\bigr\rangle$). Nevertheless,
this normalization breaks both the symmetry and the curvature properties
of $F$. All other properties of $F$ discussed above remain valid
for $F_{\mathrm{N}}$.

\subsection*{3.3 Trace distance}

\addcontentsline{toc}{subsection}{3.3 Trace distance}

\noindent Now, we evaluate the suitability of the quantum trace distance
in the context of API. To this end, we should describe two distinct
anbits ($i=1,2$):
\begin{equation}
\bigr|\psi_{i}\bigr\rangle=r_{i}\left(\cos\frac{\theta_{i}}{2}\bigl|0\bigr\rangle+e^{\mathrm{j}\varphi_{i}}\sin\frac{\theta_{i}}{2}\bigl|1\bigr\rangle\right),\tag{S3.8}\label{eq:S3.8}
\end{equation}
as a function of their associated \emph{classical} density operators
(see Supplementary Note 2):
\begin{align}
\widehat{\rho}_{i} & =\bigr|\psi_{i}\bigr\rangle\bigl\langle\psi_{i}\bigr|\nonumber \\
 & =r_{i}^{2}\left[\cos^{2}\frac{\theta_{i}}{2}\left|0\right\rangle \left\langle 0\right|+\cos\frac{\theta_{i}}{2}\sin\frac{\theta_{i}}{2}\left(e^{-\mathrm{j}\varphi_{i}}\left|0\right\rangle \left\langle 1\right|+e^{\mathrm{j}\varphi_{i}}\left|1\right\rangle \left\langle 0\right|\right)+\sin^{2}\frac{\theta_{i}}{2}\left|1\right\rangle \left\langle 1\right|\right],\tag{S3.9}\label{eq:S3.9}
\end{align}

\noindent which correspond to the following density matrices in the
standard vector basis $\left\{ \left|0\right\rangle ,\left|1\right\rangle \right\} $:\footnote{In API, the density matrix $\rho_{i}$ is obtained from the density
operator $\widehat{\rho}_{i}$ expressed in the standard vector basis
$\left\{ \left|0\right\rangle ,\left|1\right\rangle \right\} $ as
in QI. The matrix element $\left(\rho_{i}\right)_{nm}$, corresponding
to the $n$-th row and $m$-th column, is given by $\left(\rho_{i}\right)_{nm}=\bigl\langle n\bigr|\widehat{\rho}_{i}\bigr|m\bigr\rangle$,
with $n,m\in\left\{ 0,1\right\} $.}
\begin{equation}
\rho_{i}=r_{i}^{2}\left(\begin{array}{cc}
\cos^{2}\frac{\theta_{i}}{2} & e^{-\mathrm{j}\varphi_{i}}\cos\frac{\theta_{i}}{2}\sin\frac{\theta_{i}}{2}\\
e^{\mathrm{j}\varphi_{i}}\cos\frac{\theta_{i}}{2}\sin\frac{\theta_{i}}{2} & \sin^{2}\frac{\theta_{i}}{2}
\end{array}\right)\equiv\frac{r_{i}}{2}\left(r_{i}I+\mathbf{r}_{i}\cdot\boldsymbol{\sigma}\right),\tag{S3.10}\label{eq:S3.10}
\end{equation}
where: 
\begin{equation}
\mathbf{r}_{i}=r_{i}\left(\sin\theta_{i}\cos\varphi_{i}\hat{\mathbf{x}}+\sin\theta_{i}\sin\varphi_{i}\hat{\mathbf{y}}+\cos\theta_{i}\hat{\mathbf{z}}\right),\tag{S3.11}\label{eq:S3.11}
\end{equation}
is the position vector (or Bloch vector) in the GBS and $\boldsymbol{\sigma}=\left(\sigma_{x},\sigma_{y},\sigma_{z}\right)$
are the Pauli matrices. Here, we define the \emph{trace distance}
in API using the same expression as in QI \cite{key-S6}:
\begin{equation}
D\left(\rho_{1},\rho_{2}\right)\coloneqq\frac{1}{2}\mathrm{Tr}\bigl|\rho_{1}-\rho_{2}\bigr|=\frac{1}{2}\mathrm{Tr}\sqrt{\left(\rho_{1}-\rho_{2}\right)^{2}}.\tag{S3.12}\label{eq:S3.12}
\end{equation}

In QI, the utility of this parameter lies in the fact that it defines
a metric that quantifies the distinguishability between the states
$\rho_{1}$ and $\rho_{2}$, reflecting their Euclidean distance in
the Bloch sphere. Accordingly, the trace distance will be meaningful
in the context of API if and only if $D\left(\rho_{1},\rho_{2}\right)$
reflects the Euclidean distance in the GBS between the anbits defined
in Eq.\,(\ref{eq:S3.10}). To verify this condition, we begin by
rewriting $\rho_{1}-\rho_{2}$ in the form:
\begin{equation}
\rho_{1}-\rho_{2}=\frac{1}{2}\left(r_{1}^{2}-r_{2}^{2}\right)I+\frac{1}{2}\left(r_{1}\mathbf{r}_{1}-r_{2}\mathbf{r}_{2}\right)\cdot\boldsymbol{\sigma},\tag{S3.13}\label{eq:S3.13}
\end{equation}
which leads to:
\begin{align}
\bigl|\rho_{1}-\rho_{2}\bigr| & =\sqrt{\left(\rho_{1}-\rho_{2}\right)^{2}}\nonumber \\
 & =\sqrt{\frac{1}{4}\left(r_{1}^{2}-r_{2}^{2}\right)^{2}I+\frac{1}{2}\left(r_{1}^{2}-r_{2}^{2}\right)\left(r_{1}\mathbf{r}_{1}-r_{2}\mathbf{r}_{2}\right)\cdot\boldsymbol{\sigma}+\frac{1}{4}d^{2}\left(r_{1}\mathbf{r}_{1},r_{2}\mathbf{r}_{2}\right)I},\tag{S3.14}\label{eq:S3.14}
\end{align}
where $d$ is the Euclidean distance. If $r_{1}=r_{2}=1$, the GBS
reduces to the Bloch sphere and, hence, the above equation becomes:
\begin{equation}
\bigl|\rho_{1}-\rho_{2}\bigr|=\frac{1}{2}d\left(\mathbf{r}_{1},\mathbf{r}_{2}\right)I,\tag{S3.15}\label{eq:S3.15}
\end{equation}
ensuring that:
\begin{equation}
D\left(\rho_{1},\rho_{2}\right)=\frac{1}{2}\mathrm{Tr}\bigl|\rho_{1}-\rho_{2}\bigr|\equiv\frac{1}{2}d\left(\mathbf{r}_{1},\mathbf{r}_{2}\right),\tag{S3.16}\label{eq:S3.16}
\end{equation}
with the factor 1/2 accounting for the geometric scaling inherent
to the construction of the Bloch sphere \cite{key-S6}. However, in
API, the radius of the GBS may differ from unity; in general, $r_{1}\neq r_{2}\neq1$.
Therefore, the trace distance does not accurately represent the Euclidean
distance between the anbits $\bigr|\psi_{1}\bigr\rangle$ and $\bigr|\psi_{2}\bigr\rangle$,
that is:

\newpage{}

\begin{equation}
D\left(\rho_{1},\rho_{2}\right)\neq\frac{1}{2}d\left(\mathbf{r}_{1},\mathbf{r}_{2}\right).\tag{S3.17}\label{eq:S3.17}
\end{equation}
Consequently, the trace distance is not an appropriate metric in API
for quantifying the similarity between different anbits. Instead,
a distinct metric should be introduced: \emph{the GBS distance}.

\subsection*{3.4 GBS distance}

\addcontentsline{toc}{subsection}{3.4 GBS distance}

\noindent The GBS distance serves as the analogue of the quantum trace
distance within the API framework. Considering that the trace distance
quantifies the Euclidean distance between quantum bits (qubits) on
the Bloch sphere {[}Eq.\,(\ref{eq:S3.16}){]}, we introduce the GBS
distance as an intuitive and formally analogous metric for evaluating
the Euclidean distance between anbits within the GBS:
\begin{equation}
D_{\mathrm{GBS}}\left(\bigl|\psi_{1}\bigr\rangle,\bigl|\psi_{2}\bigr\rangle\right)\coloneqq\frac{1}{2}d\bigl(\mathbf{r}_{1},\mathbf{r}_{2}\bigr)=\frac{1}{2}\left\Vert \mathbf{r}_{1}-\mathbf{r}_{2}\right\Vert =\frac{1}{2}\sqrt{\bigl\langle\mathbf{r}_{1}-\mathbf{r}_{2},\mathbf{r}_{1}-\mathbf{r}_{2}\bigr\rangle},\tag{S3.18}\label{eq:S3.18}
\end{equation}
being $d$ the Euclidean distance - defined via the Euclidean norm
in $\mathbb{R}^{3}$ - and being $\mathbf{r}_{1,2}$ the position
vectors associated with the anbits $\bigl|\psi_{1,2}\bigr\rangle$
in the GBS, given by Eq.\,(\ref{eq:S3.11}). In addition, note that
the factor 1/2 is included to account for the geometric scaling inherent
to the construction of the GBS (see Supporting Information of ref.\,\cite{key-S2}).
Likewise, it is noteworthy that $D_{\mathrm{GBS}}$ can also be applied
to mixed classical states, enabling a comparison between different
encoders. Specifically, the GBS distance satisfies the following properties:
\begin{enumerate}
\item \emph{Extremal values}. The GBS distance reaches its minimum value
$D_{\mathrm{GBS}}=0$ when the anbits are the same, and its maximum
value $D_{\mathrm{GBS}}=\left(r_{1}+r_{2}\right)/2$ when the anbits
are orthogonal (opposite points on the GBS). Unlike the trace distance,
this metric is not normalized, reflecting the arbitrary radius of
the GBS.
\item \emph{Metric}. It is straightforward to verify that $D_{\mathrm{GBS}}$
is a metric (positivity, symmetry, and triangle inequality are satisfied).
This property is directly inherited from the Euclidean distance in
Eq.\,(\ref{eq:S3.18}).
\item \emph{Base independent}. This property emerges from the Euclidean
distance, which is a base-independent application.
\item \emph{Unitary invariance}. The GBS distance is invariant under unitary
operations:
\begin{equation}
D_{\mathrm{GBS}}\bigl(\widehat{\mathrm{U}}\bigl|\psi_{1}\bigr\rangle,\widehat{\mathrm{U}}\bigl|\psi_{2}\bigr\rangle\bigr)=D_{\mathrm{GBS}}\bigl(\bigl|\psi_{1}\bigr\rangle,\bigl|\psi_{2}\bigr\rangle\bigr).\tag{S3.19}\label{eq:S3.19}
\end{equation}
A unitary anbit operation is geometrically equivalent to a rotation
of the anbits $\bigl|\psi_{1}\bigr\rangle$ and $\bigl|\psi_{2}\bigr\rangle$
on the GBS \cite{key-S2}. Hence, this operation does not modify the
Euclidean distance.
\item \emph{Monotonicity}. Within the APC framework, an anbit operation
can reduce, preserve, or increase the Euclidean distance between states
on the GBS. Consequently, $D_{\mathrm{GBS}}$ does not satisfy a specific
monotonicity criterion.
\item \emph{Composite systems}. The metric $D_{\mathrm{GBS}}$ does not
apply to multi-anbit states, constructed via the tensor or the Cartesian
products. This limitation can, in principle, be overcome by generalizing
the definition of $D_{\mathrm{GBS}}$ to accommodate comparisons between
multi-anbit states, an extension that lies beyond the scope of the
present work.
\end{enumerate}

\subsection*{3.5 State distance}

\addcontentsline{toc}{subsection}{3.5 State distance}

\noindent An alternative parameter for comparing anbits is the state
distance, which is applicable to both simple and composite systems
within the API framework, and is defined as follows:
\begin{align}
D_{\mathrm{S}}\left(\bigl|\psi_{1}\bigr\rangle,\bigl|\psi_{2}\bigr\rangle\right) & \coloneqq\left\Vert \bigl|\psi_{1}\bigr\rangle-\bigl|\psi_{2}\bigr\rangle\right\Vert =\sqrt{\bigl\langle\psi_{1}-\psi_{2}\bigr|\psi_{1}-\psi_{2}\bigr\rangle}\nonumber \\
 & =\sqrt{\bigl\langle\psi_{1}\bigr|\psi_{1}\bigr\rangle+\bigl\langle\psi_{2}\bigr|\psi_{2}\bigr\rangle-2\textrm{Re}\left\{ \bigl\langle\psi_{1}\bigr|\psi_{2}\bigr\rangle\right\} }.\tag{S3.20}\label{eq:S3.20}
\end{align}
In contrast to the GBS distance, the state distance is defined in
terms of the norm induced by the inner product of the Hilbert space
to which the states belong. This definition allows the state distance
to be applied not only to single-anbit states (simple systems), but
also to multi-anbit states (composite systems). Concretely, the state
distance fulfills the following properties:
\begin{enumerate}
\item \emph{Extremal values}. The state distance reaches its minimum value
$D_{\mathrm{S}}=0$ when the anbits are the same, and its maximum
value $D_{\mathrm{S}}=\sqrt{r_{1}^{2}+r_{2}^{2}}$ when the anbits
are orthogonal (opposite points on the GBS). 
\item \emph{Metric}. It is direct to verify that $D_{\mathrm{S}}$ is a
metric, as positivity, symmetry, and triangle inequality are satisfied.
These properties are inherited from the norm of the Hilbert space.
\item \emph{Base independent}. This property directly emerges from the norm
of the Hilbert space, which is a base-independent application.
\item \emph{Unitary invariance}. The state distance is invariant under U-gates,
as any unitary operator preserves the norm:
\begin{align}
D_{\mathrm{S}}\bigl(\widehat{\mathrm{U}}\bigl|\psi_{1}\bigr\rangle,\widehat{\mathrm{U}}\bigl|\psi_{2}\bigr\rangle\bigr) & =\left\Vert \widehat{\mathrm{U}}\bigl|\psi_{1}\bigr\rangle-\widehat{\mathrm{U}}\bigl|\psi_{2}\bigr\rangle\right\Vert =\left\Vert \widehat{\mathrm{U}}\left(\bigl|\psi_{1}\bigr\rangle-\bigl|\psi_{2}\bigr\rangle\right)\right\Vert \nonumber \\
 & =\left\Vert \bigl|\psi_{1}\bigr\rangle-\bigl|\psi_{2}\bigr\rangle\right\Vert =D_{\mathrm{S}}\bigl(\bigl|\psi_{1}\bigr\rangle,\bigl|\psi_{2}\bigr\rangle\bigr).\tag{S3.21}\label{eq:S3.21}
\end{align}
\item \emph{Monotonicity}. Within the APC framework, a computational operation
can reduce, preserve, or increase the norm of an arbitrary state $\bigl|\chi\bigr\rangle$.
Hence, setting $\bigl|\chi\bigr\rangle\equiv\bigl|\psi_{1}\bigr\rangle-\bigl|\psi_{2}\bigr\rangle$,
we demonstrate that $D_{\mathrm{S}}$ does not satisfy a specific
monotonicity criterion.
\item \emph{Composite systems}. In multi-anbit computational systems composed
by using the tensor product, it is direct to demonstrate that:
\begin{equation}
D_{\mathrm{S}}\bigl(\bigl|\psi_{X}\bigr\rangle\otimes\bigl|\alpha_{Y}\bigr\rangle,\bigl|\varphi_{X}\bigr\rangle\otimes\bigl|\alpha_{Y}\bigr\rangle\bigr)=\left\Vert \bigl|\alpha_{Y}\bigr\rangle\right\Vert D_{\mathrm{S}}\bigl(\bigl|\psi_{X}\bigr\rangle,\bigl|\varphi_{X}\bigr\rangle\bigr).\tag{S3.22}\label{eq:S3.22}
\end{equation}
Contrariwise, in multi-anbit systems composed via the Cartesian product,
the state distance satisfies the triangle inequality (see Appendix
A, on p.\,\pageref{sec:APPENDIX_A}):
\begin{equation}
D_{\mathrm{S}}\bigl(\bigl|\psi_{X}\bigr\rangle\times\bigl|\psi_{Y}\bigr\rangle,\bigl|\varphi_{X}\bigr\rangle\times\bigl|\varphi_{Y}\bigr\rangle\bigr)\leq D_{\mathrm{S}}\bigl(\bigl|\psi_{X}\bigr\rangle,\bigl|\varphi_{X}\bigr\rangle\bigr)+D_{\mathrm{S}}\bigl(\bigl|\psi_{Y}\bigr\rangle,\bigl|\varphi_{Y}\bigr\rangle\bigr).\tag{S3.23}\label{eq:S3.23}
\end{equation}
\item \emph{Relation between $D_{\mathrm{S}}$ and $F$}. Operating with
orthogonal states, it follows that:
\begin{equation}
D_{\mathrm{S}}\left(\bigl|\psi_{1}\bigr\rangle,\bigl|\psi_{2}\bigr\rangle\right)=\sqrt{F\left(\bigl|\psi_{1}\bigr\rangle,\bigl|\psi_{1}\bigr\rangle\right)+F\left(\bigl|\psi_{2}\bigr\rangle,\bigl|\psi_{2}\bigr\rangle\right)}.\tag{S3.24}\label{eq:S3.24}
\end{equation}
The proof of this equation is straightforward from the definitions
of fidelity {[}Eq.\,(\ref{eq:S3.2}){]} and the state distance {[}Eq.\,(\ref{eq:S3.20}){]}.
\end{enumerate}

\subsection*{3.6 KL divergence}

\addcontentsline{toc}{subsection}{3.6 KL divergence}

\noindent As in DI, in API, the Kullback-Leibler (KL) divergence -
or relative entropy - allows us to quantify how distinguishable two
different pmfs $\mathbf{p}=\left(p_{1},\ldots,p_{M}\right)$ and $\mathbf{q}=\left(q_{1},\ldots,q_{M}\right)$
are prior to the encoder block \cite{key-S7}:
\begin{equation}
D\left(\mathbf{p}\bigr\Vert\mathbf{q}\right)\coloneqq\sum_{i}p_{i}\log_{2}\frac{p_{i}}{q_{i}},\ \ \ \textrm{(bits)}.\tag{S3.25}\label{eq:S3.25}
\end{equation}
In addition, the KL divergence enables a comparison between the mixed
classical states $\bigl|\psi_{X}\bigr\rangle=\sum_{i}p_{i}\bigl|\psi_{i}\bigr\rangle$
and $\bigl|\varphi_{X}\bigr\rangle=\sum_{i}q_{i}\bigl|\varphi_{i}\bigr\rangle$,
each associated with a specific encoder, by means of a large number
($m$) of anbit measurements performed on the system.\footnote{The theory of anbit measurement is detailed in Supplementary Note
5.} In particular, the probability of misidentifying $\bigl|\varphi_{X}\bigr\rangle$
as $\bigl|\psi_{X}\bigr\rangle$ after $m$ anbit measurements is
asymptotically equivalent to the probability of misidentifying $\mathbf{q}$
as $\mathbf{p}$ \cite{key-S6,key-S7}:
\begin{equation}
p\left(\bigl|\psi_{X}\bigr\rangle\bigr|\bigl|\varphi_{X}\bigr\rangle\right)=p\left(\mathbf{p}\bigr|\mathbf{q}\right)\simeq2^{-m\cdot D(\mathbf{p}\bigl\Vert\mathbf{q})}.\tag{S3.26}\label{eq:S3.26}
\end{equation}
Although the KL divergence is not a true metric - being neither symmetric
nor satisfying the triangle inequality - it is maybe the most suitable
parameter for comparing mixed classical states. By virtue of the positivity
property of $D\left(\mathbf{p}\bigr\Vert\mathbf{q}\right)$ \cite{key-S7},
a larger KL divergence implies greater statistical distinguishability
between the encoders and their associated analog constellations. 

\subsection*{3.7 Applications}

\addcontentsline{toc}{subsection}{3.7 Applications}

\noindent As discussed throughout the main text, the primary applications
of the state-comparative parameters introduced above are: 1) establishing
design criteria for conceiving analog constellations in the GBS, 2)
quantifying the error associated with an anbit estimation at the receiver,
3) characterizing the non-ideal behavior of basic PIP components (see Supplementary Note~6).

\newpage{}

\section*{Supplementary Note 4: differential opto-electrical converters\label{sec:4}}

\addcontentsline{toc}{section}{Supplementary Note 4: differential opto-electrical converters}

\noindent In this section, we analyze in detail the differential opto-electrical
(O/E) converters shown in Fig.\,4 of the main text, which has been
replicated in Fig.\,S2 by incorporating the intermediate complex
envelopes generated at the outputs of the optical devices, subsequently
employed in the mathematical discussions.
\noindent \begin{center}
\includegraphics[width=15.6cm,height=12cm,keepaspectratio]{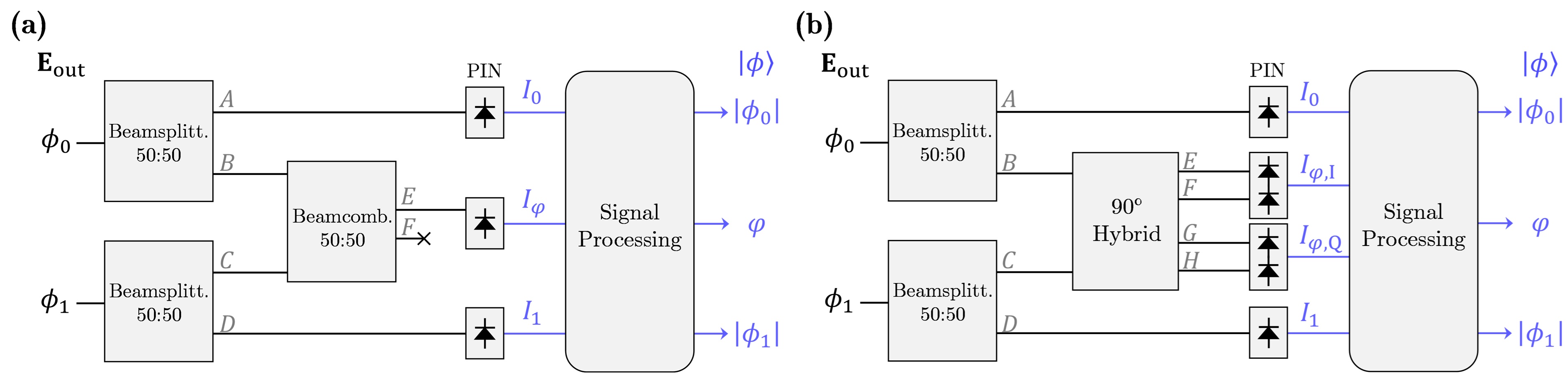}
\par\end{center}

\noindent {\small{}Figure S2. Differential O/E converters. (a) Unbalanced
architecture. (b) Quadrature architecture. Here, we detail the complex
envelopes generated at the outputs of the optical devices, denoted
by capital letters $A$, $B$, $C$, etc.\\ }{\small \par}

The electric field $\mathbf{E}_{\mathrm{out}}$ at the channel output
is composed of two complex envelopes $\phi_{0}$ and $\phi_{1}$ with
a phase shift $\varphi=\arg\left(\phi_{1}\right)-\arg\left(\phi_{0}\right)$
between them, where ``$\arg$'' is the argument of a complex number.
Here, the purpose of the differential O/E converters is to recover
the effective degrees of freedom (EDFs) $\left|\phi_{0}\right|$,
$\left|\phi_{1}\right|$, and $\varphi$ that constitute the received
anbit $\left|\phi\right\rangle =\left|\phi_{0}\right|\left|0\right\rangle +e^{\mathrm{j}\varphi}\left|\phi_{1}\right|\left|1\right\rangle $,
using photocurrents (denoted by the letter $I$). These photocurrents
are derived from the transfer matrices of the 50:50 beam splitter
(BS), implemented via a Y-junction \cite{key-S8}, the 50:50 beam
combiner (BC) \cite{key-S9}, and the 90-degree hybrid \cite{key-S10}:
\begin{align}
T_{\mathrm{BS}} & =\frac{e^{\mathrm{j}\delta}}{\sqrt{2}}\left(\begin{array}{c}
1\\
1
\end{array}\right),\ \ \ T_{\mathrm{BC}}=\frac{e^{\mathrm{j}\delta}}{\sqrt{2}}\left(\begin{array}{cc}
1 & \mathrm{j}\\
\mathrm{j} & 1
\end{array}\right),\ \ \ T_{\mathrm{hybrid}}=\frac{e^{\mathrm{j}\delta}}{2}\left(\begin{array}{cc}
1 & 1\\
1 & -1\\
1 & \mathrm{j}\\
1 & -\mathrm{j}
\end{array}\right).\tag{S4.1}\label{eq:S4.1}
\end{align}
In some references \cite{key-S11}, the global phase term $e^{\mathrm{j}\delta}$
is written of the form $\mathrm{j}e^{\mathrm{j}\delta}$, but both
expressions are physically equivalent since $\delta$ (or $\delta+\pi/2$)
describes an unknown design parameter that depends on the manufacturing
process. Accordingly, the global phase $\delta$ should be assumed
different in each device.

\paragraph{Unbalanced architecture.}

The envelopes at the outputs of the beam splitters are:
\begin{align}
\left(\begin{array}{c}
A\\
B
\end{array}\right) & =\frac{e^{\mathrm{j}\delta_{0}}}{\sqrt{2}}\left(\begin{array}{c}
1\\
1
\end{array}\right)\phi_{0}=\left(\begin{array}{c}
e^{\mathrm{j}\delta_{0}}\frac{1}{\sqrt{2}}\phi_{0}\\
e^{\mathrm{j}\delta_{0}}\frac{1}{\sqrt{2}}\phi_{0}
\end{array}\right),\tag{S4.2}\label{eq:S4.2}\\
\left(\begin{array}{c}
C\\
D
\end{array}\right) & =\frac{e^{\mathrm{j}\delta_{1}}}{\sqrt{2}}\left(\begin{array}{c}
1\\
1
\end{array}\right)\phi_{0}=\left(\begin{array}{c}
e^{\mathrm{j}\delta_{1}}\frac{1}{\sqrt{2}}\phi_{1}\\
e^{\mathrm{j}\delta_{1}}\frac{1}{\sqrt{2}}\phi_{1}
\end{array}\right),\tag{S4.3}\label{eq:S4.3}
\end{align}
and the envelopes at the outputs of the beam combiner are:
\begin{equation}
\left(\begin{array}{c}
E\\
F
\end{array}\right)=\frac{e^{\mathrm{j}\delta_{\varphi}}}{\sqrt{2}}\left(\begin{array}{cc}
1 & \mathrm{j}\\
\mathrm{j} & 1
\end{array}\right)\left(\begin{array}{c}
B\\
C
\end{array}\right)=\frac{e^{\mathrm{j}\delta_{\varphi}}}{2}\left(\begin{array}{c}
e^{\mathrm{j}\delta_{0}}\phi_{0}+\mathrm{j}e^{\mathrm{j}\delta_{1}}\phi_{1}\\
\mathrm{j}e^{\mathrm{j}\delta_{0}}\phi_{0}+e^{\mathrm{j}\delta_{1}}\phi_{1}
\end{array}\right).\tag{S4.4}\label{eq:S4.4}
\end{equation}
Hence, the photocurrents are of the form:
\begin{align}
I_{0} & =\mathcal{R}\left|A\right|^{2}=\frac{1}{2}\mathcal{R}\left|\phi_{0}\right|^{2},\tag{S4.5}\label{eq:S4.5}\\
I_{1} & =\mathcal{R}\left|D\right|^{2}=\frac{1}{2}\mathcal{R}\left|\phi_{1}\right|^{2},\tag{S4.6}\label{eq:S4.6}\\
I_{\varphi} & =\mathcal{R}\left|E\right|^{2}=\frac{1}{4}\mathcal{R}\left[\left|\phi_{0}\right|^{2}+\left|\phi_{1}\right|^{2}-2\left|\phi_{0}\right|\left|\phi_{1}\right|\sin\left(\delta_{1}-\delta_{0}+\varphi\right)\right].\tag{S4.7}\label{eq:S4.7}
\end{align}
where $\mathcal{R}$ is the responsivity of the PIN photodiodes. The
EDFs $\left|\phi_{0}\right|$, $\left|\phi_{1}\right|$, and $\varphi$
can be directly calculated from the above equations. 

Alternatively, the received anbit $\left|\phi\right\rangle =\left|\phi_{0}\right|\left|0\right\rangle +e^{\mathrm{j}\varphi}\left|\phi_{1}\right|\left|1\right\rangle $
can be expressed of the form:
\begin{equation}
\left|\phi\right\rangle =r\left(\cos\frac{\theta}{2}\left|0\right\rangle +e^{\mathrm{j}\varphi}\sin\frac{\theta}{2}\left|1\right\rangle \right),\tag{S4.8}\label{eq:S4.8}
\end{equation}
where the EDFs $r$, $\theta$, and $\varphi$ are also extracted
from the photocurrents using the following expressions in the unbalanced
architecture:
\begin{align}
r & =\sqrt{\frac{2\left(I_{0}+I_{1}\right)}{\mathcal{R}}},\tag{S4.9}\label{eq:S4.9}\\
\theta & =2\arctan\sqrt{\frac{I_{1}}{I_{0}}},\tag{S4.10}\label{eq:S4.10}\\
\varphi & =\arcsin\frac{I_{0}+I_{1}-2I_{\varphi}}{2\sqrt{I_{0}I_{1}}}+\delta_{0}-\delta_{1}.\tag{S4.11}\label{eq:S4.11}
\end{align}
In particular, Eqs.\,(\ref{eq:S4.9})-(\ref{eq:S4.11}) are employed
in the Materials and Methods section of the main text to recover the
anbits from the channel by assuming $\delta_{0}=\delta_{1}$ for simplicity.

\paragraph{Quadrature architecture.}

In this scheme, the envelopes $A$, $B$, $C$, and $D$ at the outputs
of the beam splitters are the same. Therefore, the envelopes at the
outputs of the hybrid are:
\begin{equation}
\left(\begin{array}{c}
E\\
F\\
G\\
H
\end{array}\right)=\frac{e^{\mathrm{j}\delta_{\varphi}}}{2}\left(\begin{array}{cc}
1 & 1\\
1 & -1\\
1 & \mathrm{j}\\
1 & -\mathrm{j}
\end{array}\right)\left(\begin{array}{c}
B\\
C
\end{array}\right)=\frac{e^{\mathrm{j}\delta_{\varphi}}}{2\sqrt{2}}\left(\begin{array}{c}
e^{\mathrm{j}\delta_{0}}\phi_{0}+e^{\mathrm{j}\delta_{1}}\phi_{1}\\
e^{\mathrm{j}\delta_{0}}\phi_{0}-e^{\mathrm{j}\delta_{1}}\phi_{1}\\
e^{\mathrm{j}\delta_{0}}\phi_{0}+\mathrm{j}e^{\mathrm{j}\delta_{1}}\phi_{1}\\
e^{\mathrm{j}\delta_{0}}\phi_{0}-\mathrm{j}e^{\mathrm{j}\delta_{1}}\phi_{1}
\end{array}\right),\tag{S4.12}\label{eq:S4.12}
\end{equation}
which yield the photocurrents:
\begin{align}
I_{0} & =\mathcal{R}\left|A\right|^{2}=\frac{1}{2}\mathcal{R}\left|\phi_{0}\right|^{2},\tag{S4.13}\label{eq:S4.13}\\
I_{1} & =\mathcal{R}\left|D\right|^{2}=\frac{1}{2}\mathcal{R}\left|\phi_{1}\right|^{2},\tag{S4.14}\label{eq:S4.14}\\
I_{\varphi,\mathrm{I}} & =\mathcal{R}\left\{ \left|E\right|^{2}-\left|F\right|^{2}\right\} =\frac{1}{2}\mathcal{R}\left|\phi_{0}\right|\left|\phi_{1}\right|\cos\left(\delta_{1}-\delta_{0}+\varphi\right),\tag{S4.15}\label{eq:S4.15}\\
I_{\varphi,\mathrm{Q}} & =\mathcal{R}\left\{ \left|G\right|^{2}-\left|H\right|^{2}\right\} =-\frac{1}{2}\mathcal{R}\left|\phi_{0}\right|\left|\phi_{1}\right|\sin\left(\delta_{1}-\delta_{0}+\varphi\right).\tag{S4.16}\label{eq:S4.16}
\end{align}
The EDFs $\left|\phi_{0}\right|$, $\left|\phi_{1}\right|$, and $\varphi$
(or equivalently $r$, $\theta$, and $\varphi$) are directly found
from the above photocurrents. 

\newpage{}

\section*{Supplementary Note 5: anbit measurement\label{sec:5}}

\addcontentsline{toc}{section}{Supplementary Note 5: anbit measurement}

\noindent In this section, we detail the theory of anbit measurement.
To this end, let us start by reviewing the information system shown
in Fig.\,1 of the main text. The originator source $X$ can generate
$M$ different symbols $x_{i}$ encoded into $M$ different anbits
$\bigl|\psi_{i}\bigr\rangle$ ($i=1,\ldots,M$), physically implemented
by a space-anbit modulator (Fig.\,3). These anbits are sent through
the channel, composed of the PIP circuits implementing the computational
gates. In ideal transmissions, free from noise and hardware imperfections,
we can receive $N$ different anbits $\bigl|\phi_{j}\bigr\rangle$,
which are later decoded into $N$ different symbols $y_{j}$ of the
recipient source ($j=1,\ldots,N$). The discretization of the GBS
by selecting the set of anbits allowed at the transmitter and receiver
is determined by the accuracy required to solve a specific computational
problem in APC. In non-ideal transmissions, each emitted anbit $\bigl|\psi_{i}\bigr\rangle$
is transformed into a noisy anbit $\left|\phi\right\rangle $, which
encodes an ideal anbit $\bigl|\phi_{j}\bigr\rangle$ along with system
noise and hardware imperfections. At the output of the O/E conversion,
we must recover the ideal anbit $\bigl|\phi_{j}\bigr\rangle$ by performing
a \emph{measurement} on the noisy anbit $\left|\phi\right\rangle $.

Before delving into the theory of anbit measurement, we should take
into account that the channel can be composed of reversible or irreversible
gates, described by non-singular or singular matrices, respectively
\cite{key-S2}. Therefore, we should distinguish between \emph{bijective}
and \emph{non-bijective} channels, respectively (see Fig.\,S3). For
the sake of simplicity, we first detail the theory of anbit measurement
for bijective channels and, subsequently, we extend the framework
for non-bijective channels.
\noindent \begin{center}
\includegraphics[width=15.6cm,height=8cm,keepaspectratio]{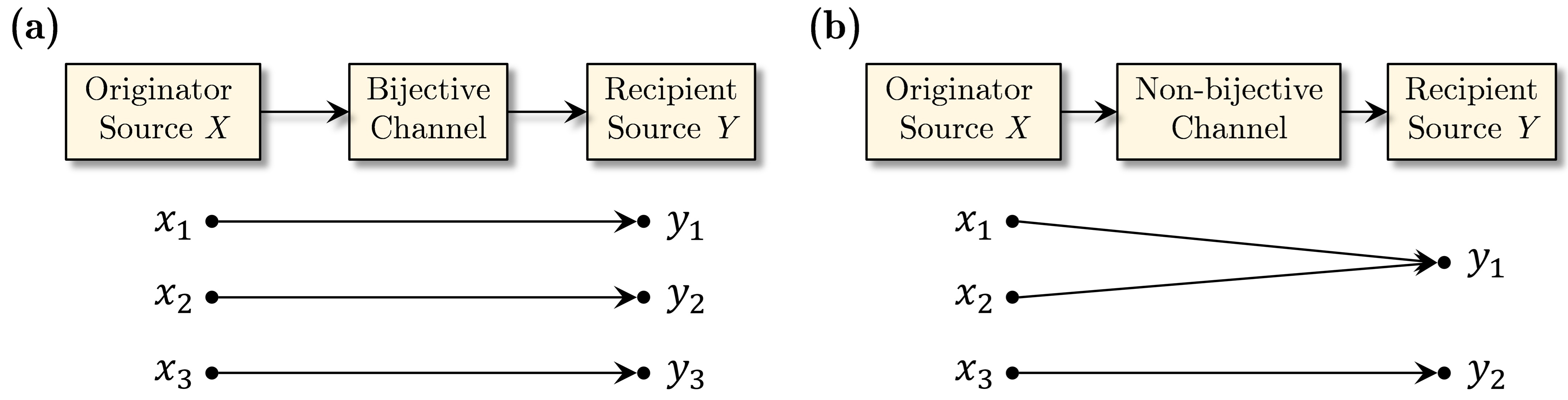}
\par\end{center}

\noindent {\small{}Figure S3. Illustrative example of symbol correspondence
between originator and recipient sources in ideal communication channels.
Symbol mappings are shown for (a) a bijective channel and (b) a non-bijective
channel. Both channels are assumed to be ideal, with no noise or hardware
imperfections.}{\small \par}

\subsection*{5.1 Bijective channels: reversible gates}

\addcontentsline{toc}{subsection}{5.1 Bijective channels: reversible gates}

\noindent In bijective channels, assuming ideal conditions without
noise or hardware imperfections, there is a one-to-one correspondence
between the emitted anbit $\bigl|\psi_{i}\bigr\rangle$ and the ideal
anbit $\bigl|\phi_{j}\bigr\rangle$ that should be received. This
scenario can be modeled by using the same subindex to describe the
mapping $\bigl|\psi_{i}\bigr\rangle\overset{1:1}{\leftrightarrow}\bigl|\phi_{i}\bigr\rangle$
(or $x_{i}\overset{1:1}{\leftrightarrow}y_{i}$), as the analog constellations
of the GBS at both the transmitter and receiver contain an equal number
of anbits (or symbols), i.e., $M=N$. The goal of an \emph{optimal}
anbit measurement is to select the ideal anbit $\bigl|\phi_{i}\bigr\rangle$
from the discrete set $\left\{ \bigl|\phi_{1}\bigr\rangle,\bigl|\phi_{2}\bigr\rangle,\ldots,\bigl|\phi_{N}\bigr\rangle\right\} $
by performing a decision on the noisy anbit $\left|\phi\right\rangle $
that should minimize the error probability or \emph{Symbol Error Rate}
(SER). 

The probability of error-free transmission is:
\begin{equation}
p_{\textrm{error-free}}=p\left[\bigcup_{i=1}^{N}\left\{ \left(x_{i},y_{i}\right)\right\} \right]=\sum_{i=1}^{N}p\left(x_{i},y_{i}\right).\tag{S5.1}\label{eq:S5.1}
\end{equation}
Thus, the SER may be defined as the complementary probability:
\begin{equation}
\textrm{SER}\coloneqq1-\sum_{i}p\left(x_{i},y_{i}\right)=1-\sum_{i}p\left(x_{i}\right)p\left(y_{i}|x_{i}\right),\tag{S5.2}\label{eq:S5.2}
\end{equation}
which is consistent with the definition employed in DI and QI \cite{key-S6,key-S7}.
The minimization of the SER is performed through the following three-step
procedure.

\paragraph{Step 1.}

As commented in the main text, the GBS does not preserve the linear
perturbation induced by an additive noise on the anbits. Therefore,
as a first step, it is necessary to select an appropriate vector space
$\mathbb{S}$ in which the anbits can be represented geometrically.
This space must preserve the form in which the channel introduces
noise and hardware imperfections in the transformation $\left|\psi_{i}\right\rangle \rightarrow\left|\phi\right\rangle $.
We denote this transformation in $\mathbb{S}$ as $\mathbf{r}_{i}\rightarrow\mathbf{r}$,
where $\mathbf{r}_{i}$ ($\mathbf{r}$) represents $\left|\psi_{i}\right\rangle $
($\left|\phi\right\rangle $) in $\mathbb{S}$. For instance, if the
channel induces the transformation $\left|\phi\right\rangle =\left|\psi_{i}\right\rangle +\left|n\right\rangle $,
where $\left|n\right\rangle $ denotes the system noise, then the
corresponding transformation in $\mathbb{S}$ must take the form $\mathbf{r}=\mathbf{r}_{i}+\mathbf{n}$,
with $\mathbf{n}$ being the representation of the noise ket $\left|n\right\rangle $
in $\mathbb{S}$.

Here, it is \emph{fundamental} that the mapping $x_{i}\rightarrow\left|\psi_{i}\right\rangle \rightarrow\mathbf{r}_{i}$
is one-to-one (i.e., distinct source symbols must correspond to distinct
representations in the chosen $\mathbb{S}$-space) in order to enable
the application of the law of total probability {[}Eq.\,(\ref{eq:S5.3});
see below{]}. Within this geometric representation, the channel is
described through the conditional pdfs $f\left(\mathbf{r}|\mathbf{r}_{i}\right)$.
The specific form of $f\left(\mathbf{r}|\mathbf{r}_{i}\right)$ is
determined by the computational operation of the channel along with
the system noise and hardware imperfections of the PIP circuits (Supplementary Note 6). Finally, in this framework, the ideal anbit
$\left|\phi_{i}\right\rangle $ that should be measured from $\left|\phi\right\rangle $
is denoted as $\mathbf{r}_{i}^{\prime}$ in $\mathbb{S}$. 

\paragraph{Step 2.}

Second, we must define an \emph{optimal} \emph{decision function }$g_{\mathrm{opt}}\bigl(\left|\phi\right\rangle \bigr)=\left|\phi_{i}\right\rangle $,
which should minimize the SER. From the continuous version of the
law of total probability \cite{key-S1}, we can write $p\left(y_{i}|x_{i}\right)$
of the form (see Appendix A, on p.\,\pageref{sec:APPENDIX_B}):
\begin{equation}
p\left(y_{i}|x_{i}\right)=\int_{\mathbb{S}}p\left(y_{i}|\mathbf{r}\right)f\left(\mathbf{r}|\mathbf{r}_{i}\right)\mathrm{d}^{s}r,\tag{S5.3}\label{eq:S5.3}
\end{equation}
with $s=\dim\left(\mathbb{S}\right)$. Hence, the SER becomes:
\begin{equation}
\textrm{SER}=1-\sum_{i}p\left(x_{i}\right)\int_{\mathbb{S}}p\left(y_{i}|\mathbf{r}\right)f\left(\mathbf{r}|\mathbf{r}_{i}\right)\mathrm{d}^{s}r.\tag{S5.4}\label{eq:S5.4}
\end{equation}
It follows that the minimization of the SER is achieved \emph{if and
only if} the probabilities $p\left(y_{i}|\mathbf{r}\right)$ are maximized,
as they constitute the only component of the subtrahend determined
by the measurement process.

\newpage{}

How can we maximize the probabilities $p\left(y_{i}|\mathbf{r}\right)$?
Inspired by DI \cite{key-S12}, we can establish that an optimal decision
function ($g_{\mathrm{opt}}$) selects the correct symbol $y_{i}$
(or anbit $\left|\phi_{i}\right\rangle $) by maximizing the ``a
posteriori'' probability $p\left(x_{i}|\mathbf{r}\right)$, where
$x_{i}$ is the transmitted symbol that corresponds to the symbol
$y_{i}$ at the receiver. This decision rule is referred to as the
\emph{maximum a posteriori probability} (MAP) criterion within the
context of DI, which may be mathematically formulated in API as:
\begin{equation}
\left|\phi_{i}\right\rangle =g_{\mathrm{opt}}\bigl(\left|\phi\right\rangle \bigr)=\arg\max_{i}\left\{ p\left(x_{i}|\mathbf{r}\right)\right\} .\tag{S5.5}\label{eq:S5.5}
\end{equation}
Since the probabilities $p\left(x_{i}|\mathbf{r}\right)$ are unknown,
we can take advantage of the continuous version of the \emph{Bayes
theorem} \cite{key-S1} to connect such probabilities with the conditional
pdfs, accounting for the channel properties:
\begin{equation}
p\left(x_{i}|\mathbf{r}\right)=p\left(x_{i}\right)\frac{f\left(\mathbf{r}|\mathbf{r}_{i}\right)}{f\left(\mathbf{r}\right)}.\tag{S5.6}\label{eq:S5.6}
\end{equation}
Thus, the MAP decision may be restated as:
\begin{equation}
\left|\phi_{i}\right\rangle =g_{\mathrm{opt}}\bigl(\left|\phi\right\rangle \bigr)=\arg\max_{i}\left\{ p\left(x_{i}\right)f\left(\mathbf{r}|\mathbf{r}_{i}\right)\right\} ,\tag{S5.7}\label{eq:S5.7}
\end{equation}
where the function $f\left(\mathbf{r}\right)$ can be omitted, as
it does not depend on the subindex $i$.

\paragraph{Step 3.}

Third, we introduce a set of decision regions $D_{1},\ldots,D_{N}\subseteq\mathbb{S}$
that allow us to optimize the decision function using the MAP criterion
via a geometrical approach. Here, any decision function $g$ (regardless
of whether it is optimal) is defined as:
\begin{equation}
\left|\phi_{i}\right\rangle =g\bigl(\left|\phi\right\rangle \bigr)\overset{\textrm{def}}{\Longleftrightarrow}\mathbf{r}\in D_{i}.\tag{S5.8}\label{eq:S5.8}
\end{equation}
The measurement is optimal ($g\equiv g_{\mathrm{opt}}$) if and only
if the decision regions minimize the SER. Using the MAP criterion,
the decision regions should be defined as:
\begin{equation}
D_{i}\coloneqq\left\{ \mathbf{r}\in\mathbb{S}/\ p\left(x_{i}|\mathbf{r}\right)>p\left(x_{j}|\mathbf{r}\right),\ \forall j\in\left\{ 1,\ldots,N\right\} /j\neq i\right\} ,\tag{S5.9}\label{eq:S5.9}
\end{equation}
with $D_{i}\cap D_{j}=\emptyset$. Therefore, the optimal region $D_{i}$
contains the vectors $\mathbf{r}\in\mathbb{S}$ where $p\left(x_{i}|\mathbf{r}\right)$
dominates over all other a posteriori probabilities. Once again, using
the Bayes theorem given by Eq.\,(\ref{eq:S5.6}) and noting that
$f\left(\mathbf{r}|\mathbf{r}_{i}\right)\equiv f\left(\mathbf{r}|\mathbf{r}_{i}^{\prime}\right)$
(as demonstrated below in Supplementary Note 6), the previous expression
becomes:
\begin{equation}
D_{i}=\left\{ \mathbf{r}\in\mathbb{S}/\ p\left(x_{i}\right)f\left(\mathbf{r}|\mathbf{r}_{i}^{\prime}\right)>p\left(x_{j}\right)f\left(\mathbf{r}|\mathbf{r}_{j}^{\prime}\right),\ \forall j\in\left\{ 1,\ldots,N\right\} /j\neq i\right\} ,\tag{S5.10}\label{eq:S5.10}
\end{equation}
which is Eq.\,(3) of the main text.\hspace*{10cm}$\boxempty$\\

All in all, we can derive the final expression of the SER in bijective
channels. From Eq.\,(\ref{eq:S5.8}), it follows that the probability
$p\left(y_{i}|\mathbf{r}\right)$ is given by the piecewise function:
\begin{equation}
p\left(y_{i}|\mathbf{r}\right)=\begin{cases}
1 & \mathbf{r}\in D_{i}\\
0 & \mathbf{r}\notin D_{i}
\end{cases}.\tag{S5.11}\label{eq:S5.11}
\end{equation}
Accordingly, Eq.\,(\ref{eq:S5.4}) reduces to the expression:
\begin{equation}
\textrm{SER}=1-\sum_{i}p\left(x_{i}\right)\int_{D_{i}}f\left(\mathbf{r}|\mathbf{r}_{i}^{\prime}\right)\mathrm{d}^{s}r,\tag{S5.12}\label{eq:S5.12}
\end{equation}
that is, Eq.\,(4) of the paper. In the next section, Supplementary
Note 6, we discuss the conditional pdfs, accounting for the computational
operation of the channel along with the system\textquoteright s physical
impairments (noise and hardware imperfections of the PIP devices). 

\subsection*{5.2 Non-bijective channels: irreversible gates\label{subsec:5.2-Non-bijective-channels}}

\addcontentsline{toc}{subsection}{5.2 Non-bijective channels: irreversible gates}

\noindent In non-bijective channels, assuming ideal conditions without
noise or hardware imperfections, two different anbits (or symbols)
at the transmitter may correspond to the same anbit (or symbol) at
the receiver, see Fig.\,S3(b). In APC, this situation is found when
the channel integrates an irreversible gate, described by a singular
matrix \cite{key-S2}. Figure S4 depicts two different mathematical
notations to describe a non-bijective correspondence between symbols
of the originator and recipient sources. The notation presented on
the right-hand side of Fig.\,S4 is of particular relevance.
\noindent \begin{center}
\includegraphics[width=15.6cm,height=8cm,keepaspectratio]{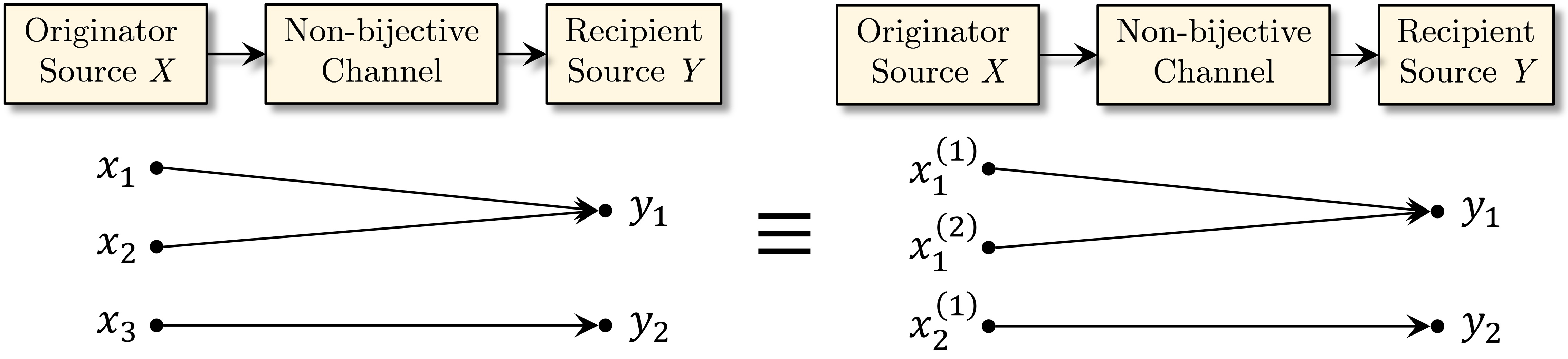}
\par\end{center}

\noindent \begin{center}
{\small{}Figure S4. Equivalent description of an ideal non-bijective
channel using different notations.\\}
\par\end{center}{\small \par}

Under ideal conditions, we can receive $N$ different anbits $\bigl\{\bigl|\phi_{i}\bigr\rangle\bigr\}_{i=1}^{N}$,
which are subsequently decoded into $N$ distinct symbols $\bigl\{ y_{i}\bigr\}_{i=1}^{N}$
of the recipient source. Each symbol $y_{i}$ corresponds to $g_{i}$
different symbols $\bigl\{ x_{i}^{\left(k\right)}\bigr\}_{k=1}^{g_{i}}$
of the originator source, which are encoded into $g_{i}$ anbits $\bigl\{\bigl|\psi_{i}^{\left(k\right)}\bigr\rangle\bigr\}_{k=1}^{g_{i}}$
in the GBS. In such a scenario, the parameter $g_{i}$ is termed as
the \emph{degree of degeneracy} of the symbol $y_{i}$ (or anbit $\bigl|\phi_{i}\bigr\rangle$).
Therefore, the number of distinct symbols that can be emitted by the
transmitter is:
\begin{equation}
M=\sum_{i=1}^{N}g_{i}>N.\tag{S5.13}\label{eq:S5.13}
\end{equation}
In non-bijective channels, there exits at least one symbol $y_{i}$
for which $g_{i}>1$. In contrast, in bijective channels, it follows
that $g_{i}=1$ for all $i=1,\ldots,N$, resulting in an equal number
of transmitted and received symbols (and anbits), that is, $M=N$.

Consequently, information is computed following the next flowchart
of transformations in the single-anbit Hilbert space $\mathscr{E}_{1}=\mathrm{span}\left\{ \left|0\right\rangle ,\left|1\right\rangle \right\} $:
\begin{equation}
x_{i}^{\left(1,\ldots,g_{i}\right)}\overset{g_{i}:g_{i}}{\leftrightarrow}\bigl|\psi_{i}^{\left(1,\ldots,g_{i}\right)}\bigr\rangle\underset{\textrm{channel}}{\longrightarrow}\bigl|\phi\bigr\rangle\underset{\textrm{measurement}}{\longrightarrow}\bigl|\phi_{i}\bigr\rangle\overset{1:1}{\leftrightarrow}y_{i},\tag{S5.14}\label{eq:S5.14}
\end{equation}
being the end-to-end symbol correspondence $x_{i}^{\left(k\right)}\rightarrow y_{i}$
a non-bijective mapping ($g_{i}:1$). This flowchart of transformations
can also be described in the $\mathbb{S}$-space:
\begin{equation}
x_{i}^{\left(1,\ldots,g_{i}\right)}\overset{g_{i}:g_{i}}{\leftrightarrow}\mathbf{r}_{i}^{\left(1,\ldots,g_{i}\right)}\underset{\textrm{channel}}{\longrightarrow}\mathbf{r}\underset{\textrm{measurement}}{\longrightarrow}\mathbf{r}_{i}^{\prime}\overset{1:1}{\leftrightarrow}y_{i}.\tag{S5.15}\label{eq:S5.15}
\end{equation}

Here, the probability of error-free transmission is:
\begin{equation}
p_{\textrm{error-free}}=p\left[\bigcup_{i=1}^{N}\bigcup_{k=1}^{g_{i}}\left\{ \bigl(x_{i}^{\left(k\right)},y_{i}\bigr)\right\} \right]=\sum_{i=1}^{N}\sum_{k=1}^{g_{i}}p\bigl(x_{i}^{\left(k\right)},y_{i}\bigr),\tag{S5.16}\label{eq:S5.16}
\end{equation}
and, therefore, the SER should be defined as:
\begin{equation}
\textrm{SER}\coloneqq1-\sum_{i,k}p\bigl(x_{i}^{\left(k\right)},y_{i}\bigr)=1-\sum_{i,k}p\bigl(x_{i}^{\left(k\right)}\bigr)p\bigl(y_{i}|x_{i}^{\left(k\right)}\bigr).\tag{S5.17}\label{eq:S5.17}
\end{equation}
The SER is minimized using the same three-step procedure applied in
the case of bijective channels.

\paragraph{Step 1.}

Selection of the vector space $\mathbb{S}$ in which the anbits are
represented geometrically. This step proceeds identically to the bijective
case and requires no further clarification.

\paragraph{Step 2.}

Second, we must define an optimal decision function ($g_{\mathrm{opt}}$)
that should minimize the SER. From the continuous version of the law
of total probability \cite{key-S1}, $p\bigl(y_{i}|x_{i}^{\left(k\right)}\bigr)$
can be recast of the form:
\begin{equation}
p\bigl(y_{i}|x_{i}^{\left(k\right)}\bigr)=\int_{\mathbb{S}}p\bigl(y_{i}|\mathbf{r}\bigr)f\bigl(\mathbf{r}|\mathbf{r}_{i}^{\left(k\right)}\bigr)\mathrm{d}^{s}r,\tag{S5.18}\label{eq:S5.18}
\end{equation}
which leads to the following expression for the SER:
\begin{equation}
\textrm{SER}=1-\sum_{i,k}p\bigl(x_{i}^{\left(k\right)}\bigr)\int_{\mathbb{S}}p\bigl(y_{i}|\mathbf{r}\bigr)f\bigl(\mathbf{r}|\mathbf{r}_{i}^{\left(k\right)}\bigr)\mathrm{d}^{s}r.\tag{S5.19}\label{eq:S5.19}
\end{equation}
This formulation implies that the SER is minimized if and only if
the conditional probabilities $p\left(y_{i}|\mathbf{r}\right)$ are
maximized. Consequently, an optimal decision function selects the
correct symbol $y_{i}$ (or anbit $\left|\phi_{i}\right\rangle $)
by maximizing the ``a posteriori'' probability: 
\begin{equation}
p\left[\bigcup_{k=1}^{g_{i}}\left\{ x_{i}^{\left(k\right)}|\mathbf{r}\right\} \right]=\sum_{k=1}^{g_{i}}p\bigl(x_{i}^{\left(k\right)}|\mathbf{r}\bigr),\tag{S5.20}\label{eq:S5.20}
\end{equation}
being $x_{i}^{\left(1,\ldots,g_{i}\right)}$ the symbols at the transmitter
that correspond to the symbol $y_{i}$ at the receiver. Consequently,
the MAP decision criterion is:
\begin{equation}
\left|\phi_{i}\right\rangle =g_{\mathrm{opt}}\bigl(\left|\phi\right\rangle \bigr)=\arg\max_{i}\left\{ \sum_{k=1}^{g_{i}}p\bigl(x_{i}^{\left(k\right)}|\mathbf{r}\bigr)\right\} .\tag{S5.21}\label{eq:S5.21}
\end{equation}
Next, using the Bayes theorem:
\begin{equation}
p\bigl(x_{i}^{\left(k\right)}|\mathbf{r}\bigr)=p\bigl(x_{i}^{\left(k\right)}\bigr)\frac{f\bigl(\mathbf{r}|\mathbf{r}_{i}^{\left(k\right)}\bigr)}{f\left(\mathbf{r}\right)},\tag{S5.22}\label{eq:S5.22}
\end{equation}
Eq.\,(\ref{eq:S5.21}) can be recast of the form:
\begin{equation}
\left|\phi_{i}\right\rangle =g_{\mathrm{opt}}\bigl(\left|\phi\right\rangle \bigr)=\arg\max_{i}\left\{ \sum_{k=1}^{g_{i}}p\bigl(x_{i}^{\left(k\right)}\bigr)f\bigl(\mathbf{r}|\mathbf{r}_{i}^{\left(k\right)}\bigr)\right\} ,\tag{S5.23}\label{eq:S5.23}
\end{equation}
omitting the function $f\left(\mathbf{r}\right)$, as it does not
depend on the subindex $i$.

\paragraph{Step 3.}

Third, we introduce a set of decision regions $D_{1},\ldots,D_{N}\subseteq\mathbb{S}$
that allow us to optimize the decision function using the MAP criterion
via a geometrical approach. Specifically, we select the anbit $\left|\phi_{i}\right\rangle $
if and only if $\mathbf{r}\in D_{i}$, which should be defined as:
\begin{equation}
D_{i}\coloneqq\left\{ \mathbf{r}\in\mathbb{S}/\sum_{k=1}^{g_{i}}p\bigl(x_{i}^{\left(k\right)}|\mathbf{r}\bigr)>\sum_{k=1}^{g_{j}}p\bigl(x_{j}^{\left(k\right)}|\mathbf{r}\bigr),\ \forall j\in\left\{ 1,\ldots,N\right\} /j\neq i\right\} ,\tag{S5.24}\label{eq:S5.24}
\end{equation}
with $D_{i}\cap D_{j}=\emptyset$. Using the Bayes theorem {[}Eq.\,(\ref{eq:S5.22}){]},
the decision regions may be described via the conditional pdfs, accounting
for the channel properties:
\begin{equation}
D_{i}=\left\{ \mathbf{r}\in\mathbb{S}\Bigr/\sum_{k=1}^{g_{i}}p\bigl(x_{i}^{\left(k\right)}\bigr)f\bigl(\mathbf{r}|\mathbf{r}_{i}^{\left(k\right)}\bigr)>\sum_{k=1}^{g_{j}}p\bigl(x_{j}^{\left(k\right)}\bigr)f\bigl(\mathbf{r}|\mathbf{r}_{j}^{\left(k\right)}\bigr),\ \forall j\in\left\{ 1,\ldots,N\right\} /j\neq i\right\} .\tag{S5.25}\label{eq:S5.25}
\end{equation}
The conditional pdfs $f\bigl(\mathbf{r}|\mathbf{r}_{i}^{\left(k\right)}\bigr)$
can be calculated from the theory reported in Supplementary Note 6,
where we demonstrate that:
\begin{equation}
f\bigl(\mathbf{r}|\mathbf{r}_{i}^{\left(k\right)}\bigr)=f\bigl(\mathbf{r}|\mathbf{r}_{i}^{\prime}\bigr),\ \ \ \forall k\in\left\{ 1,\ldots,g_{i}\right\} .\tag{S5.26}\label{eq:S5.26}
\end{equation}
This property allows us to connect the expression of the decision
regions between the bijective and non-bijective cases. It is worthy
to note that Eq.\,(\ref{eq:S5.25}) reduces to Eq.\,(\ref{eq:S5.10})
when $g_{i}=1$ for all $i=1,\ldots,N$.\hspace*{10cm}$\boxempty$\\

All in all, we can derive the final expression of the SER in non-bijective
channels. Using Eq.\,(\ref{eq:S5.11}), which also applies to non-bijective
channels, Eq.\,(\ref{eq:S5.19}) becomes:
\begin{equation}
\textrm{SER}=1-\sum_{i,k}p\bigl(x_{i}^{\left(k\right)}\bigr)\int_{D_{i}}f\bigl(\mathbf{r}|\mathbf{r}_{i}^{\left(k\right)}\bigr)\mathrm{d}^{s}r.\tag{S5.27}\label{eq:S5.27}
\end{equation}
Likewise, using Eq.\,(\ref{eq:S5.26}), note that the above equation
reduces to the expression of the SER in bijective channels {[}Eq.\,(\ref{eq:S5.12}){]}
when $g_{i}=1$ for all $i=1,\ldots,N$. 

Moreover, combining Eqs.\,(\ref{eq:S5.26}) and (\ref{eq:S5.27}),
it is straightforward to demonstrate the \emph{anbit measurement theorem}
in non-bijective channels. If the 3D regions defined by the noisy
anbits in the received constellation are not overlapped, then the
conditional pdfs are disjoint. Consequently, it is always possible
to define a set of decision regions $\left\{ D_{1},\ldots,D_{N}\right\} $
fulfilling the condition:
\begin{equation}
\int_{D_{i}}f\bigl(\mathbf{r}|\mathbf{r}_{i}^{\prime}\bigr)\mathrm{d}^{s}r=\int_{D_{i}}f\bigl(\mathbf{r}|\mathbf{r}_{i}^{\left(k\right)}\bigr)\mathrm{d}^{s}r=1,\tag{S5.28}\label{eq:S5.28}
\end{equation}
for all $i=1,\ldots,N$ and $k=1,\ldots,g_{i}$. This scenario ensures
a zero SER:
\begin{align}
\textrm{SER} & =1-\sum_{i,k}p\bigl(x_{i}^{\left(k\right)}\bigr)\int_{D_{i}}f\bigl(\mathbf{r}|\mathbf{r}_{i}^{\left(k\right)}\bigr)\mathrm{d}^{s}r=1-\sum_{i,k}p\bigl(x_{i}^{\left(k\right)}\bigr)=0.\tag{S5.29}\label{eq:S5.29}
\end{align}

\newpage{}

\section*{Supplementary Note 6: noise and hardware imperfections\label{sec:6}}

\addcontentsline{toc}{section}{Supplementary Note 6: noise and hardware imperfections}

\noindent The theory of anbit measurement, presented in Supplementary
Note 5, requires the calculation of the conditional pdfs $f\bigl(\mathbf{r}|\mathbf{r}_{i}^{\prime}\bigr)$
for both bijective and non-bijective channels. As discussed in the
main text, these conditional pdfs are governed by the statistical
properties of the main system\textquoteright s physical impairments:
noise and hardware imperfections in the PIP circuits.

In this supplementary note, we therefore examine the calculation of
these conditional pdfs for different classes of noise, while also
accounting for the non-ideal behavior of PIP devices. First, we investigate
additive noise on the EDFs of the anbits. Second, we outline how to
assess non-additive noise, i.e., non-additive perturbations acting
on the EDFs, which can emerge when optical non-linear effects are
stimulated in the PIP circuits. Third, we classify the noise sources
in PIP platforms as either additive or non-additive, and we identify
the dominant noise mechanisms in passive linear PIP circuits (i.e.,
circuits performing linear wave transformations and without integrated
optical amplifiers). Finally, we complete the theoretical framework
by incorporating hardware imperfections of the PIP circuitry together
with the system noise.

\subsection*{6.1 Additive anbit-amplitude noise\label{subsec:6.1}}

\addcontentsline{toc}{subsection}{6.1 Additive anbit-amplitude noise}

\noindent Here, we study \emph{additive noise} affecting either the
\emph{field} or the \emph{power} of an optical wave, as typically
encountered in information-processing systems exhibiting \emph{linear}
electromagnetic propagation within the channel \cite{key-S7}. Remarkably,
the main noise sources in API systems are of this type, as discussed
in detail on p.\,\pageref{subsec:6.5}. For example, the amplified
spontaneous emission (ASE) noise of an optical amplifier can be modeled
as an additive perturbation on the complex envelope of the electric
field \cite{key-S13}. In contrast, shot and thermal noises present
in an O/E conversion are typically modeled as additive perturbations
on the photocurrents, which are proportional to the power of the optical
field \cite{key-S14}. Hence, shot and thermal noises are examples
of additive noise acting on the power of an electromagnetic wave.

Specifically, an additive noise perturbing the field or the power
of a two-dimensional (2D) electromagnetic wave - with complex envelopes
$\psi_{0}$ and $\psi_{1}$ implementing an anbit\linebreak{}
$\left|\psi\right\rangle =\psi_{0}\left|0\right\rangle +\psi_{1}\left|1\right\rangle $
- can be described by a ket $\left|n\right\rangle =n_{0}\left|0\right\rangle +n_{1}\left|1\right\rangle $
acting on the anbit as $\left|\psi\right\rangle +\left|n\right\rangle $.
If the additive noise perturbs the \emph{field} of the electromagnetic
wave, it is direct to see that the resulting field is composed of
complex envelopes of the form $\psi_{k}+n_{k}$ ($k=0,1$). Accordingly,
the perturbed field is equivalent to an anbit of the form $\left|\psi\right\rangle +\left|n\right\rangle $.
Nevertheless, if the additive noise perturbs the \emph{power} of the
electromagnetic wave, the perturbed field is also equivalent to an
anbit of the form $\left|\psi\right\rangle +\left|n\right\rangle $,
provided that the phases of $n_{0}$ and $n_{1}$ (degrees of freedom
in the problem) are adequately selected to fulfill a specific \emph{phase
condition}, see Appendix C on p.\,\pageref{sec:APPENDIX_C}. Consequently,
both field and power perturbations can be commonly regarded as an
\emph{additive anbit-amplitude (AA) noise}, inducing an additive perturbation
on the anbit amplitudes. In this subsection, the goal is to analyze
how to calculate the conditional pdfs for an AA noise.

As a starting point, assume that we have a dominant AA noise - represented
by a random ket $\left|n\right\rangle $ - perturbing the ideal anbit
$\left|\phi_{i}\right\rangle $ that is expected to be measured. This
anbit is the error-free computational result of a single-anbit M-gate
applied to the transmitted anbit $\left|\psi_{i}\right\rangle $,
such that $\left|\phi_{i}\right\rangle =\widehat{\mathrm{M}}\left|\psi_{i}\right\rangle $
\cite{key-S2}. The gate may be a reversible or irreversible operation
and, therefore, the channel may be bijective or non-bijective. Both
cases are considered in our analysis. In this scenario, the channel
generates a noisy anbit $\left|\phi\right\rangle $ (i.e. the anbit
obtained at the output of the O/E conversion) of the form:
\begin{equation}
\left|\phi\right\rangle =\left|\phi_{i}\right\rangle +\left|n\right\rangle =\widehat{\mathrm{M}}\left|\psi_{i}\right\rangle +\left|n\right\rangle ,\tag{S6.1}\label{eq:S6.1}
\end{equation}
with $\left|n\right\rangle $ being statistically independent of the
API transmitter.

Following the anbit-measurement theory (see Supplementary Note 5),
we must represent the kets of the above equation in a vector space
$\mathbb{S}$ that preserves the form in which the channel transforms
the transmitted anbit $\left|\psi_{i}\right\rangle $ in Eq.\,(\ref{eq:S6.1}).
Accordingly, if the kets are represented by the following vectors
belonging to $\mathbb{S}$: $\left|\phi\right\rangle \rightarrow\mathbf{r}$,
$\left|\phi_{i}\right\rangle \rightarrow\mathbf{r}_{i}^{\prime}$,
$\left|\psi_{i}\right\rangle \rightarrow\mathbf{r}_{i}$, and $\left|n\right\rangle \rightarrow\mathbf{n}$;
then the representation of Eq.\,(\ref{eq:S6.1}) in the $\mathbb{S}$-space
must be of the form:
\begin{equation}
\mathbf{r}=\mathbf{r}_{i}^{\prime}+\mathbf{n}=\mathcal{M}\left(\mathbf{r}_{i}\right)+\mathbf{n},\ \ \ \textrm{[affine map]}\tag{S6.2}\label{eq:S6.2}
\end{equation}
with $\mathbf{r}_{i}^{\prime}=\mathcal{M}\left(\mathbf{r}_{i}\right)$
accounting for the computational operation $\left|\phi_{i}\right\rangle =\widehat{\mathrm{M}}\left|\psi_{i}\right\rangle $.
In such conditions, as demonstrated in Appendix D (p.\,\pageref{sec:APPENDIX_D}),
the conditional pdfs are governed by the expression:
\begin{equation}
f\left(\mathbf{r}|\mathbf{r}_{i}^{\prime}\right)=f\left(\mathbf{r}|\mathbf{r}_{i}\right)=f_{\mathbf{N}}\left(\mathbf{n}=\mathbf{r}-\mathbf{r}_{i}^{\prime}\right),\tag{S6.3}\label{eq:S6.3}
\end{equation}
being $f_{\mathbf{N}}\left(\mathbf{n}\right)$ the pdf of the AA noise
in the $\mathbb{S}$-space. Equation (\ref{eq:S6.3}) applies to both
bijective and non-bijective channels. In bijective channels, the mapping
$\mathbf{r}_{i}\rightarrow\mathbf{r}_{i}^{\prime}$ is $1:1$. In
non-bijective channels, the mapping $\mathbf{r}_{i}\rightarrow\mathbf{r}_{i}^{\prime}$
is $g_{i}:1$, that is, there are $g_{i}$ distinct vectors $\mathbf{r}_{i}^{\left(1,\ldots,g_{i}\right)}$
at the transmitter that correspond to $\mathbf{r}_{i}^{\prime}$ at
the receiver. In this case, Eq.\,(\ref{eq:S6.3}) can equivalently
be expressed as $f\left(\mathbf{r}|\mathbf{r}_{i}^{\prime}\right)=f\bigl(\mathbf{r}|\mathbf{r}_{i}^{\left(k\right)}\bigr)=f_{\mathbf{N}}\left(\mathbf{n}=\mathbf{r}-\mathbf{r}_{i}^{\prime}\right)$,
for all $k=1,\ldots,g_{i}$.

Now, the next task is to find a suitable $\mathbb{S}$-space satisfying
Eq.\,(\ref{eq:S6.3}). Let us start by discussing the natural candidate:
the GBS. As commented in the paper, the GBS does not constitute a
suitable candidate for the $\mathbb{S}$-space. The Bloch vector $\mathbf{r}$
associated with $\left|\phi\right\rangle $ cannot be calculated by
summing the Bloch vectors of $\left|\phi_{i}\right\rangle $ and $\left|n\right\rangle $,
denoted $\mathbf{r}_{i}^{\prime}$ and $\mathbf{n}$, respectively.
This can be verified by considering, e.g., $\left|\phi_{i}\right\rangle \equiv\left|0\right\rangle $
and $\left|n\right\rangle \equiv\left|1\right\rangle $, with $\mathbf{r}_{i}^{\prime}=\left(0,0,1\right)$
and $\mathbf{n}=\left(0,0,-1\right)$ (Cartesian coordinates). In
this example, we observe that the Bloch vector associated with $\left|\phi\right\rangle =\left|0\right\rangle +\left|1\right\rangle $
is found to be $\mathbf{r}=\left(\sqrt{2},0,0\right)\neq\mathbf{r}_{i}^{\prime}+\mathbf{n}$.
Consequently, \emph{the GBS is not a valid $\mathbb{S}$-space candidate}.
However, this conclusion \emph{does not preclude} the use of the GBS
for geometrically representing the anbits and the system noise. It
merely indicates that the GBS is not suitable for optimizing anbit
measurements.

Outstandingly, a valid (but not unique) geometric representation to
optimize an anbit measurement is identified in the so-called \emph{half-angle
GBS}. While the GBS corresponds to a hypersphere embedded in $\mathbb{C}^{2}$,
the half-angle GBS is its counterpart in $\mathbb{R}^{3}$ (see Fig.\,S5).
Consequently, \emph{a valid $\mathbb{S}$-space is $\mathbb{S}\equiv\mathbb{R}^{3}$},
as demonstrated below. To verify this statement, note that the position
vector $\mathbf{r}=\left(x,y,z\right)$ in the half-angle GBS associated
with an anbit $\left|\phi\right\rangle =\left|\phi_{0}\right|\left|0\right\rangle +e^{\mathrm{j}\varphi}\left|\phi_{1}\right|\left|1\right\rangle $
can be calculated by performing the identification (see Supporting
Information of ref.\,\cite{key-S2}):
\begin{equation}
\left|\phi\right\rangle =\left|\phi_{0}\right|\left|0\right\rangle +e^{\mathrm{j}\varphi}\left|\phi_{1}\right|\left|1\right\rangle \equiv z\left|0\right\rangle +\left(x+\mathrm{j}y\right)\left|1\right\rangle .\tag{S6.4}\label{eq:S6.4}
\end{equation}
Thus, it is straightforward to infer that:
\begin{equation}
\mathbf{r}=\left|\phi_{1}\right|\cos\varphi\hat{\mathbf{x}}+\left|\phi_{1}\right|\sin\varphi\hat{\mathbf{y}}+\left|\phi_{0}\right|\hat{\mathbf{z}}.\tag{S6.5}\label{eq:S6.5}
\end{equation}
The right-hand side of Eq.\,(\ref{eq:S6.4}) is especially significant,
as it allows us to verify that an anbit of the form $\left|\phi\right\rangle =\left|\phi_{i}\right\rangle +\left|n\right\rangle $,
with $\left|\phi_{i}\right\rangle \equiv z_{i}\left|0\right\rangle +\left(x_{i}+\mathrm{j}y_{i}\right)\left|1\right\rangle $
and $\left|n\right\rangle \equiv n_{z}\left|0\right\rangle +\left(n_{x}+\mathrm{j}n_{y}\right)\left|1\right\rangle $,
takes the explicit form:
\begin{align}
\left|\phi\right\rangle  & =\left(z_{i}+n_{z}\right)\left|0\right\rangle +\left[\left(x_{i}+n_{x}\right)+\mathrm{j}\left(y_{i}+n_{y}\right)\right]\left|1\right\rangle .\tag{S6.6}\label{eq:S6.6}
\end{align}
By comparing Eqs.\,(\ref{eq:S6.4}) and (\ref{eq:S6.6}), we conclude
that the position vector $\mathbf{r}$ corresponding to $\left|\phi\right\rangle $
satisfies the required condition $\mathbf{r}=\mathbf{r}_{i}^{\prime}+\mathbf{n}$,
being $\mathbf{r}_{i}^{\prime}=\left(x_{i},y_{i},z_{i}\right)$ and
$\mathbf{n}=\left(n_{x},n_{y},n_{z}\right)$ the position vectors
in the half-angle GBS corresponding to $\left|\phi_{i}\right\rangle $
and $\left|n\right\rangle $, respectively. 
\noindent \begin{center}
\includegraphics[width=12cm,height=9cm,keepaspectratio]{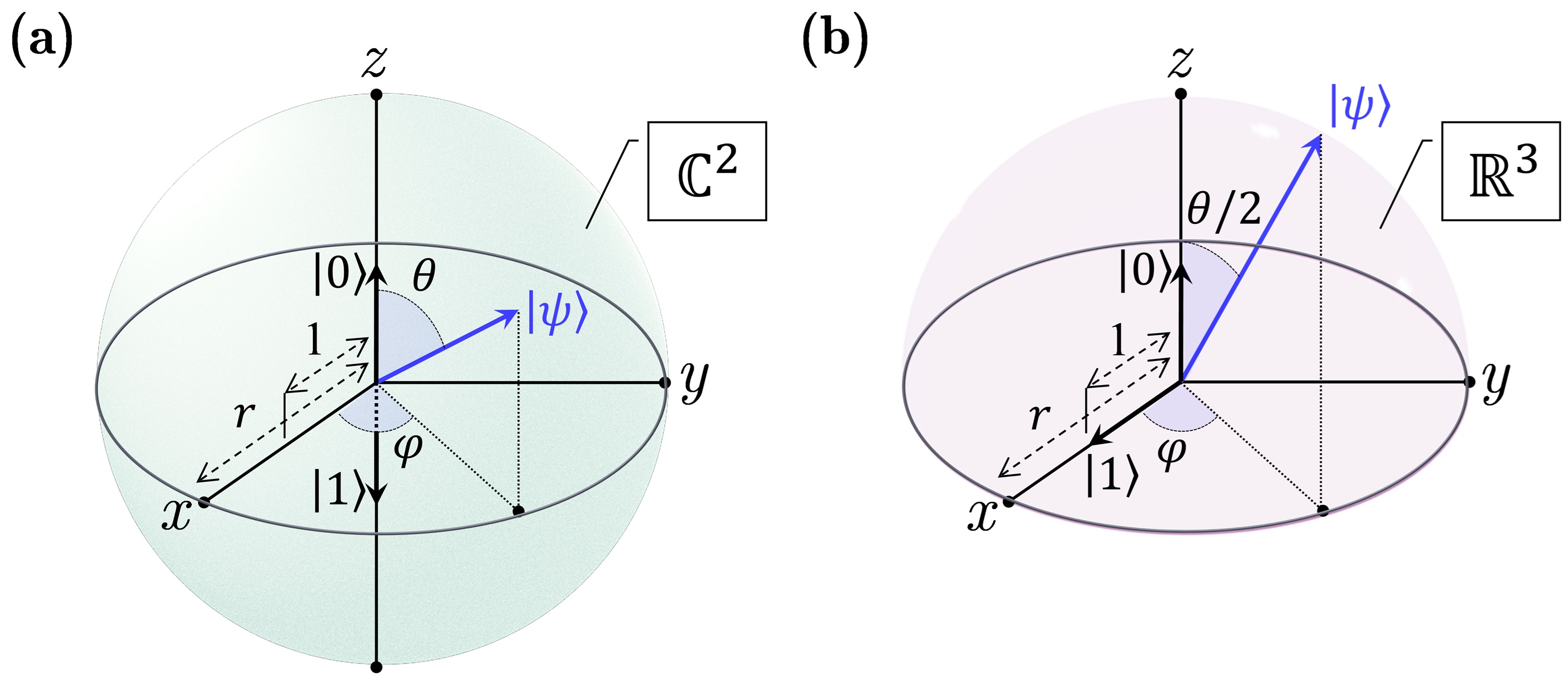}
\par\end{center}

\noindent {\small{}Figure S5. Equivalent geometric representations
of an anbit. (a) GBS representation. (b) Half-angle GBS representation.}{\small \par}

\subsection*{6.2 Additive anbit-phase noise}

\addcontentsline{toc}{subsection}{6.2 Additive anbit-phase noise}

\noindent Now, we examine \emph{additive noise} affecting the \emph{phase}
of an optical wave, as represents computational errors induced, for
example, by phase shifters in a PIP platform (see p.\,\pageref{subsec:6.5}).
Within the API framework, this class of noise is equivalent to a \emph{linear}
random perturbation $\eta$ on the differential phase $\varphi_{i}^{\prime}$
of the ideal anbit $\left|\phi_{i}\right\rangle $ that is expected
to be measured. Therefore, if the dominant system noise is an \emph{additive
anbit-phase (AP) noise}, then the ideal anbit:
\begin{equation}
\left|\phi_{i}\right\rangle =\left|\phi_{i,0}\right|\left|0\right\rangle +e^{\mathrm{j}\varphi_{i}^{\prime}}\left|\phi_{i,1}\right|\left|1\right\rangle ,\tag{S6.7}\label{eq:S6.7}
\end{equation}
represents the error-free result of an anbit measurement performed
on a noisy state (obtained at the output of the O/E conversion) of
the form:
\begin{equation}
\left|\phi\right\rangle =\left|\phi_{0}\right|\left|0\right\rangle +e^{\mathrm{j}\varphi}\left|\phi_{1}\right|\left|1\right\rangle ,\tag{S6.8}\label{eq:S6.8}
\end{equation}
where the differential phase:
\begin{equation}
\varphi=\varphi_{i}^{\prime}+\eta,\tag{S6.9}\label{eq:S6.9}
\end{equation}
accounts for the phase perturbation $\eta$ applied to $\varphi_{i}^{\prime}$.
Here, $\eta$ belongs to the range of a real random variable $\mathcal{N}$
with pdf $f_{\mathcal{N}}\left(\eta\right)$.

Proceeding analogously to the case of AA noise, we now seek a suitable
$\mathbb{S}$-space that simplifies the calculation of the conditional
pdfs. Here, bearing in mind that: (\emph{i}) the noise perturbs only
a single EDF, and (\emph{ii}) the relation between $\mathbf{r}_{i}^{\prime}$
and $\mathbf{r}$ in the $\mathbb{S}$-space (representing the relation
between $\left|\phi_{i}\right\rangle $ and $\left|\phi\right\rangle $
in the state space) must preserve the form in which the system introduces
noise in the differential phase; then it is natural to choose $\mathbb{S}=\mathbb{R}$
with:
\begin{equation}
\mathbf{r}_{i}^{\prime}=\varphi_{i}^{\prime}\hat{\mathbf{x}},\ \ \ \mathbf{n}=\eta\hat{\mathbf{x}},\ \ \ \mathbf{r}=\varphi\hat{\mathbf{x}}.\tag{S6.10}\label{eq:S6.10}
\end{equation}
Interestingly, this geometric representation ensures that the mapping
$\mathbf{r}_{i}^{\prime}\rightarrow\mathbf{r}$ is additive:
\begin{equation}
\mathbf{r}=\varphi\hat{\mathbf{x}}=\left(\varphi_{i}^{\prime}+\eta\right)\hat{\mathbf{x}}\equiv\mathbf{r}_{i}^{\prime}+\mathbf{n},\tag{S6.11}\label{eq:S6.11}
\end{equation}
preserving the form of Eq.\,(\ref{eq:S6.9}). As seen, the transformation
given by Eq.\,(\ref{eq:S6.11}) is mathematically the same as that
of Eq.\,(\ref{eq:S6.2}) in the case of AA noise (affine map). Hence,
the conditional pdfs can be calculated by particularizing Eq.\,(\ref{eq:S6.3})
to this scenario:
\begin{equation}
f\left(\mathbf{r}|\mathbf{r}_{i}^{\prime}\right)=f_{\mathbf{N}}\left(\mathbf{n}=\mathbf{r}-\mathbf{r}_{i}^{\prime}\right)\equiv f_{\mathcal{N}}\left(\eta=\varphi-\varphi_{i}^{\prime}\right).\tag{S6.12}\label{eq:S6.12}
\end{equation}

Outstandingly, the formalism of AP noise, which describes the noisy
anbit $\left|\phi\right\rangle $ as indicated by Eqs.\,(\ref{eq:S6.8})
and (\ref{eq:S6.9}), also enables the modeling of \emph{phase perturbations
induced by AA noise}\footnote{AA noise also induces a (non-linear) perturbation in the phases of
the anbit amplitudes. Assuming a noisy anbit $\left|\phi\right\rangle =\left|\psi\right\rangle +\left|n\right\rangle $,
where $\left|n\right\rangle $ represents the AA noise, then it follows
that the anbit amplitudes $\phi_{k}=\psi_{k}+n_{k}$ ($k=0,1$) exhibit
a phase of the form:
\[
\arg\left(\phi_{k}\right)=\arctan\frac{\textrm{Im}\bigl(\phi_{k}\bigr)}{\textrm{Re}\bigl(\phi_{k}\bigr)}=\arctan\frac{\textrm{Im}\bigl(\psi_{k}\bigr)+\textrm{Im}\bigl(n_{k}\bigr)}{\textrm{Re}\bigl(\psi_{k}\bigr)+\textrm{Re}\bigl(n_{k}\bigr)}.
\]
If $n_{k}=0$, then $\arg\left(\phi_{k}\right)=\arg\left(\psi_{k}\right)$.
However, in general, this expression shows that $\arg\left(\phi_{k}\right)\neq\arg\left(\psi_{k}\right)$,
and that the phase perturbation induced by $n_{k}$ is inherently
non-linear. Here, we can alternatively model such a phase pertubation
using the AP noise formalism by introducing random variables $\eta_{k}\coloneqq\arg\left(\phi_{k}\right)-\arg\left(\psi_{k}\right)$.
In this way, the perturbation on the differential phase of $\left|\psi\right\rangle $
is described via the random variable $\eta\coloneqq\eta_{1}-\eta_{0}$.} when user information is encoded solely in the differential phase.
Specifically, this approach is used in the Materials and Methods section
of the paper to optimize the anbit measurement in the analog constellation
located at the equator of the GBS, see Figs.\,7(e, f) and Supplementary
Note 9.

\subsection*{6.3 Additive anbit-amplitude and anbit-phase noise}

\addcontentsline{toc}{subsection}{6.3 Additive anbit-amplitude and anbit-phase noise}

\noindent So far, we have assumed a dominant AA or AP noise. Nonetheless,
how should we proceed when the dominant physical impairment arises
from a combination of both AA and AP noises? Such a scenario may occur,
for instance, when ASE noise affects the EDFs of the anbits to a degree
comparable to the phase noise introduced by the phase shifters.

In such conditions, the mathematical formalism can be simplified by
assuming that both AA and AP noise contributions are \emph{independent}
perturbations. Under this assumption, we can independently represent
the AA and AP noise in different $\mathbb{S}$-spaces, for example,
the AA noise in the half-angle GBS (denoted $\mathbb{S}_{\mathrm{AA}}$)
and the AP noise in the real line (denoted $\mathbb{S}_{\mathrm{AP}}$).
These representations can then be combined via the \emph{Cartesian
product}, yielding the global $\mathbb{S}$-space:
\begin{equation}
\mathbb{S}\coloneqq\mathbb{S}_{\mathrm{AA}}\times\mathbb{S}_{\mathrm{AP}}.\tag{S6.13}\label{eq:S6.13}
\end{equation}

We now examine this formalism in greater detail. Let us assume that
the ideal state that should be recovered at the receiver is:
\begin{equation}
\left|\phi_{i}\right\rangle =\left|\phi_{i,0}\right|\left|0\right\rangle +\left|\phi_{i,1}\right|e^{\mathrm{j}\varphi_{i}^{\prime}}\left|1\right\rangle ,\tag{S6.14}\label{eq:S6.14}
\end{equation}
which is the error-free result of an anbit measurement performed on
a noisy state of the form:
\begin{equation}
\left|\phi\right\rangle =\left[\left|\phi_{i,0}\right|\left|0\right\rangle +\left|\phi_{i,1}\right|e^{\mathrm{j}\left(\varphi_{i}^{\prime}+\eta_{\mathrm{AP}}\right)}\left|1\right\rangle \right]+\left|n_{\mathrm{AA}}\right\rangle ,\tag{S6.15}\label{eq:S6.15}
\end{equation}
where $\left|n_{\mathrm{AA}}\right\rangle $ and $\eta_{\mathrm{AP}}$
describe the contributions of AA and AP noise, respectively. As indicated
above, we should represent the AA and AP noises in distinct vector
spaces. Concretely, the AA noise should be represented in the $\mathbb{S}_{\mathrm{AA}}$-space
(the half-angle GBS) by omitting the contribution of the AP noise
in $\left|\phi\right\rangle $ (i.e. setting $\eta_{\mathrm{AP}}\equiv0$):
\begin{equation}
\left|\phi_{i}\right\rangle \rightarrow\mathbf{r}_{i,\mathrm{AA}}^{\prime},\ \ \ \left|n_{\mathrm{AA}}\right\rangle \rightarrow\mathbf{n}_{\mathrm{AA}},\ \ \ \left.\left|\phi\right\rangle \right\rfloor _{\eta_{\mathrm{AP}}\equiv0}\rightarrow\mathbf{r}_{\mathrm{AA}},\tag{S6.16}\label{eq:S6.16}
\end{equation}
and denoting the pdf of $\mathbf{n}_{\mathrm{AA}}$ in the $\mathbb{S}_{\mathrm{AA}}$-space
as $f_{\mathrm{AA}}\left(\mathbf{n}_{\mathrm{AA}}\right)$. Likewise,
the AP noise should be represented in the $\mathbb{S}_{\mathrm{AP}}$-space
(the real line) by omitting the contribution of the AA noise in $\left|\phi\right\rangle $
(i.e. taking $\left|n_{\mathrm{AA}}\right\rangle \equiv\left|\mathbf{0}\right\rangle $,
with $\left|\mathbf{0}\right\rangle $ being the null ket):
\begin{equation}
\left|\phi_{i}\right\rangle \rightarrow\mathbf{r}_{i,\mathrm{AP}}^{\prime}=\varphi_{i}^{\prime}\hat{\mathbf{x}},\ \ \ \eta_{\mathrm{AP}}\rightarrow\mathbf{n}_{\mathrm{AP}}=\eta_{\mathrm{AP}}\hat{\mathbf{x}},\ \ \ \left.\left|\phi\right\rangle \right\rfloor _{\left|n_{\mathrm{AA}}\right\rangle \equiv\left|\mathbf{0}\right\rangle }\rightarrow\mathbf{r}_{\mathrm{AP}}=\left(\varphi_{i}^{\prime}+\eta_{\mathrm{AP}}\right)\hat{\mathbf{x}},\tag{S6.17}\label{eq:S6.17}
\end{equation}
and denoting the pdf of $\mathbf{n}_{\mathrm{AP}}$ in the $\mathbb{S}_{\mathrm{AP}}$-space
as $f_{\mathrm{AP}}\left(\mathbf{n}_{\mathrm{AP}}\right)$. Next,
we construct the global vector space $\mathbb{S}=\mathbb{S}_{\mathrm{AA}}\times\mathbb{S}_{\mathrm{AP}}$,
where the ideal anbit, the AA+AP noise, and the noisy anbit are respectively
represented by the vectors:
\begin{equation}
\mathbf{r}_{i}^{\prime}=\left(\mathbf{r}_{i,\mathrm{AA}}^{\prime},\mathbf{r}_{i,\mathrm{AP}}^{\prime}\right),\ \ \ \mathbf{n}=\left(\mathbf{n}_{\mathrm{AA}},\mathbf{n}_{\mathrm{AP}}\right),\ \ \ \mathbf{r}=\left(\mathbf{r}_{\mathrm{AA}},\mathbf{r}_{\mathrm{AP}}\right).\tag{S6.18}\label{eq:S6.18}
\end{equation}
Notably, this geometric representation ensures that the mapping $\mathbf{r}_{i}^{\prime}\rightarrow\mathbf{r}$
is additive:
\begin{align}
\mathbf{r} & =\left(\mathbf{r}_{\mathrm{AA}},\mathbf{r}_{\mathrm{AP}}\right)=\left(\mathbf{r}_{i,\mathrm{AA}}^{\prime}+\mathbf{n}_{\mathrm{AA}},\mathbf{r}_{i,\mathrm{AP}}^{\prime}+\mathbf{n}_{\mathrm{AP}}\right)\nonumber \\
 & =\left(\mathbf{r}_{i,\mathrm{AA}}^{\prime},\mathbf{r}_{i,\mathrm{AP}}^{\prime}\right)+\left(\mathbf{n}_{\mathrm{AA}},\mathbf{n}_{\mathrm{AP}}\right)\equiv\mathbf{r}_{i}^{\prime}+\mathbf{n},\tag{S6.19}\label{eq:S6.19}
\end{align}
preserving the form in which the channel introduces both AA and AP
noises. Here, the transformation $\mathbf{r}_{i}^{\prime}\rightarrow\mathbf{r}$
given by Eq.\,(\ref{eq:S6.19}) is mathematically the same as that
of Eq.\,(\ref{eq:S6.2}). Accordingly, the conditional pdfs can be
calculated by particularizing Eq.\,(\ref{eq:S6.3}) to this case:
\begin{align}
f\left(\mathbf{r}|\mathbf{r}_{i}^{\prime}\right) & =f_{\mathbf{N}}\left(\mathbf{n}=\mathbf{r}-\mathbf{r}_{i}^{\prime}\right)\nonumber \\
 & \equiv f_{\mathrm{AA}}\left(\mathbf{n}_{\mathrm{AA}}=\mathbf{r}_{\mathrm{AA}}-\mathbf{r}_{i,\mathrm{AA}}^{\prime}\right)\cdot f_{\mathrm{AP}}\left(\mathbf{n}_{\mathrm{AP}}=\mathbf{r}_{\mathrm{AP}}-\mathbf{r}_{i,\mathrm{AP}}^{\prime}\right),\tag{S6.20}\label{eq:S6.20}
\end{align}
which are directly obtained by multiplying the marginal pdfs of the
AA and AP noises, as we have assumed independent noise sources.

\subsection*{6.4 Non-additive anbit noise}

\addcontentsline{toc}{subsection}{6.4 Non-additive anbit noise}

\noindent In this subsection, we briefly discuss how to characterize
\emph{non-adddive anbit (NA) noise}, which can arise when an APC system
includes non-linear anbit gates \cite{key-S2} or when undesired non-linear
effects are stimulated within the PIP circuits of the channel. Since
these scenarios are atypical within the APC paradigm, we limit our
discussion to a preliminary theoretical treatment, reserving a comprehensive
analysis for future contributions to the principles of API theory.

\newpage{}

As a basic example, consider a channel with a second-order non-linear
effect. An NA noise perturbing the complex envelopes that implement
an anbit $\left|\psi\right\rangle =\psi_{0}\left|0\right\rangle +\psi_{1}\left|1\right\rangle $
can be described via a noise ket $\left|n\right\rangle =n_{0}\left|0\right\rangle +n_{1}\left|1\right\rangle $
that interacts with $\left|\psi\right\rangle $ of the form $\psi_{k}n_{l}$
($k,l\in\left\{ 0,1\right\} $). This wave mixing between the amplitudes
of $\left|\psi\right\rangle $ and $\left|n\right\rangle $ resembles
the structure of the \emph{tensor product} state $\left|\psi\right\rangle \otimes\left|n\right\rangle $.

Consequently, we can infer that an NA noise might be characterized
through the tensor product. In particular, a computational channel
composed of an M-gate along with a second-order NA noise might be
described in the form:
\begin{equation}
\left|\phi\right\rangle =\left|\phi_{i}\right\rangle \otimes\left|n\right\rangle =\left(\widehat{\mathrm{M}}\left|\psi_{i}\right\rangle \right)\otimes\left|n\right\rangle .\tag{S6.21}\label{eq:S6.21}
\end{equation}
A suitable strategy to calculate the conditional pdfs requires representing
the kets in a vector space $\mathbb{S}$ whose properties remain to
be investigated in future work. Alternatively, noting the role of
the tensor product in NA noise, an extrapolation of the Kraus operators
from the formalism of open quantum systems \cite{key-S3} could offer
a viable pathway. In any case, the proper theoretical treatment of
NA noise demands further basic research.

\subsection*{6.5 Classification of PIP noise sources in the API framework\label{subsec:6.5}}

\addcontentsline{toc}{subsection}{6.5 Classification of PIP noise sources in the API framework}

\noindent As discussed in the main text, system noise introduced by
a PIP platform originates from multiple sources, including the laser,
phase shifters, optical amplifiers, non-linear effects within the
channel, and the O/E converter. Here, we categorize these contributions
as AA noise, AP noise, or NA noise within the API framework.

\paragraph{Laser noise.}

\noindent A laser generates relative intensity noise (RIN) and phase
noise \cite{key-S15,key-S16}. The \emph{RIN} of the laser can be
modeled as an additive perturbation on the complex envelope of the
emitted electric field. Hence, the RIN is an \emph{AA noise} within
the API context. Moreover, the \emph{phase noise} of the laser may
be described through an additive perturbation on the phase of the
emitted electric field. Accordingly, the phase noise is an \emph{AP
noise} using the API terminology. In the SAM hardware depicted in
Fig.\,3 of the paper, we use a single laser, which introduces a global
phase noise on the generated anbit. Such a global phase is not observable
in the GBS. Under these conditions, the phase noise of the laser can
be safely neglected.

\paragraph{Phase-shifter noise.}

\noindent A thermo-optic phase shifter introduces additive (thermal)
noise on the phase of the electric field that propagates through the
device \cite{key-S11}. Therefore, within the API paradigm, such a
class of physical impairment is an \emph{AP noise}.

\paragraph{Optical-amplifier noise.}

\noindent \emph{ASE noise} in optical amplifiers arises from spontaneously
emitted photons that are added to the optical field and subsequently
amplified \cite{key-S13}. Such additive perturbation on the field
is an \emph{AA noise} within the API theory. Indeed, ASE noise was
previously discussed as a representative example of AA noise on p.\,\pageref{subsec:6.1}.

\paragraph{Channel non-linearities.}

\noindent Non-linear noise within the channel of an API system can
emerge from: (\emph{i}) undesired non-linear effects that are stimulated
within the waveguides of PIP circuits \cite{key-S17}, (\emph{ii})
computational errors introduced during non-linear anbit operations
\cite{key-S2}. Both scenarios belong to the class of perturbation
termed as \emph{NA noise} in API\emph{.}

\paragraph{O/E conversion noise.}

\noindent As in classical optical communication systems, the O/E converter
in an API system introduces both shot noise and thermal noise \cite{key-S14}.
These noise sources can be modeled as additive perturbations on the
photocurrents generated by the O/E converter, regardless of we use
coherent or differential O/E converters. In the following, we examine
how these noise contributions manifest in each type of converter.

\emph{Coherent O/E conversion}. Consider that we generate an anbit
$\left|\psi\right\rangle =\psi_{0}\left|0\right\rangle +\psi_{1}\left|1\right\rangle $
that is propagated through a noiseless channel without gates. Hence,
the ideal anbit that should be recovered is $\left|\psi\right\rangle $.
Using the coherent O/E converter reported in Supporting Information
of ref.\,\cite{key-S2}, we will be able to retrieve, in the electrical
domain, the moduli and phases of $\psi_{0}$ and $\psi_{1}$. In noiseless
conditions, the converter generates four photocurrents ($k=0,1$):
\begin{align}
I_{k,\mathrm{I}}\left(t\right) & =\mathcal{R}\left|\psi_{k}\left(t\right)\right|\cos\angle_{k},\ \ \ I_{k,\mathrm{Q}}\left(t\right)=\mathcal{R}\left|\psi_{k}\left(t\right)\right|\sin\angle_{k},\tag{S6.22}\label{eq:S6.22}
\end{align}
where $\mathcal{R}$ is the responsivity of the photodiodes and $\angle_{k}=\arg\left(\psi_{k}\left(t\right)\right)$.\footnote{Anbit amplitudes are physically implemented as complex envelopes with
time-independent phases \cite{key-S2}.} Now, we include the shot+thermal noise, which can be regarded as
an additive perturbation to each photocurrent:
\begin{align}
I_{k,\mathrm{I}}\left(t\right) & =\mathcal{R}\left|\psi_{k}\left(t\right)\right|\cos\angle_{k}+\mathcal{N}_{k,\mathrm{I}}\left(t\right),\ \ \ I_{k,\mathrm{Q}}\left(t\right)=\mathcal{R}\left|\psi_{k}\left(t\right)\right|\sin\angle_{k}+\mathcal{N}_{k,\mathrm{Q}}\left(t\right),\tag{S6.23}\label{eq:S6.23}
\end{align}
being $\mathcal{N}_{k,\mathrm{I}}$ and $\mathcal{N}_{k,\mathrm{Q}}$
the corresponding perturbation to the in-phase (I) and quadrature
(Q) photocurrents, respectively. The resulting photocurrents can be
reinterpreted by the signal processing module (integrated at the output
of the O/E converter) as follows:
\begin{equation}
I_{k}\left(t\right)\coloneqq I_{k,\mathrm{I}}\left(t\right)+\mathrm{j}I_{k,\mathrm{Q}}\left(t\right)=\mathcal{R}\left|\psi_{k}\left(t\right)\right|e^{\mathrm{j}\angle_{k}}+\mathcal{N}_{k,\mathrm{I}}\left(t\right)+\mathrm{j}\mathcal{N}_{k,\mathrm{Q}}\left(t\right).\tag{S6.24}\label{eq:S6.24}
\end{equation}
As a result, by restating the noise terms as $\mathcal{N}_{k,\mathrm{I}}+\mathrm{j}\mathcal{N}_{k,\mathrm{Q}}\equiv\mathcal{R}n_{k}$,
we can introduce a noise ket $\left|n\right\rangle \coloneqq\sum_{k}n_{k}\left|k\right\rangle $,
which simplifies the above equation to:
\begin{equation}
I_{k}\left(t\right)=\mathcal{R}\left(\psi_{k}\left(t\right)+n_{k}\left(t\right)\right).\tag{S6.25}\label{eq:S6.25}
\end{equation}
Hence, when using a coherent O/E converter, both \emph{shot and thermal
noise} are equivalent to \emph{AA noise} within the API model.

\emph{Differential O/E conversion}. Using the unbalanced differential
O/E converter shown in Fig.\,4(a) of the main text, we will be able
to recover, in the electrical domain, the moduli $\left|\psi_{0}\right|$
and $\left|\psi_{1}\right|$ along with the differential phase $\varphi^{\prime}$
of the anbit $\left|\psi\right\rangle =\left|\psi_{0}\right|\left|0\right\rangle +e^{\mathrm{j}\varphi^{\prime}}\left|\psi_{1}\right|\left|1\right\rangle $.
In noiseless conditions, the converter generates three photocurrents
$I_{0}$, $I_{1}$, and $I_{\varphi}$ given by Eqs.\,(\ref{eq:S4.5})-(\ref{eq:S4.7}).
Including shot+thermal noise, these expressions should be restated
as:
\begin{align}
I_{0}\left(t\right) & =\frac{1}{2}\mathcal{R}\left|\psi_{0}\left(t\right)\right|^{2}+\mathcal{N}_{0}\left(t\right),\tag{S6.26}\label{eq:S6.26}\\
I_{1}\left(t\right) & =\frac{1}{2}\mathcal{R}\left|\psi_{1}\left(t\right)\right|^{2}+\mathcal{N}_{1}\left(t\right),\tag{S6.27}\label{eq:S6.27}\\
I_{\varphi}\left(t\right) & =\frac{1}{4}\mathcal{R}\left[\left|\psi_{0}\left(t\right)\right|^{2}+\left|\psi_{1}\left(t\right)\right|^{2}-2\left|\psi_{0}\left(t\right)\right|\left|\psi_{1}\left(t\right)\right|\sin\varphi^{\prime}\right]+\mathcal{N}_{\varphi}\left(t\right),\tag{S6.28}\label{eq:S6.28}
\end{align}
where $\mathcal{N}_{0,1,\varphi}$ are the shot+thermal noise contributions
to each photocurrent. As shown below, both shot and thermal noise
can also be reinterpreted as AA noise in this case. To demonstrate
this statement, let us assume that we are interested in performing
the O/E conversion of the state $\left|\psi\right\rangle +\left|n\right\rangle $,
where $\left|n\right\rangle =n_{0}\left|0\right\rangle +n_{1}\left|1\right\rangle $
is a noise ket that should be calculated to correctly describe the
shot+thermal noise contributions present in the above photocurrents.
By performing the identification $\mathcal{N}_{0}\equiv\mathcal{R}\left|n_{0}\right|^{2}/2$
and $\mathcal{N}_{1}\equiv\mathcal{R}\left|n_{1}\right|^{2}/2$, Eqs.\,(\ref{eq:S6.26})
and (\ref{eq:S6.27}) become:
\begin{equation}
I_{0}\left(t\right)=\frac{1}{2}\mathcal{R}\left(\left|\psi_{0}\left(t\right)\right|^{2}+\left|n_{0}\left(t\right)\right|^{2}\right),\ \ \ I_{1}\left(t\right)=\frac{1}{2}\mathcal{R}\left(\left|\psi_{1}\left(t\right)\right|^{2}+\left|n_{1}\left(t\right)\right|^{2}\right).\tag{S6.29}\label{eq:S6.29}
\end{equation}
Here, as commented in Appendix C on p.\,\pageref{sec:APPENDIX_C},
note that the phases of $n_{0}$ and $n_{1}$ are degrees of freedom
in the problem, that is, the value of $\left|\psi_{k}\right|^{2}+\left|n_{k}\right|^{2}$
is invariant under changes in $\arg\left(n_{k}\right)$. Therefore,
we can select a specific phase $\arg\left(n_{k}\right)$ satisfying
the phase condition Eq.\,(\ref{eq:SC.2}), which ensures the identity
$\left|\psi_{k}\right|^{2}+\left|n_{k}\right|^{2}\equiv\left|\psi_{k}+n_{k}\right|^{2}$.
Following this approach, we can identify the desired noise ket $\left|n\right\rangle $.
Finally, assuming that $\left|n_{k}\right|^{2}\ll\left|\psi_{k}\right|^{2}$
in general, we can safely assume that we introduce a negligible error
by approximating $I_{\varphi}$ as (we omit the time variable for
simplicity):
\begin{equation}
I_{\varphi}\simeq\frac{1}{4}\mathcal{R}\left[\left|\psi_{0}+n_{0}\right|^{2}+\left|\psi_{1}+n_{1}\right|^{2}-2\left|\psi_{0}+n_{0}\right|\left|\psi_{1}+n_{1}\right|\sin\varphi\right],\tag{S6.30}\label{eq:S6.30}
\end{equation}
being $\varphi$ the differential phase of $\left|\psi\right\rangle +\left|n\right\rangle $.
Consequently, we demonstrate that, using the differential O/E converter,
both \emph{shot and thermal noise} can also be regarded as \emph{AA
noise} within the API theory. By repeating the same discussion for
the quadrature differential O/E converter {[}Fig.\,4(b){]}, we reach
the same conclusions as for the unbalanced architecture. 

\paragraph{\label{par:Dominant-noise}Dominant noise source in passive linear
PIP circuits.}

\noindent In passive, linear PIP platforms - i.e., circuits performing
linear wave transformations and without integrated optical amplifiers
- the dominant noise stems from the RIN of the laser together with
the combined shot- and thermal-noise contributions of the O/E converter.
We substantiated this conclusion numerically by simulating the API
system of Fig.\,7(a) using \emph{OptSim software}, which confirmed
that phase-shifter noise is negligible relative to RIN, shot, and
thermal noises. Therefore, according to the classification of Table
S1, these dominant noises fall into the AA category. Moreover, since
\emph{RIN, shot, and thermal noises} can each be independently modeled
as \emph{white noise sources}, specifically, as \emph{zero-mean wide-sense
stationary Gaussian processes} \cite{key-S14,key-S15}, the combined
contribution of these AA noises can be described within the anbit-measurement
formalism by a noise ket $\left|n\right\rangle =n_{0}\left|0\right\rangle +n_{1}\left|1\right\rangle $
with amplitudes $n_{0}$ and $n_{1}$ that should be considered as
independent, zero-mean Gaussian random variables.
\noindent \begin{center}
{\small{}}%
\begin{tabular}{cccc}
\toprule 
\textbf{\small{}Noise source/API terminology} & \textbf{\small{}AA noise} & \textbf{\small{}AP noise} & \textbf{\small{}NA noise}\tabularnewline
\midrule
\midrule 
{\small{}RIN} & \textcolor{blue}{\small{}$\checkmark$} & {\small{}\textendash{}} & {\small{}\textendash{}}\tabularnewline
\midrule 
{\small{}Laser phase noise} & {\small{}\textendash{}} & \textcolor{blue}{\small{}$\checkmark$} & {\small{}\textendash{}}\tabularnewline
\midrule 
{\small{}Phase-shifter noise} & {\small{}\textendash{}} & \textcolor{blue}{\small{}$\checkmark$} & {\small{}\textendash{}}\tabularnewline
\midrule 
{\small{}ASE noise} & \textcolor{blue}{\small{}$\checkmark$} & {\small{}\textendash{}} & {\small{}\textendash{}}\tabularnewline
\midrule 
{\small{}Non-linear noise} & {\small{}\textendash{}} & {\small{}\textendash{}} & \textcolor{blue}{\small{}$\checkmark$}\tabularnewline
\midrule 
{\small{}Shot+thermal noise} & \textcolor{blue}{\small{}$\checkmark$} & {\small{}\textendash{}} & {\small{}\textendash{}}\tabularnewline
\bottomrule
\end{tabular}
\par\end{center}{\small \par}

\noindent {\small{}Table S1. Classification of noise sources in a
PIP platform according to the terminology of API theory. (AA: additive
anbit-amplitude noise. AP: additive anbit-phase noise. NA: non-additive
anbit noise).}{\small \par}

\subsection*{6.6 Hardware imperfections in PIP circuits}

\addcontentsline{toc}{subsection}{6.6 Hardware imperfections in PIP circuits}

\noindent In the preceding subsections, we examined the system noise.
Now, we evaluate an additional system's physical impairment: the non-ideal
operation of PIP devices, arising from manufacturing imperfections.
To this end, consider a noiseless API system composed of non-ideal
PIP components. In such a scenario, the anbit retrieved at the output
of the O/E converter is:
\begin{equation}
\left|\phi\right\rangle =\left|\phi_{i}\right\rangle +\left|e\right\rangle ,\tag{S6.31}\label{eq:S6.31}
\end{equation}
where $\left|\phi_{i}\right\rangle $ is the ideal anbit that is expected
to be measured and the (deterministic) state $\left|e\right\rangle $
quantifies the \emph{static error} induced by hardware imperfections
of the PIP circuits. A priori, the value of $\left|e\right\rangle $
is unknown. However, the above equation is deterministic and does
not involve any random variables. Thus, it follows that $\widehat{\mathrm{E}}\bigl(\left|\phi\right\rangle \bigr)=\left|\phi_{i}\right\rangle +\left|e\right\rangle $,
where $\widehat{\mathrm{E}}$ is the expectation operator.

Now, we include the system noise. As discussed above, the dominant
noise sources can be modeled as an AA noise described by a \emph{zero-mean}
random ket, here denoted $\left|n_{\mathrm{AA}}\right\rangle $. Therefore,
Eq.\,(\ref{eq:S6.31}) becomes:
\begin{equation}
\left|\phi\right\rangle =\left|\phi_{i}\right\rangle +\left|n_{\mathrm{AA}}\right\rangle +\left|e\right\rangle ,\tag{S6.32}\label{eq:S6.32}
\end{equation}
with:
\begin{equation}
\widehat{\mathrm{E}}\bigl(\left|\phi\right\rangle \bigr)=\left|\phi_{i}\right\rangle +\widehat{\mathrm{E}}\bigl(\left|n_{\mathrm{AA}}\right\rangle \bigr)+\left|e\right\rangle =\left|\phi_{i}\right\rangle +\left|e\right\rangle .\tag{S6.33}\label{eq:S6.33}
\end{equation}
This expression indicates that the random distributions of the EDFs
of $\left|\phi\right\rangle $ have, as their statistical mean, the
EDFs of $\left|\phi_{i}\right\rangle +\left|e\right\rangle $. In
other words, the statistical mean of the EDFs of $\left|\phi\right\rangle $
provides direct information about the error induced by the non-ideal
behavior of PIP circuits. Consequently, to experimentally characterize
such hardware imperfections and decouple them from system noise, it
suffices to measure multiple random samples of $\left|\phi\right\rangle $
and compare the average state $\widehat{\mathrm{E}}\bigl(\left|\phi\right\rangle \bigr)$
with the ideal state $\left|\phi_{i}\right\rangle $, for example,
using the GBS distance or any other state-comparative parameter introduced
in Supplementary Note 3. This approach is used in Fig.\,7(d) of the
paper to characterize hardware imperfections in the poles of the GBS.

Finally, note that the \emph{combined effect} of the dominant noise
sources and the non-ideal operation of PIP devices can be jointly
modeled as an \emph{equivalent AA noise} $\left|n\right\rangle =\left|n_{\mathrm{AA}}\right\rangle +\left|e\right\rangle $
inducing random perturbations on the EDFs of $\left|\phi_{i}\right\rangle $,
with the average state $\widehat{\mathrm{E}}\bigl(\left|n\right\rangle \bigr)=\left|e\right\rangle $
accounting for the hardware imperfections.

\noindent 

\newpage{}

\section*{Supplementary Note 7: channel capacity\label{sec:7}}

\addcontentsline{toc}{section}{Supplementary Note 7: channel capacity}

\noindent Here, we report the mathematical formalism in API to calculate the channel capacity in single-anbit (or simple) systems, considering two distinct scenarios at the receiver: (\emph{i}) anbit\linebreak{} estimation or (\emph{ii}) anbit measurement. We should examine how to calculate the channel\linebreak{} capacity in both cases, as information recovery relies on distinct signal-filtering strategies.

\subsection*{7.1 General properties of channel capacity}

\addcontentsline{toc}{subsection}{7.1 General properties of channel capacity}

\noindent The API paradigm only deals with classical information.
Accordingly, the definition of channel capacity in API systems is
provided by Shannon\textquoteright s theory, as discussed in the main
text. Therefore, before delving into tedious mathematical discussions,
we first revisit the definition of Shannon\textquoteright s channel
capacity along with its general properties, contextualized within
the API framework.

Consider a simple API system comprising an originator source $X$,
capable of generating $M$ distinct symbols, and a recipient source
$Y$, which can receive $N$ different symbols. The definition of
the channel capacity ($C$) established by Shannon\textquoteright s
theory is \cite{key-S7}:
\begin{equation}
C\coloneqq\max_{p\left(x\right)}\left\{ H\left(X;Y\right)\right\} \ \ \ \textrm{(bits)},\tag{S7.1}\label{eq:S7.1}
\end{equation}
where $H\left(X;Y\right)$ is the mutual information between the originator
and recipient sources and $p\left(x\right)$ is the pmf of $X$. Next,
we briefly revisit the \emph{general properties} of $C$ and the \emph{channel-coding
theorem}.

\subsubsection*{General properties}

\noindent Shannon\textquoteright s channel capacity satisfies the
following fundamental properties \cite{key-S18}:
\begin{enumerate}
\item \emph{Positivity or lower bound}. $C\geq0$ given that $H\left(X;Y\right)\geq0$.
\item \emph{Upper bound}. $C\leq\min\left\{ H\left(X\right),H\left(Y\right)\right\} $
and it is useful to distinguish between two scenarios:
\begin{enumerate}
\item Bijective channels (reversible gates, $M=N$). In this case, the minimum
of the two entropies is usually given by $H\left(X\right)$, since
the entropy of $Y$ tends to be higher due to the noise introduced
by the channel. Thus, in general, the channel capacity satisfies $C\leq H\left(X\right)$.
\item Non-bijective channels (irreversible gates, $M>N$). Note that the
two entropies satisfy that $H\left(X\right)\leq\log_{2}M$ and $H\left(Y\right)\leq\log_{2}N$.
Hence, in this case, the minimum of the two entropies is typically
determined by $H\left(Y\right)$. Thus, the channel capacity generally
satisfies $C\leq H\left(Y\right)$.
\end{enumerate}
\item \emph{Continuity}. $H\left(X;Y\right)$ is a continuous function of
$\mathbf{p}=\left(p\left(x_{1}\right),\ldots,p\left(x_{M}\right)\right)$.
\item \emph{Uniqueness}. $H\left(X;Y\right)$ is a concave function of the
random distribution $\mathbf{p}$. Hence, any local maximum within
a closed subset of $\mathbb{R}^{M}$ is also a global maximum. As
a result, there exists a unique pmf $\mathbf{p}$ for the originator
source that maximizes the mutual information.
\item \emph{Data processing inequality}. A computational operation (or anbit
gate) $X\overset{g}{\rightarrow}Z$ cannot increase the mutual information
of the system:
\begin{equation}
H\left(X;Y\right)\geq H\left(Z;Y\right).\tag{S7.2}\label{eq:S7.2}
\end{equation}
Equality holds only for reversible gates, since in that case $g$
defines a one-to-one mapping that preserves the entropy of $X$ in
$Z$. Consequently, the channel capacity in an API system will generally
be higher for reversible gates (bijective channels) than for irreversible
gates (non-bijective channels).
\end{enumerate}

\subsubsection*{Channel-coding theorem (CCT)}

\noindent The information rate of the transmitter, $R$ (in bits),
is defined as the average amount of information actually emitted by
the originator source $X$ \cite{key-S7}. When information is transmitted
in symbol strings encoded into classical states of the GBS (assuming
a one-to-one codification between symbols and anbits), the information
rate measured in anbits, $\widetilde{R}$, corresponds to the average
string length in anbits. Accordingly, the information rate in bits
is given by $R=\mathrm{BAR}_{X}\cdot\widetilde{R}$, where $\mathrm{BAR}_{X}\coloneqq H\left(X\right)/M$
(bits/anbit) is the bit-anbit ratio of the encoder (defined in Subsection
2.1 of the paper).

The CCT states that $C$ is the limit on the maximum amount of information
generated by $X$ that can be reliably transmitted over the channel,
that is \cite{key-S7}:
\begin{equation}
R\leq C\leq\min\left\{ H\left(X\right),H\left(Y\right)\right\} \ \ \ \textrm{(bits)}.\tag{S7.3}\label{eq:S7.3}
\end{equation}
Since the entropies $H\left(X\right)$ and $H\left(Y\right)$ may
exceed 1 bit, then simple API systems can exhibit an information rate
and a channel capacity exceeding 1 bit. This constitutes a fundamental
distinction between API and QI, where the channel capacity in single-qubit
systems is limited to 1 bit due to the Holevo bound \cite{key-S6}.

In API, the CCT can alternatively be expressed in terms of anbits
by introducing the BAR parameter into the previous equation. For instance,
assuming that $H\left(X\right)<H\left(Y\right)$ - the most common
scenario, as we typically work with noisy bijective channels (reversible
computational operations) - the CCT may be rewritten as follows:
\begin{equation}
\widetilde{R}\leq\widetilde{C}\leq M\ \ \ \textrm{(anbits)},\tag{S7.4}\label{eq:S7.4}
\end{equation}
where $\widetilde{C}=C/\mathrm{BAR}_{X}$ is the channel capacity
quantified in anbits.

Remarkably, the BAR parameter not only allows us to rewrite the CCT
in terms of anbits; it also provides a useful tool for establishing
an \emph{upper bound} - measured in anbits - on the mutual information.
In bijective channels ($M=N$), the mutual information satisfies the
inequality:\footnote{Note that $H\left(X;Y\right)=H\left(X\right)-H\left(X|Y\right)$,
where $H\left(X|Y\right)$ is the conditional entropy or equivocation
\cite{key-S7}.}
\begin{equation}
\widetilde{H}\left(X;Y\right)\coloneqq\frac{H\left(X;Y\right)}{\mathrm{BAR}_{X}}=M\left(1-\frac{H\left(X|Y\right)}{H\left(X\right)}\right)\leq M\ \ \ \textrm{(anbits)}.\tag{S7.5}\label{eq:S7.5}
\end{equation}
Since $H\left(X|Y\right)\leq H\left(X\right)$, it follows that $\widetilde{H}\left(X;Y\right)\leq M$
anbits. In contrast, in non-bijective channels ($M>N$), normalizing
the mutual information by the decoder's BAR parameter $\mathrm{BAR}_{Y}\coloneqq H\left(Y\right)/N$
(bits/anbit), we find that:
\begin{equation}
\widetilde{H}\left(X;Y\right)\coloneqq\frac{H\left(X;Y\right)}{\mathrm{BAR}_{Y}}=N\left(1-\frac{H\left(Y|X\right)}{H\left(Y\right)}\right)\leq N\ \ \ \textrm{(anbits)}.\tag{S7.6}\label{eq:S7.6}
\end{equation}
Since $H\left(Y|X\right)\leq H\left(Y\right)$, we infer that $\widetilde{H}\left(X;Y\right)\leq N$
anbits. In summary, the upper bounds given by Eqs.\,(\ref{eq:S7.5})
and (\ref{eq:S7.6}) indicate that, in both bijective and non-bijective
channels, the limit on the maximum amount of information generated
by $X$ that can be reliably computed within the channel and recovered
at the receiver is $\log_{2}N$ bits, or equivalently, $N$ anbits.
Note that this result is consistent with the upper bound of \emph{accessible
information} in API, discussed in the main text.

\subsection*{7.2 Channel capacity in single-anbit \emph{estimated} systems}

\addcontentsline{toc}{subsection}{7.2 Channel capacity in single-anbit \emph{estimated} systems}

\noindent The channel capacity in simple API systems based on anbit
estimation at the receiver can be calculated by directly analyzing
the mutual information between the originator ($X$) and recipient
($Y$) sources. To this end, we should first describe the anbit transformation
of the channel as a function of the random variables $X$ and $Y$.

\subsubsection*{Preliminary remarks}
\begin{enumerate}
\item As discussed in the main text and Supplementary Note 6, the dominant
noise sources in combination with hardware imperfections can be commonly
modeled as an equivalent AA noise within the API context, described
by a noise ket $\left|n\right\rangle $. Consequently, the general
expression governing the anbit transformation performed by the channel
is given by Eq.\,(\ref{eq:S6.1}).
\item By recasting Eq.\,(\ref{eq:S6.1}) as a random-variable relation
between the originator and recipient sources, it is natural to assume
an expression of the form:
\begin{equation}
Y=g\left(X\right)+\mathcal{N}.\tag{S7.7}\label{eq:S7.7}
\end{equation}
Here, the random variable $\mathcal{N}$ describes the main physical
impairments of the system (dominant noises and hardware imperfections).
In addition, the $g$-function accounts for the computational operation
of the channel {[}the $\widehat{\mathrm{M}}$ operator in Eq.\,(\ref{eq:S6.1}){]}
along with the mapping implemented by the encoder and decoder between
the symbols of the sources and the anbits that define the transmitted
and received constellations. Accordingly, note that $g$ may be a
non-linear function, as it represents the noiseless (or ideal) correspondence
between the symbols of $X$ and $Y$, which may be a non-linear mapping.
Nevertheless, this function is assumed to be either bijective or non-bijective,
depending on whether the underlying computational operation of the
channel is reversible or irreversible.
\item It is reasonable to assume that $\mathcal{N}$ is independent of $X$
and Gaussian-distributed, reflecting the statistical properties of
the noise ket $\left|n\right\rangle $ (see p.\,\pageref{par:Dominant-noise}).
Likewise, a Gaussian distribution for $\mathcal{N}$ may be justified
from an alternative perspective. In the absence of prior knowledge
about the distribution of the dominant noise sources, the principle
of maximum entropy (PME) provides a direct strategy for inferring
the distribution of $\mathcal{N}$. Thus, applying PME under constraints
on the first and second moments of $\mathcal{N}$ (imposed by optical
power limitations to prevent non-linear effects in the channel), the
distribution that maximizes entropy is Gaussian. In this case, the
maximum entropy is given by $H_{\mathrm{max}}\left(\mathcal{N}\right)=\log_{2}\sqrt{2\pi e\sigma_{\mathcal{N}}^{2}}$,
where $\sigma_{\mathcal{N}}^{2}$ is the noise variance \cite{key-S1,key-S7}.
\end{enumerate}
\newpage{}

\subsubsection*{Bijective channels: reversible gates}

\noindent From the data processing inequality {[}Eq.\,(\ref{eq:S7.2}){]},
we know that $H\left(X;Y\right)=H\left(g\left(X\right);Y\right)$
when $g$ describes a reversible computational operation. Consequently,
the $g$-function does not modify the channel capacity. This implies
that the channel capacity of the information system $Y=g\left(X\right)+\mathcal{N}$
is identical to that of the system $Y=X+\mathcal{N}$. For simplicity,
and without loss of generality, we analyze the second case. Here,
the channel capacity is:\footnote{As demonstrated in Appendix D (on p.\,\pageref{sec:APPENDIX_D}),
the distribution of $Y|X$ is given by the distribution of $\mathcal{N}$.
As a result, it is direct to verify that $H\left(Y|X\right)=H\left(\mathcal{N}\right)$.}
\begin{equation}
C=\max_{p\left(x\right)}\left\{ H\left(X;Y\right)\right\} =\max_{p\left(x\right)}\left\{ H\left(Y\right)-H\left(Y|X\right)\right\} =\max_{p\left(x\right)}\left\{ H\left(Y\right)\right\} -H\left(\mathcal{N}\right).\tag{S7.8}\label{eq:S7.8}
\end{equation}
The entropy of $Y$ is here maximized by maximizing the entropy of
$X$. Considering constraints on the first and second moments of $X$
(to prevent non-linear effects in the channel), we know that the Gaussian
distribution maximizes the entropy of both random variables. As a
result, we obtain that $H_{\mathrm{max}}\left(Y\right)=\log_{2}\sqrt{2\pi e\sigma_{Y}^{2}}$,
where $\sigma_{Y}^{2}=\sigma_{X}^{2}+\sigma_{\mathcal{N}}^{2}$ is
the variance of $Y$, which can be calculated from the variances of
$X$ and $\mathcal{N}$. Finally, we derive a closed-form expression
for channel capacity in this scenario, which corresponds to the Shannon-Hartley
theorem \cite{key-S7}:
\begin{equation}
C=\frac{1}{2}\log_{2}\left(1+\frac{\sigma_{X}^{2}}{\sigma_{\mathcal{N}}^{2}}\right)\ \ \ \textrm{(bits)}.\tag{S7.9}\label{eq:S7.9}
\end{equation}

\subsubsection*{Non-bijective channels: irreversible gates}

\noindent From the data processing inequality {[}Eq.\,(\ref{eq:S7.2}){]},
we know that $H\left(X;Y\right)>H\left(g\left(X\right);Y\right)$
when $g$ describes an irreversible computational operation. This
directly implies that the Shannon-Hartley theorem establishes an upper
bound to the channel capacity in this scenario:
\begin{equation}
C<\frac{1}{2}\log_{2}\left(1+\frac{\sigma_{X}^{2}}{\sigma_{\mathcal{N}}^{2}}\right)\ \ \ \textrm{(bits)}.\tag{S7.10}\label{eq:S7.10}
\end{equation}

\subsection*{7.3 Channel capacity in single-anbit \emph{measured} systems}

\addcontentsline{toc}{subsection}{7.3 Channel capacity in single-anbit \emph{measured} systems}

\noindent As commented in the main text, the Shannon-Hartley theorem
does not capture the influence of the decision regions utilized to
optimize the anbit measurement in the calculation of the channel capacity.
Therefore, the analysis of the channel capacity in simple API systems
based on anbit measurement requires a distinct mathematical formalism.
We begin by considering bijective channels and subsequently extend
the theory to encompass non-bijective channels.

\subsubsection*{Bijective channels: reversible gates}

\noindent The mutual information is given by the general expression
\cite{key-S7}:
\begin{equation}
H\left(X;Y\right)=\sum_{i=1}^{M}\sum_{j=1}^{N}p\left(x_{i},y_{j}\right)\log_{2}\frac{p\left(x_{i},y_{j}\right)}{p\left(x_{i}\right)p\left(y_{j}\right)}=\sum_{i,j}p\left(x_{i}\right)p\left(y_{j}|x_{i}\right)\log_{2}\frac{p\left(y_{j}|x_{i}\right)}{p\left(y_{j}\right)},\tag{S7.11}\label{eq:S7.11}
\end{equation}
with $M=N$ in bijective channels. The probability terms on the right-hand
side of the above equation can be expressed as a function of the pmf
$p\left(x_{i}\right)\equiv p_{i}$, the conditional pdfs $f\left(\mathbf{r}|\mathbf{r}_{i}^{\prime}\right)$
of the system, and the decision regions $D_{j}$ employed to perform
the anbit measurement (see Supplementary Note 5.1):

\newpage{}

\noindent 
\begin{align}
p\left(y_{j}|x_{i}\right) & =\int_{\mathbb{S}}p\left(y_{j}|\mathbf{r}\right)f\left(\mathbf{r}|\mathbf{r}_{i}\right)\mathrm{d}^{s}r=\int_{D_{j}}f\left(\mathbf{r}|\mathbf{r}_{i}^{\prime}\right)\mathrm{d}^{s}r,\tag{S7.12}\label{eq:S7.12}\\
p\left(y_{j}\right) & =\sum_{i}p\left(x_{i}\right)p\left(y_{j}|x_{i}\right)=\sum_{i}p\left(x_{i}\right)\int_{D_{j}}f\left(\mathbf{r}|\mathbf{r}_{i}^{\prime}\right)\mathrm{d}^{s}r=\int_{D_{j}}f\left(\mathbf{r}\right)\mathrm{d}^{s}r,\tag{S7.13}\label{eq:S7.13}
\end{align}
with $s=\dim\left(\mathbb{S}\right)$, $f\left(\mathbf{r}|\mathbf{r}_{i}^{\prime}\right)=f\left(\mathbf{r}|\mathbf{r}_{i}\right)$,
and $f\left(\mathbf{r}\right)=\sum_{i}p\left(x_{i}\right)f\left(\mathbf{r}|\mathbf{r}_{i}^{\prime}\right)$.
Consequently, the channel capacity is given by the expression:
\begin{equation}
C=\max_{p_{i},\mathbf{r}_{i}^{\prime},D_{j}}\left\{ \sum_{i,j}p_{i}\int_{D_{j}}f\left(\mathbf{r}|\mathbf{r}_{i}^{\prime}\right)\mathrm{d}^{s}r\log_{2}\frac{\int_{D_{j}}f\left(\mathbf{r}|\mathbf{r}_{i}^{\prime}\right)\mathrm{d}^{s}r}{\int_{D_{j}}f\left(\mathbf{r}\right)\mathrm{d}^{s}r}\right\} \ \ \ \textrm{(bits)},\tag{S7.14}\label{eq:S7.14}
\end{equation}
which is Eq.\,(8) of the paper. Interestingly, Eq.\,(\ref{eq:S7.14})
mirrors the mathematical structure of the quantum channel capacity
(see Appendix E on p.\,\pageref{sec:APPENDIX_E}), suggesting a deeper
underlying connection between API and QI.

\subsubsection*{Non-bijective channels: irreversible gates}

\noindent In order to derive the channel capacity in single-anbit
measured systems with non-bijective channels, we make use of the notation
introduced on p.\,\pageref{subsec:5.2-Non-bijective-channels} to
describe such channels. Here, let us remember that information is
computed following the next flowchart of transformations in the vector
space $\mathbb{S}$ used to optimize the anbit measurement:
\begin{equation}
x_{i}^{\left(1,\ldots,g_{i}\right)}\overset{g_{i}:g_{i}}{\leftrightarrow}\mathbf{r}_{i}^{\left(1,\ldots,g_{i}\right)}\underset{\textrm{channel}}{\longrightarrow}\mathbf{r}\underset{\textrm{measurement}}{\longrightarrow}\mathbf{r}_{i}^{\prime}\overset{1:1}{\leftrightarrow}y_{i},\tag{S7.15}\label{eq:S7.15}
\end{equation}
where $g_{i}$ is referred to as the degree of degeneracy of the symbol
$y_{i}$ in the recipient source, indicating that $y_{i}$ corresponds
to $g_{i}$ different symbols $\bigl\{ x_{i}^{\left(k\right)}\bigr\}_{k=1}^{g_{i}}$
in the originator source. Hence, the number of distinct symbols that
can be emitted by the transmitter is $M=\sum_{i=1}^{N}g_{i}>N$.

In such a scenario, the mutual information is given by the general
expression:
\begin{equation}
H\left(X;Y\right)=\sum_{i=1}^{N}\sum_{k=1}^{g_{i}}\sum_{j=1}^{N}p\bigl(x_{i}^{\left(k\right)}\bigr)p\bigl(y_{j}|x_{i}^{\left(k\right)}\bigr)\log_{2}\frac{p\bigl(y_{j}|x_{i}^{\left(k\right)}\bigr)}{p\left(y_{j}\right)}.\tag{S7.16}\label{eq:S7.16}
\end{equation}
The probability terms on the right-hand side of the above equation
can be expressed as a function of the pmf $p\bigl(x_{i}^{\left(k\right)}\bigr)\equiv p_{i}^{\left(k\right)}$,
the conditional pdfs $f\left(\mathbf{r}|\mathbf{r}_{i}^{\prime}\right)=f\bigl(\mathbf{r}|\mathbf{r}_{i}^{\left(k\right)}\bigr)$,
and the decision regions $D_{j}$ defined to implement the anbit measurement
(see Supplementary Note 5.2):
\begin{align}
p\bigl(y_{j}|x_{i}^{\left(k\right)}\bigr) & =\int_{\mathbb{S}}p\left(y_{j}|\mathbf{r}\right)f\bigl(\mathbf{r}|\mathbf{r}_{i}^{\left(k\right)}\bigr)\mathrm{d}^{s}r=\int_{D_{j}}f\bigl(\mathbf{r}|\mathbf{r}_{i}^{\left(k\right)}\bigr)\mathrm{d}^{s}r,\tag{S7.17}\label{eq:S7.17}\\
p\left(y_{j}\right) & =\sum_{i,k}p\bigl(x_{i}^{\left(k\right)}\bigr)p\bigl(y_{j}|x_{i}^{\left(k\right)}\bigr)=\sum_{i,k}p\bigl(x_{i}^{\left(k\right)}\bigr)\int_{D_{j}}f\bigl(\mathbf{r}|\mathbf{r}_{i}^{\left(k\right)}\bigr)\mathrm{d}^{s}r=\int_{D_{j}}f\left(\mathbf{r}\right)\mathrm{d}^{s}r,\tag{S7.18}\label{eq:S7.18}
\end{align}
with $f\left(\mathbf{r}\right)=\sum_{i,k}p\bigl(x_{i}^{\left(k\right)}\bigr)f\bigl(\mathbf{r}|\mathbf{r}_{i}^{\left(k\right)}\bigr)$.
As a result, the channel capacity can be calculated as:
\begin{equation}
C=\max_{p_{i}^{\left(k\right)},\mathbf{r}_{i}^{\left(k\right)},D_{j}}\left\{ \sum_{i,k,j}p_{i}^{\left(k\right)}\int_{D_{j}}f\bigl(\mathbf{r}|\mathbf{r}_{i}^{\left(k\right)}\bigr)\mathrm{d}^{s}r\log_{2}\frac{\int_{D_{j}}f\bigl(\mathbf{r}|\mathbf{r}_{i}^{\left(k\right)}\bigr)\mathrm{d}^{s}r}{\int_{D_{j}}f\left(\mathbf{r}\right)\mathrm{d}^{s}r}\right\} \ \ \ \textrm{(bits)}.\tag{S7.19}\label{eq:S7.19}
\end{equation}
It is worth highlighting that the above expression can also be applied
to bijective channels by setting $g_{i}\equiv1$ for all $i=1,\ldots,N$.
In such a case, Eq.\,(\ref{eq:S7.19}) reduces to Eq.\,(\ref{eq:S7.14}).

Given the generality of Eq.\,(\ref{eq:S7.19}), this expression can
be employed to explore the \emph{upper bound} of the channel capacity
in single-anbit measured systems by evaluating a noiseless channel.
In such a scenario, the received anbits do not overlap at the output
of the O/E converter. Hence, the anbit measurement theorem ensures
that one can always define a set of decision regions $\left\{ D_{j}\right\} _{j=1}^{N}$
fulfilling the condition:
\begin{equation}
\int_{D_{j}}f\left(\mathbf{r}|\mathbf{r}_{i}^{\prime}\right)\mathrm{d}^{s}r=\int_{D_{j}}f\bigl(\mathbf{r}|\mathbf{r}_{i}^{\left(k\right)}\bigr)\mathrm{d}^{s}r=\delta_{ij}.\tag{S7.20}\label{eq:S7.20}
\end{equation}
As a result, we find that:
\begin{equation}
\int_{D_{j}}f\left(\mathbf{r}\right)\mathrm{d}^{s}r=\sum_{i,k}p_{i}^{\left(k\right)}\int_{D_{j}}f\bigl(\mathbf{r}|\mathbf{r}_{i}^{\left(k\right)}\bigr)\mathrm{d}^{s}r=\sum_{i,k}p_{i}^{\left(k\right)}\delta_{ij}=\sum_{k}p_{j}^{\left(k\right)},\tag{S7.21}\label{eq:S7.21}
\end{equation}
and:\footnote{Note that $p\left(y_{i}\right)=\sum_{k}p\bigl(x_{i}^{\left(k\right)}\bigr)\equiv\sum_{k}p_{i}^{\left(k\right)}$
in a noiseless channel.}
\begin{align}
C & =\max_{p_{i}^{\left(k\right)}}\left(\sum_{i,k,j}p_{i}^{\left(k\right)}\delta_{ij}\log_{2}\frac{\delta_{ij}}{\sum_{k}p_{j}^{\left(k\right)}}\right)=\max_{p_{i}^{\left(k\right)}}\left(-\sum_{i,k}p_{i}^{\left(k\right)}\log_{2}\sum_{k}p_{i}^{\left(k\right)}\right)\nonumber \\
 & =\max_{p\left(y_{i}\right)}\left(-\sum_{i}p\left(y_{i}\right)\log_{2}p\left(y_{i}\right)\right)\equiv H_{\mathrm{max}}\left(Y\right)=\log_{2}N\ \ \ \textrm{(bits)},\tag{S7.22}\label{eq:S7.22}
\end{align}
or, equivalently, $\widetilde{C}=C/\mathrm{BAR}_{Y}=N$ anbits. This
result is consistent with the upper bound of the mutual information
given by Eq.\,(\ref{eq:S7.6}). Moreover, as commented in the main
text, the same bound emerges for noisy channels when there is no overlap
in the received constellation at the output of the O/E converter,
as Eq.\,(\ref{eq:S7.20}) is also found to be valid by virtue of
the anbit measurement theorem. 

\subsection*{7.4 Channel Capacity in bits/s or anbits/s}

\addcontentsline{toc}{subsection}{7.4 Channel Capacity in bits/s or anbits/s}

\noindent The waves used to physically implement anbits at the modulator
are classical in nature. Consequently, we can combine the Nyquist-Shannon
sampling theorem with the expressions derived in the previous subsections
for the channel capacity. This implies that the channel capacity $C$
in bits (or $\widetilde{C}$ in anbits) can be expressed in bits per
second (or anbits per second) by multiplying the corresponding equations
by the Nyquist sampling rate \cite{key-S7}, $f_{\textrm{S}}=2B$,
where $B$ is the maximum baseband frequency of the complex envelopes
implementing the anbit amplitudes. For simplicity, we can assume that
both anbit amplitudes are implemented by using envelopes with the
same bandwidth.

\newpage{}

\section*{Supplementary Note 8: numerical example on measurement and channel
capacity\label{sec:8}}

\addcontentsline{toc}{section}{Supplementary Note 8: numerical example on measurement\protect\\ and channel capacity}

\noindent In this supplementary note, we present a didactical example
on the \emph{optimization }of anbit measurement and channel capacity
in a basic API system perturbed by \emph{Gaussian AA noise}. In particular,
we provide a detailed solution to the numerical example discussed
in Sections 2.2 and 2.3 of the main text.

We begin by describing the system under analysis. The originator source
$X$ emits two \emph{equiprobable} symbols, $x_{1}$ and $x_{2}$,
which are encoded into the anbits:
\begin{align}
\bigl|\psi_{1}\bigr\rangle & =\cos\frac{\theta}{2}\left|0\right\rangle +\sin\frac{\theta}{2}\left|1\right\rangle ,\tag{S8.1}\label{eq:S8.1}\\
\bigl|\psi_{2}\bigr\rangle & =\cos\frac{\theta}{2}\left|0\right\rangle -\sin\frac{\theta}{2}\left|1\right\rangle ,\tag{S8.2}\label{eq:S8.2}
\end{align}
with $0<\theta\leq\pi/2$. The states are non-orthogonal for all values
of $\theta$, except when $\theta=\pi/2$, see Fig.\,5(c).
This analog constellation is propagated through a channel that does
not execute any computational operation, i.e., the ideal anbits
to be measured are $\bigl|\phi_{i}\bigr\rangle=\bigl|\psi_{i}\bigr\rangle$,
for all $i=1,2$. Nonetheless, the channel introduces an AA noise
modeled by a ket $\left|n\right\rangle =n_{0}\left|0\right\rangle +n_{1}\left|1\right\rangle $,
where $n_{0}$ and $n_{1}$ are assumed independent and identically
distributed Gaussian random variables with zero mean and variance
$\sigma^{2}$. These assumptions about the system noise align with
the dominant noise sources identified in passive linear PIP circuits
(see Supplementary Note 6). As a result, the noisy anbits $\bigl|\phi\bigr\rangle$
at the output of the O/E converter are of the form $\bigl|\phi\bigr\rangle=\bigl|\phi_{i}\bigr\rangle+\left|n\right\rangle $,
which define the received constellation, represented in the half-angle
GBS, see Fig.\,5(d).\footnote{In this example, for simplicity, we assume that the same noise ket
$\left|n\right\rangle $ perturbs each transmitted anbit. In practice,
however, noise may affect each anbit differently, requiring a distinct
noise ket to be considered for each anbit of the constellation.}

The goal is to optimize the anbit measurement at the output of the
O/E converter by designing decision regions ($D_{1}$ and $D_{2}$)
in the half-angle GBS using the MAP criterion {[}Eq.\,(\ref{eq:S5.10}){]}.
Specifically, in the half-angle GBS, the kets $\bigl|\phi_{1}\bigr\rangle$,
$\bigl|\phi_{2}\bigr\rangle$, and $\left|n\right\rangle $ are respectively
described by the position vectors (we use Cartesian coordinates):\footnote{These position vectors can be calculated from the corresponding kets
as detailed on p.\,\pageref{eq:S6.4}.}
\begin{align}
\mathbf{r}_{1}^{\prime} & =\left(\sin\frac{\theta}{2},0,\cos\frac{\theta}{2}\right),\tag{S8.3}\label{eq:S8.3}\\
\mathbf{r}_{2}^{\prime} & =\left(-\sin\frac{\theta}{2},0,\cos\frac{\theta}{2}\right),\tag{S8.4}\label{eq:S8.4}\\
\mathbf{n} & =\left(n_{1},0,n_{0}\right).\tag{S8.5}\label{eq:S8.5}
\end{align}
In addition, the noisy anbit $\bigl|\phi\bigr\rangle$, over which
the measurement must be optimized, is represented by the arbitrary
position vector $\mathbf{r}=\left(x,y,z\right)$. 

Next, we should calculate the noise distribution in the half-angle
GBS, that is, the pdf $f_{\mathbf{N}}\left(\mathbf{n}\right)$. Bearing
in mind that $n_{0}$ and $n_{1}$ are independent and identically
distributed Gaussian random variables (with zero mean and variance
$\sigma^{2}$), it is direct to find $f_{\mathbf{N}}\left(\mathbf{n}\right)$
from the product of the marginal pdfs of $n_{0}$ and $n_{1}$:
\begin{equation}
f_{\mathbf{N}}\left(n_{1},0,n_{0}\right)=f_{\mathcal{N}_{1}}\left(n_{1}\right)f_{\mathcal{N}_{0}}\left(n_{0}\right)=\frac{1}{2\pi\sigma^{2}}\exp\left(-\frac{n_{1}^{2}+n_{0}^{2}}{2\sigma^{2}}\right).\tag{S8.6}\label{eq:S8.6}
\end{equation}
Hence, by using Eq.\,(\ref{eq:S6.3}) of the AA noise model, we obtain
the conditional pdfs required to determine the optimal decision regions:
\begin{align}
f_{1}\left(\mathbf{r}\right) & \coloneqq f\left(\mathbf{r}|\mathbf{r}_{1}^{\prime}\right)=f_{\mathbf{N}}\left(\mathbf{n}=\mathbf{r}-\mathbf{r}_{1}^{\prime}\right)=f_{\mathbf{N}}\left(x-\sin\frac{\theta}{2},y,z-\cos\frac{\theta}{2}\right),\tag{S8.7}\label{eq:S8.7}\\
f_{2}\left(\mathbf{r}\right) & \coloneqq f\left(\mathbf{r}|\mathbf{r}_{2}^{\prime}\right)=f_{\mathbf{N}}\left(\mathbf{n}=\mathbf{r}-\mathbf{r}_{2}^{\prime}\right)=f_{\mathbf{N}}\left(x+\sin\frac{\theta}{2},y,z-\cos\frac{\theta}{2}\right),\tag{S8.8}\label{eq:S8.8}
\end{align}
that is:
\begin{align}
f_{1}\left(\mathbf{r}\right) & =\frac{1}{2\pi\sigma^{2}}\exp\left(-\frac{\left(x-\sin\frac{\theta}{2}\right)^{2}+\left(z-\cos\frac{\theta}{2}\right)^{2}}{2\sigma^{2}}\right),\tag{S8.9}\label{eq:S8.9}\\
f_{2}\left(\mathbf{r}\right) & =\frac{1}{2\pi\sigma^{2}}\exp\left(-\frac{\left(x+\sin\frac{\theta}{2}\right)^{2}+\left(z-\cos\frac{\theta}{2}\right)^{2}}{2\sigma^{2}}\right).\tag{S8.10}\label{eq:S8.10}
\end{align}
Accordingly, the optimal decision regions based on MAP criterion are:
\begin{align}
D_{1} & =\left\{ \mathbf{r}\in\mathbb{R}^{3}/f_{1}\left(\mathbf{r}\right)>f_{2}\left(\mathbf{r}\right)\right\} =\left\{ x>0\right\} ,\tag{S8.11}\label{eq:S8.11}\\
D_{2} & =\left\{ \mathbf{r}\in\mathbb{R}^{3}/f_{1}\left(\mathbf{r}\right)<f_{2}\left(\mathbf{r}\right)\right\} =\left\{ x<0\right\} .\tag{S8.12}\label{eq:S8.12}
\end{align}
Figure 5(e) in the main text illustrates these regions. Therefore,
using Eq.\,(\ref{eq:S5.12}), we obtain a closed-form expression
for the SER:
\begin{align}
\textrm{SER} & =1-\frac{1}{2}\int_{D_{1}}f_{1}\left(\mathbf{r}\right)\mathrm{d}x\mathrm{d}z-\frac{1}{2}\int_{D_{2}}f_{2}\left(\mathbf{r}\right)\mathrm{d}x\mathrm{d}z\nonumber \\
 & =1-\frac{1}{2}\int_{x=0}^{\infty}\int_{z=-\infty}^{\infty}\frac{1}{2\pi\sigma^{2}}\exp\left(-\frac{\left(x-\sin\frac{\theta}{2}\right)^{2}+\left(z-\cos\frac{\theta}{2}\right)^{2}}{2\sigma^{2}}\right)\mathrm{d}x\mathrm{d}z\nonumber \\
 & -\frac{1}{2}\int_{x=-\infty}^{0}\int_{z=-\infty}^{\infty}\frac{1}{2\pi\sigma^{2}}\exp\left(-\frac{\left(x+\sin\frac{\theta}{2}\right)^{2}+\left(z-\cos\frac{\theta}{2}\right)^{2}}{2\sigma^{2}}\right)\mathrm{d}x\mathrm{d}z\nonumber \\
 & =\frac{1}{2}\left[1-\mathrm{erf}\left(\frac{1}{\sqrt{2}\sigma}\sin\frac{\theta}{2}\right)\right]=\frac{1}{2}\mathrm{erfc}\left(\frac{1}{\sqrt{2}\sigma}\sin\frac{\theta}{2}\right),\tag{S8.13}\label{eq:S8.13}
\end{align}
where erf is the error function, defined as $\mathrm{erf}\left(z\right)\coloneqq\left(2/\sqrt{\pi}\right)\int_{0}^{z}e^{-w^{2}}\mathrm{d}w$,
and erfc is the complementary error function, $\mathrm{erfc}\left(z\right)\coloneqq1-\mathrm{erf}\left(z\right)$
\cite{key-S19}.\footnote{In particular, the integrals in Eq.\,(\ref{eq:S8.13}) have been
calculated using the following property of the error function \cite{key-S19}:
\begin{equation}
\sqrt{\frac{\alpha}{\pi}}\int_{a}^{b}e^{-\alpha w^{2}}\mathrm{d}w=\frac{1}{2}\left[\mathrm{erf}\left(b\sqrt{\alpha}\right)-\mathrm{erf}\left(a\sqrt{\alpha}\right)\right],\tag{S8.14}\label{eq:S8.14}
\end{equation}
with $\alpha\in\left(-\infty,\infty\right)$ and $a,b\in\left[-\infty,\infty\right]$.} Note that Eq.\,(\ref{eq:S8.13}) corresponds to Eq.\,(7) in the
main text.

Once the anbit measurement has been optimized, the next step is to
optimize the channel capacity. Since the pmf of the originator source
and the analog constellation are fixed parameters in our problem,
the channel capacity can be directly calculated substituting the above
decision regions into Eq.\,(\ref{eq:S7.14}), which reduces to:
\begin{equation}
C=\frac{1}{2}\sum_{i,j=1}^{2}\int_{D_{j}}f_{i}\left(\mathbf{r}\right)\mathrm{d}x\mathrm{d}z\log_{2}\frac{\int_{D_{j}}f_{i}\left(\mathbf{r}\right)\mathrm{d}x\mathrm{d}z}{\int_{D_{j}}f\left(\mathbf{r}\right)\mathrm{d}x\mathrm{d}z}\ \ \ \textrm{(bits)},\tag{S8.15}\label{eq:S8.15}
\end{equation}
with $f\left(\mathbf{r}\right)=0\textrm{.}5\left[f_{1}\left(\mathbf{r}\right)+f_{2}\left(\mathbf{r}\right)\right]$.
Concretely, the integrals $\int_{D_{j}}f_{i}\left(\mathbf{r}\right)\mathrm{d}x\mathrm{d}z$
in Eq.\,(\ref{eq:S8.15}) may be calculated using Eq.\,(\ref{eq:S8.14}).
After some tedious but straightforward algebraic manipulations, we
find that:
\begin{equation}
C=\frac{1}{2}\sum_{k=1}^{2}\mathrm{erfc}\left(\frac{\left(-1\right)^{k}}{\sqrt{2}\sigma}\sin\frac{\theta}{2}\right)\log_{2}\left[\mathrm{erfc}\left(\frac{\left(-1\right)^{k}}{\sqrt{2}\sigma}\sin\frac{\theta}{2}\right)\right]\ \ \ \textrm{(bits)},\tag{S8.16}\label{eq:S8.16}
\end{equation}
which is Eq.\,(9) of the paper.

\paragraph{SER and channel capacity in quantum information.}

Consider a \emph{noiseless} quantum channel where the emitted quantum
states are the same as the classical states of this example. The optimization
of the quantum measurement is detailed on pp.\,101 and 102 of ref.\,\cite{key-S6},
which leads to a SER:
\begin{equation}
\textrm{SER}_{\textrm{QI}}=\frac{1}{2}\left(1-\sin\theta\right).\tag{S8.17}\label{eq:S8.17}
\end{equation}
Equation (\ref{eq:S8.17}) corresponds to the blue line shown in Fig.\,5(f)
of the main text. Likewise, the calculation of the channel capacity
is reported on p.\,216 of ref.\,\cite{key-S6}:
\begin{align}
C_{\textrm{QI}} & =\frac{1}{2}\left(1+\sin\theta\right)\log_{2}\left(1+\sin\theta\right)\nonumber \\
 & +\frac{1}{2}\left(1-\sin\theta\right)\log_{2}\left(1-\sin\theta\right)\ \ \ \textrm{(bits)},\tag{S8.18}\label{eq:S8.18}
\end{align}
which corresponds to the blue line depicted in Fig.\,6 of the paper.

\noindent 

\newpage{}

\section*{Supplementary Note 9: materials and methods\label{sec:9}}

\addcontentsline{toc}{section}{Supplementary Note 9: materials and methods}

\noindent In this section, we provide further details of the fabricated
chip, along with theoretical calculations of the SER and channel capacity
shown in Fig.\,7(g) of the main text.

\paragraph{Manufacturing process.}

The photonic integrated circuit was fabricated by Advanced Micro Foundry
using a standard silicon-on-insulator (SOI) process. The chip was
manufactured on an SOI wafer with a 220\,nm thick silicon slab, and
500\,nm wide single-mode waveguides were defined through deep ultraviolet
lithography at 193\,nm. Phase-shifter sections were realized by depositing
a 120\,nm layer of titanium nitride over the waveguides, forming thermo-optic
heaters. These thermo-optic phase shifters were implemented using
suspended wave-guides and etched trenches, design features that contribute
to a low power consumption of just 1.35\,mW//$\pi$. The integrated
photodetectors were implemented using germanium-on-silicon technology,
achieving responsivities of up to 0.85\,A/W.

\paragraph{Assembly process and insertion losses.}

After fabrication, the chip was mounted on a custom-designed printed
circuit board and electrically packaged to facilitate control of the
phase shifters and readout of the photodetectors, while optical input
was provided via vertical coupling through grating couplers, since
no optical packaging was implemented. These grating couplers exhibit
insertion losses of approximately 4\,dB, centered at a wavelength
of 1550\,nm. Propagation losses within the chip are considered negligible
due to the short length of the optical paths, especially when compared
to the typical waveguide propagation loss of the platform, which is
around 1.17\,dB/cm. 

\paragraph{Characterization and performance.}

Each phase shifter was individually characterized, and its response
was fitted based on the expected quadratic relationship between the
induced phase shift and the square of the applied current, using the
photodetectors as output monitors. The optical power was fitted according
to the input-output relationship, using the standard notation of a
simple Mach-Zehnder interferometer in either the bar or cross state.

\paragraph{Theoretical calculation of the SER and channel capacity.}

Here, we theoretically analyze the SER and channel capacity in a transmission
of $M$ equiprobable anbits located on the equator of the GBS, with
a differential phase ranging from 0.78 rad to 0.99 rad. In particular,
the anbits $\bigl|\psi_{i}\bigr\rangle$ transmitted through the channel
(composed of an anbit gate programmed as the identity matrix) and
the ideal anbits $\bigl|\phi_{i}\bigr\rangle$ that should be measured
are:
\begin{equation}
\bigl|\psi_{i}\bigr\rangle=\frac{1}{40}\left(\left|0\right\rangle +e^{\mathrm{j}\varphi_{i}}\left|1\right\rangle \right)\equiv\bigl|\phi_{i}\bigr\rangle,\tag{S9.1}\label{eq:S9.1}
\end{equation}
with: 
\begin{equation}
\varphi_{i}=\varphi_{1}+\left(i-1\right)\frac{\Delta\varphi}{M-1},\ \ \ \left(i=1,\ldots,M\right),\tag{S9.2}\label{eq:S9.2}
\end{equation}
$\varphi_{1}=0\textrm{.}78$ rad, $\varphi_{M}=0\textrm{.}99$ rad,
and $\Delta\varphi=\varphi_{M}-\varphi_{1}=0\textrm{.}21$ rad.

Dominant noise sources and hardware imperfections can be modeled as
an equivalent AA noise (see last paragraph of Supplementary Note 6.6).
Alternatively, these system's physical impairments can be described
using the \emph{AP noise} formalism, since the information is encoded
onto a single EDF, the differential phase (see p.\,\pageref{eq:S6.11}).
This reduces the dimensionality of the measurement optimization problem
from 3D to 1D. Accordingly, the noisy anbit $\left|\phi\right\rangle $
at the output of the O/E conversion should be described of the form:
\begin{equation}
\left|\phi\right\rangle =\frac{1}{40}\left(\left|0\right\rangle +e^{\mathrm{j}\varphi}\left|1\right\rangle \right),\tag{S9.3}\label{eq:S9.3}
\end{equation}
with the differential phase $\varphi=\varphi_{i}+\eta_{i}$ accounting
for the phase perturbation $\eta_{i}$ induced by the system's physical
impairments on the differential phase $\varphi_{i}$ of the anbit
$\bigl|\psi_{i}\bigr\rangle$. The random variable $\eta_{i}$ is
modeled as a Gaussian distribution with arbitrary mean $\mu_{i}$
and variance $\sigma_{i}^{2}$:
\begin{equation}
f_{\mathcal{N}_{i}}\left(\eta_{i}\right)=\frac{1}{\sqrt{2\pi\sigma_{i}^{2}}}\exp\left(-\frac{\left(\eta_{i}-\mu_{i}\right)^{2}}{2\sigma_{i}^{2}}\right).\tag{S9.4}\label{eq:S9.4}
\end{equation}
Thus, using the vector space $\mathbb{S}=\mathbb{R}$ to optimize
the anbit measurement through the representations:
\begin{equation}
\mathbf{r}_{i}=\mathbf{r}_{i}^{\prime}=\varphi_{i}\hat{\mathbf{x}},\ \ \ \mathbf{n}_{i}=\eta_{i}\hat{\mathbf{x}},\ \ \ \mathbf{r}=\varphi\hat{\mathbf{x}},\tag{S9.5}\label{eq:S9.5}
\end{equation}
the conditional pdfs of the problem are found to be:
\begin{equation}
f\left(\mathbf{r}|\mathbf{r}_{i}\right)=f\left(\mathbf{r}|\mathbf{r}_{i}^{\prime}\right)=f_{\mathcal{N}_{i}}\left(\eta_{i}=\varphi-\varphi_{i}\right),\tag{S9.6}\label{eq:S9.6}
\end{equation}
which are denoted as $f_{i}\left(\varphi\right)$ for simplicity.

The optimal decision regions $D_{i}$ based on MAP criterion that
minimize the SER are defined by the intersection points $\chi_{i}$
of the conditional pdfs $f_{i}\left(\varphi\right)$, such that:
\begin{equation}
D_{i}=\left\{ \varphi\in\mathbb{R}/f_{i}\left(\varphi\right)>f_{j}\left(\varphi\right),\ \forall j\in\left\{ 1,\ldots,M\right\} /j\neq i\right\} =\left\{ \chi_{i-1}<\varphi<\chi_{i}\right\} .\tag{S9.7}\label{eq:S9.7}
\end{equation}
By substituting the conditional pdfs and their corresponding decision
regions into Eqs.\,(4) and (8) of the main text, the SER and channel
capacity become:
\begin{align}
\textrm{SER} & =1-\frac{1}{2M}\sum_{i=1}^{M}\mathrm{ERF}\left(i,i\right),\tag{S9.8}\label{eq:S9.8}\\
C & =\frac{1}{2M}\sum_{i,j=1}^{M}\mathrm{ERF}\left(j,i\right)\log_{2}\frac{M\cdot\mathrm{ERF}\left(j,i\right)}{\sum_{k=1}^{M}\mathrm{ERF}\left(j,k\right)}\ \ \ \textrm{(bits)},\tag{S9.9}\label{eq:S9.9}
\end{align}
where the function ERF is defined as:
\begin{equation}
\mathrm{ERF}\left(j,i\right)\coloneqq\mathrm{erf}\left(\frac{\chi_{j}-\varphi_{i}-\mu_{i}}{\sqrt{2}\sigma_{i}}\right)-\mathrm{erf}\left(\frac{\chi_{j-1}-\varphi_{i}-\mu_{i}}{\sqrt{2}\sigma_{i}}\right).\tag{S9.10}\label{eq:S9.10}
\end{equation}

The above expressions are impractical for theoretical estimation of
the SER and channel capacity, as they rely on intersection points
$\chi_{i}$ that are difficult to determine analytically. However,
as commented in the main text, this issue is circumvented by approximating
the random variables $\eta_{i}$ as independent and identically distributed
Gaussian variables with zero mean and variance $\sigma^{2}\sim10^{-5}$.
Under these assumptions, the intersections points can be approximated
as:
\begin{equation}
\chi_{i}\simeq\frac{\varphi_{i}+\varphi_{i+1}}{2}=\varphi_{1}+\left(i-\frac{1}{2}\right)\frac{\Delta\varphi}{M-1},\tag{S9.11}\label{eq:S9.11}
\end{equation}
and the ERF function reduces to:
\begin{align}
\mathrm{ERF}\left(j,i\right) & \simeq\mathrm{erf}\left(\frac{\chi_{j}-\varphi_{i}}{\sqrt{2}\sigma}\right)-\mathrm{erf}\left(\frac{\chi_{j-1}-\varphi_{i}}{\sqrt{2}\sigma}\right)\nonumber \\
 & =\mathrm{erf}\left(\frac{\Delta\varphi\left(j-i+\nicefrac{1}{2}\right)}{\sqrt{2}\sigma\left(M-1\right)}\right)-\mathrm{erf}\left(\frac{\Delta\varphi\left(j-i-\nicefrac{1}{2}\right)}{\sqrt{2}\sigma\left(M-1\right)}\right).\tag{S9.12}\label{eq:S9.12}
\end{align}
Substituting Eq.\,(\ref{eq:S9.12}) into Eqs.\,(\ref{eq:S9.8})
and (\ref{eq:S9.9}), we can now theoretically predict the SER and
channel capacity. Concretely, Eq.\,(\ref{eq:S9.8}) reduces to Eq.\,(11)
in the main text, while Eq.\,(\ref{eq:S9.9}) corresponds to Eq.\,(12).

\noindent 

\newpage{}

\section*{Appendix A: triangle inequality of the state distance\label{sec:APPENDIX_A}}

\addcontentsline{toc}{section}{Appendix A: triangle inequality of the state distance}

\noindent Here, we demonstrate Eq.\,(\ref{eq:S3.23}). We begin by
defining the states $\bigl|\Psi_{XY}\bigr\rangle=\bigl|\psi_{X}\bigr\rangle\times\bigl|\psi_{Y}\bigr\rangle$
and $\bigl|\Phi_{XY}\bigr\rangle=\bigl|\varphi_{X}\bigr\rangle\times\bigl|\varphi_{Y}\bigr\rangle$.
Hence, it follows that:
\begin{align}
D_{\mathrm{S}}\bigl(\bigl|\Psi_{XY}\bigr\rangle,\bigl|\Phi_{XY}\bigr\rangle\bigr) & =\left\Vert \bigl|\Psi_{XY}\bigr\rangle-\bigl|\Phi_{XY}\bigr\rangle\right\Vert \nonumber \\
 & =\sqrt{\bigl\langle\Psi_{XY}\bigr|\Psi_{XY}\bigr\rangle+\bigl\langle\Phi_{XY}\bigr|\Phi_{XY}\bigr\rangle-2\textrm{Re}\left\{ \bigl\langle\Psi_{XY}\bigr|\Phi_{XY}\bigr\rangle\right\} },\tag{SA.1}\label{eq:SA.1}
\end{align}
where $\bigl\langle\Psi_{XY}\bigr|\Psi_{XY}\bigr\rangle=\bigl\langle\psi_{X}\bigr|\psi_{X}\bigr\rangle+\bigl\langle\psi_{Y}\bigr|\psi_{Y}\bigr\rangle$,
$\bigl\langle\Phi_{XY}\bigr|\Phi_{XY}\bigr\rangle=\bigl\langle\varphi_{X}\bigr|\varphi_{X}\bigr\rangle+\bigl\langle\varphi_{Y}\bigr|\varphi_{Y}\bigr\rangle$,
and\linebreak{} $\bigl\langle\Psi_{XY}\bigr|\Phi_{XY}\bigr\rangle=\bigl\langle\psi_{X}\bigr|\varphi_{X}\bigr\rangle+\bigl\langle\psi_{Y}\bigr|\varphi_{Y}\bigr\rangle$.
Thus, reordering terms, we find that:
\begin{align}
D_{\mathrm{S}}\bigl(\bigl|\Psi_{XY}\bigr\rangle,\bigl|\Phi_{XY}\bigr\rangle\bigr) & =\sqrt{D_{\mathrm{S}}^{2}\left(\bigr|\psi_{X}\bigr\rangle,\bigr|\varphi_{X}\bigr\rangle\right)+D_{\mathrm{S}}^{2}\left(\bigr|\psi_{Y}\bigr\rangle,\bigr|\varphi_{Y}\bigr\rangle\right)}\nonumber \\
 & \leq\sqrt{D_{\mathrm{S}}^{2}\left(\bigr|\psi_{X}\bigr\rangle,\bigr|\varphi_{X}\bigr\rangle\right)}+\sqrt{D_{\mathrm{S}}^{2}\left(\bigr|\psi_{Y}\bigr\rangle,\bigr|\varphi_{Y}\bigr\rangle\right)}\nonumber \\
 & =D_{\mathrm{S}}\left(\bigr|\psi_{X}\bigr\rangle,\bigr|\varphi_{X}\bigr\rangle\right)+D_{\mathrm{S}}\left(\bigr|\psi_{Y}\bigr\rangle,\bigr|\varphi_{Y}\bigr\rangle\right).\tag{SA.2}\label{eq:SA.2}
\end{align}

\noindent 

\newpage{}

\section*{Appendix B: conditional probability\label{sec:APPENDIX_B}}

\addcontentsline{toc}{section}{Appendix B: conditional probability}

\noindent In this appendix, we demonstrate Eq.\,(\ref{eq:S5.3}).
From the continuous version of the law of total probability \cite{key-S1},
we can write:
\begin{equation}
p\left(y_{i}\right)=\int_{\mathbb{S}}p\left(y_{i}|\mathbf{r}\right)f\left(\mathbf{r}\right)\mathrm{d}^{s}r.\tag{SB.1}\label{eq:SB.1}
\end{equation}
This equation describes all possible cases of $\mathbf{r}$ in $\mathbb{S}$
that should be measured as the symbol $y_{i}$. In addition, from
the discrete version of the law of total probability, we know that:
\begin{equation}
f\left(\mathbf{r}\right)=\sum_{k=1}^{M}p\left(x_{k}\right)f\left(\mathbf{r}|x_{k}\right),\tag{SB.2}\label{eq:SB.2}
\end{equation}
where $f\left(\mathbf{r}|x_{k}\right)\equiv f\left(\mathbf{r}|\mathbf{r}_{k}\right)$
\emph{if and only if} the mapping $x_{k}\rightarrow\mathbf{r}_{k}$
is 1:1. In such a case, Eq.\,(\ref{eq:SB.1}) becomes:
\begin{equation}
p\left(y_{i}\right)=\sum_{k}p\left(x_{k}\right)\int_{\mathbb{S}}p\left(y_{i}|\mathbf{r}\right)f\left(\mathbf{r}|\mathbf{r}_{k}\right)\mathrm{d}^{s}r.\tag{SB.3}\label{eq:SB.3}
\end{equation}
In addition, $p\left(y_{i}\right)$ can alternatively be expressed
of the form:
\begin{equation}
p\left(y_{i}\right)=\sum_{k}p\left(x_{k},y_{i}\right)=\sum_{k}p\left(x_{k}\right)p\left(y_{i}|x_{k}\right).\tag{SB.4}\label{eq:SB.4}
\end{equation}
Hence, by subtracting Eqs.\,(\ref{eq:SB.3}) and (\ref{eq:SB.4}),
we obtain:
\begin{equation}
\sum_{k}p\left(x_{k}\right)\left[p\left(y_{i}|x_{k}\right)-\int_{\mathbb{S}}p\left(y_{i}|\mathbf{r}\right)f\left(\mathbf{r}|\mathbf{r}_{k}\right)\mathrm{d}^{s}r\right]=0,\tag{SB.5}\label{eq:SB.5}
\end{equation}
which is fulfilled if and only if:
\begin{equation}
p\left(y_{i}|x_{k}\right)=\int_{\mathbb{S}}p\left(y_{i}|\mathbf{r}\right)f\left(\mathbf{r}|\mathbf{r}_{k}\right)\mathrm{d}^{s}r,\tag{SB.6}\label{eq:SB.6}
\end{equation}
for all $k=1,\ldots,M$. Finally, setting $k=i$, we find Eq.\,(\ref{eq:S5.3}).

\newpage{}

\section*{Appendix C: additive noise on the electromagnetic power\label{sec:APPENDIX_C}}

\addcontentsline{toc}{section}{Appendix C: additive noise on the electromagnetic power}

\noindent Consider a 2D electric field with complex envelopes $\psi_{0}$
and $\psi_{1}$. Now, assume that there is a 2D additive noise, represented
by two complex numbers $n_{0}=\left|n_{0}\right|e^{\mathrm{j}\arg\left(n_{0}\right)}$
and $n_{1}=\left|n_{1}\right|e^{\mathrm{j}\arg\left(n_{1}\right)}$,
inducing a perturbation on the power of the electric field of the
form $\left|\psi_{k}\right|^{2}+\left|n_{k}\right|^{2}$ ($k=0,1$).
Can we represent this perturbation on the power as an additive perturbation
on the field? In other words, can we represent the perturbation $\left|\psi_{k}\right|^{2}+\left|n_{k}\right|^{2}$
as a function of $\psi_{k}+n_{k}$?

To address this question, observe that the phase of $n_{k}$ constitutes
a \emph{degree of freedom} in the expression $\left|\psi_{k}\right|^{2}+\left|n_{k}\right|^{2}$,
that is, the value of $\left|\psi_{k}\right|^{2}+\left|n_{k}\right|^{2}$
is invariant under changes in $\arg\left(n_{k}\right)$. Hence, one
can always select a specific phase $\arg\left(n_{k}\right)$ that
ensures the identity $\left|\psi_{k}\right|^{2}+\left|n_{k}\right|^{2}\equiv\left|\psi_{k}+n_{k}\right|^{2}$.
Here, taking into account that:
\begin{align}
\left|\psi_{k}+n_{k}\right|^{2} & =\left|\psi_{k}\right|^{2}+\left|n_{k}\right|^{2}+2\textrm{Re}\left(\psi_{k}n_{k}^{\ast}\right)\nonumber \\
 & =\left|\psi_{k}\right|^{2}+\left|n_{k}\right|^{2}+2\left|\psi_{k}\right|\left|n_{k}\right|\cos\left(\arg\left(\psi_{k}\right)-\arg\left(n_{k}\right)\right),\tag{SC.1}\label{eq:SC.1}
\end{align}
we find that $\left|\psi_{k}\right|^{2}+\left|n_{k}\right|^{2}\equiv\left|\psi_{k}+n_{k}\right|^{2}$
\emph{if and only if} the following \emph{phase condition} is satisfied:
\begin{equation}
\arg\left(\psi_{k}\right)-\arg\left(n_{k}\right)=\left(2m+1\right)\frac{\pi}{2};\ \ \ m\in\mathbb{Z}.\tag{SC.2}\label{eq:SC.2}
\end{equation}
Remarkably, Eq.\,(\ref{eq:SC.2}) allows us to describe an additive
noise on the power of an electromagnetic field implementing an anbit
$\left|\psi\right\rangle =\psi_{0}\left|0\right\rangle +\psi_{1}\left|1\right\rangle $
through the expression: 
\begin{equation}
\left|\psi\right\rangle +\left|n\right\rangle =\left(\psi_{0}+n_{0}\right)\bigl|0\bigr\rangle+\left(\psi_{1}+n_{1}\right)\bigl|1\bigr\rangle,\tag{SC.3}\label{eq:SC.3}
\end{equation}
with $\left|n\right\rangle =n_{0}\left|0\right\rangle +n_{1}\left|1\right\rangle $.
In this scenario, the perturbed field is represented by the envelopes
$\psi_{k}+n_{k}$, with an optical power given by $\left|\psi_{k}+n_{k}\right|^{2}=\left|\psi_{k}\right|^{2}+\left|n_{k}\right|^{2}$.

\newpage{}

\section*{Appendix D: conditional pdf with additive noise\label{sec:APPENDIX_D}}

\addcontentsline{toc}{section}{Appendix D: conditional pdf with additive noise}

\noindent In this appendix, we demonstrate Eq.\,(\ref{eq:S6.3}).
To this end, we note that the vector mapping in Eq.\,(\ref{eq:S6.3})
is analogous to the following transformation of random variables:
\begin{equation}
Y=Z+\mathcal{N}=g\left(X\right)+\mathcal{N},\tag{SD.1}\label{eq:SD.1}
\end{equation}
where $X$, $\mathcal{N}$, $Z=g\left(X\right)$, and $Y$ are continuous
random variables, and the function $g$ may be either bijective or
non-bijective (both scenarios are considered). Here, we assume that
$\mathcal{N}$ describes a random noise source that is independent
of both $X$ and $Z$, in accordance with Eq.\,(\ref{eq:S6.3}).
The roadmap of the proof is to first derive the conditional pdf $f_{Y|X}\left(y|x\right)$,
and subsequently extend the result to the case of random vectors.

The starting point is the following relation between pdfs \cite{key-S1}:
\begin{equation}
f_{Y|X}\left(y|x\right)=\frac{f_{XY}\left(x,y\right)}{f_{X}\left(x\right)}.\tag{SD.2}\label{eq:SD.2}
\end{equation}
To derive the joint pdf $f_{XY}$, we must first determine the joint
cumulative distribution function:
\begin{align}
F_{XY}\left(x,y\right) & =p\left(X\leq x,Y\leq y\right)=p\left(X\leq x,g\left(X\right)+\mathcal{N}\leq y\right)\nonumber \\
 & =p\left(X\leq x,\mathcal{N}\leq y-g\left(x\right)\right)=p\left(X\leq x\right)p\left(\mathcal{N}\leq y-g\left(x\right)\right)\nonumber \\
 & =F_{X}\left(x\right)F_{\mathcal{N}}\left(n=y-g\left(x\right)\right).\tag{SD.3}\label{eq:SD.3}
\end{align}
Hence, by applying Schwarz\textquoteright s theorem and the chain
rule, we obtain:
\begin{align}
f_{XY}\left(x,y\right) & =\frac{\partial^{2}F_{XY}\left(x,y\right)}{\partial x\partial y}=\frac{\partial}{\partial y}\left\{ \frac{\partial}{\partial x}\left[F_{X}\left(x\right)F_{\mathcal{N}}\left(n\right)\right]\right\} =\frac{\partial}{\partial y}\left[f_{X}\left(x\right)F_{\mathcal{N}}\left(n\right)\right]\nonumber \\
 & =f_{X}\left(x\right)\frac{\partial F_{\mathcal{N}}\left(n\right)}{\partial y}=f_{X}\left(x\right)\left[\frac{\partial F_{\mathcal{N}}\left(n\right)}{\partial x}\frac{\mathrm{d}x}{\mathrm{d}z}\frac{\partial z}{\partial y}+\frac{\partial F_{\mathcal{N}}\left(n\right)}{\partial n}\frac{\partial n}{\partial y}\right]\nonumber \\
 & =f_{X}\left(x\right)f_{\mathcal{N}}\left(n=y-g\left(x\right)\right).\tag{SD.4}\label{eq:SD.4}
\end{align}
From this result, we find that:
\begin{equation}
f_{Y|X}\left(y|x\right)=f_{\mathcal{N}}\left(n=y-g\left(x\right)\right).\tag{SD.5}\label{eq:SD.5}
\end{equation}
In addition, it is direct to verify that $f_{Y|X}\left(y|x\right)\equiv f_{Y|Z}\left(y|z\right)$.
Next, by reasoning in a similar way for random vectors with $\mathbf{Y}=\mathbf{g}\left(\mathbf{X}\right)+\mathbf{N}$,
we find the sought result:
\begin{equation}
f_{\mathbf{Y}|\mathbf{X}}\left(\mathbf{y}|\mathbf{x}\right)=f_{\mathbf{Y}|\mathbf{Z}}\left(\mathbf{y}|\mathbf{z}\right)=f_{\mathbf{N}}\left(\mathbf{n}=\mathbf{y}-\mathbf{g}\left(\mathbf{x}\right)\right).\tag{SD.6}\label{eq:SD.6}
\end{equation}
Finally, by extrapolating these conclusions to the API formalism,
we obtain Eq.\,(\ref{eq:S6.3}).

\newpage{}

\section*{Appendix E: quantum-classical analogy of the channel capacity\label{sec:APPENDIX_E}}

\addcontentsline{toc}{section}{Appendix E: quantum-classical analogy of the channel capacity}

\noindent Here, we briefly outline a mathematical analogy between
the channel capacities of API and QI systems. For the sake of clarity,
let us first reproduce the channel capacity in API, given by Eq.\,(\ref{eq:S7.14}):
\begin{equation}
C_{\mathrm{API}}=\max_{p_{i},\mathbf{r}_{i}^{\prime},D_{j}}\left\{ \sum_{i,j}p_{i}\int_{D_{j}}f\left(\mathbf{r}|\mathbf{r}_{i}^{\prime}\right)\mathrm{d}^{s}r\log_{2}\frac{\int_{D_{j}}f\left(\mathbf{r}|\mathbf{r}_{i}^{\prime}\right)\mathrm{d}^{s}r}{\int_{D_{j}}f\left(\mathbf{r}\right)\mathrm{d}^{s}r}\right\} \ \ \ \textrm{(bits)}.\tag{SE.1}\label{eq:SE.1}
\end{equation}
Now, consider a QI system composed of an originator source $X$ that
can randomly generate classical symbols $x_{i}$ ($i=1,\ldots,M$).
These symbols are encoded into quantum states $\widehat{\rho}_{i}$,
which are propagated through a noisy quantum channel. At the channel
output, the received states $\widehat{\rho}_{i}^{\thinspace\prime}$
are measured by using a positive operator-valued measure (POVM) $\widehat{\pi}_{j}$,
giving rise to the post-measurement states \cite{key-S6}:
\begin{equation}
\widehat{\sigma}_{j}=\frac{\widehat{\pi}_{j}^{1/2}\widehat{\rho}_{i}^{\thinspace\prime}\widehat{\pi}_{j}^{1/2}}{\mathrm{Tr}\bigl(\widehat{\rho}_{i}^{\thinspace\prime}\widehat{\pi}_{j}\bigr)},\tag{SE.2}\label{eq:SE.2}
\end{equation}
which are decoded into classical symbols $y_{j}$ of the recipient
source $Y$ ($j=1,\ldots,N$). In this system, the channel capacity
is given by the expression \cite{key-S6}:
\begin{equation}
C_{\mathrm{QI}}=\max_{p_{i},\widehat{\rho}_{i}^{\thinspace\prime},\widehat{\pi}_{j}}\left\{ \sum_{i,j}p_{i}\mathrm{Tr}\bigl(\widehat{\rho}_{i}^{\thinspace\prime}\widehat{\pi}_{j}\bigr)\log_{2}\frac{\mathrm{Tr}\bigl(\widehat{\rho}_{i}^{\thinspace\prime}\widehat{\pi}_{j}\bigr)}{\mathrm{Tr}\bigl(\widehat{\rho}^{\thinspace\prime}\widehat{\pi}_{j}\bigr)}\right\} \ \ \ \textrm{(bits)},\tag{SE.3}\label{eq:SE.3}
\end{equation}
where $\widehat{\rho}^{\thinspace\prime}=\sum_{i}p_{i}\widehat{\rho}_{i}^{\thinspace\prime}$.
By comparing Eqs.\,(\ref{eq:SE.1}) and (\ref{eq:SE.3}), we observe
a mathematical analogy between both expressions. Specifically, Eq.\,(\ref{eq:SE.3})
emerges from Eq.\,(\ref{eq:SE.1}) by replacing the integral operator
with the trace operator ($\int\leftrightarrow\mathrm{Tr}$), the decision
regions with the POVM ($D_{j}\leftrightarrow\widehat{\pi}_{j}$),
and the classical states with the quantum states ($\mathbf{r}_{i}^{\prime}\leftrightarrow\widehat{\rho}_{i}^{\thinspace\prime}$).
Nonetheless, it is worth mentioning that Eq.\,(\ref{eq:SE.1}) involves
a vector-space optimization problem, while Eq.\,(\ref{eq:SE.3})
requires solving an intricate matrix-based optimization problem. 

\newpage{}


\begin{thebibliography}{10}
\bibitem[1]{key-P1} E. Desurvire, \emph{Classical and Quantum Information
Theory: An Introduction for the Telecom Scientist}, Cambridge University
Press (2009).

\bibitem[2]{key-P2} J. F. Wakerly, \emph{Digital Design: Principles
and Practices}, Pearson (2006).

\bibitem[3]{key-P3} T. C. Bartee, \emph{Digital Computer Fundamentals},
McGraw-Hill (1985).

\bibitem[4]{key-P4} C. E. Shannon, ``A mathematical theory of communication,''
\emph{Bell System Technical Journal} \textbf{27}, 379 (1948).

\bibitem[5]{key-P5}\textbf{ }M. A. Nielsen and I. L. Chuang, \emph{Quantum
Computation and Quantum Information}, Cambridge University Press (2016).

\bibitem[6]{key-P6} K. S. Mohamed, \emph{Neuromorphic Computing and
Beyond}, Springer (2020).

\bibitem[7]{key-P7} A. Borst and F. Theunissen, ``Information theory
and neural coding,'' \emph{Nature Neuroscience} \textbf{2}, 947 (1999).

\bibitem[8]{key-P8} F. Zangeneh-Nejad, D. L. Sounas, A. Alu, and
R. Fleury, ``Analogue computing with metamaterials,'' \emph{Nature
Reviews Materials} 6, 207 (2021).

\bibitem[9]{key-P9} R. K. Cavin, P. Lugli, and V. V. Zhirnov, ``Science
and engineering beyond Moore's law,'' \emph{Proceedings of the IEEE,
Special Centennial Issue} \textbf{100}, 1720 (2012). 

\bibitem[10]{key-P10} I. L. Markov, ``Limits on fundamental limits
to computation,'' \emph{Nature} \textbf{512}, 147 (2014).

\bibitem[11]{key-P11} J. Shalf, ``The future of computing beyond
Moore\textquoteright s Law,'' \emph{Phil. Trans. R. Soc. A} \textbf{378},
20190061 (2020).

\bibitem[12]{key-P12} F. Wu, H. Tian, Y. Shen, Z. Hou, J. Ren, G.
Gou, Y. Sun, Y. Yang, and T.-L. Ren, ``Vertical MoS2 transistors
with sub-1-nm gate lengths,'' \emph{Nature} \textbf{603}, 259 (2022).

\bibitem[13]{key-P13} Y. Shao, M. Pala, H. Tang, B. Wang, J. Li,
D. Esseni, and J. A. del Alamo, ``Scaled vertical-nanowire heterojunction
tunnelling transistors with extreme quantum confinement,'' \emph{Nature
Electronics} \textbf{8}, 157 (2025).

\bibitem[14]{key-P14} S. Hua, E. Divita, S. Yu, B. Peng, C. R.-Carmes,
Z. Su, Z. Chen, Y. Bai, J. Zou, Y. Zhu, Y. Xu, C.-K. Lu, Y. Di, H.
Chen, L. Jiang, L. Wang, L. Ou, C. Zhang, J. Chen, W. Zhang, H. Zhu,
W. Kuang, L. Wang, H. Meng, M. Steinman, and Y. Shen, ``An integrated
large-scale photonic accelerator with ultralow latency,'' \emph{Nature}
\textbf{640}, 361 (2025).

\bibitem[15]{key-P15} R. Santagati, A. A.-Guzik, R. Babbush, M. Degroote,
L. González, E. Kyoseva, N. Moll, M. Oppel, R. M. Parrish, N. C. Rubin,
M. Streif, C. S. Tautermann, H. Weiss, N. Wiebe, and C. U.-Utschig,
``Drug design on quantum computers,'' \emph{Nature Physics} \textbf{20},
549 (2024).

\bibitem[16]{key-P16} G. Gumpel, J. Kang, E. Blumenthal, A. Klevansky,
E. Kim, and S. H.-Gourgy, ``Analog quantum simulation of Dirac hamiltonians
in circuit QED using Rabi driven qubits,'' \emph{arXiv:2504.08944}
(2025).

\bibitem[17]{key-P17} L. Bombieri, Z. Zeng, R. Tricarico, R. Lin,
S. Notarnicola, M. Cain, M. D. Lukin, and H. Pichler, ``Quantum adiabatic
optimization with Rydberg arrays: localization phenomena and encoding
strategies,'' \emph{PRX Quantum} \textbf{6}, 020306 (2025).

\bibitem[18]{key-P18} M. Chen, X. An, S. J. Ki, X. Liu, N. Sekhon,
A. Boyarov, A. Acharya, J. Tawil, M. Bederman, and X. Liang, ``Nanoelectronics-enabled
reservoir computing hardware for real-time robotic controls,'' \emph{Science
Advances} \textbf{26}, 11 (2025).

\bibitem[19]{key-P19} S. S. Bharadwaj and K. R. Sreenivasan, ``Compact
quantum algorithms for time-dependent differential equations,'' \emph{Phys.
Rev. Research} \textbf{7}, 023262 (2025).

\bibitem[20]{key-P20} A. Acharya, R. Yalovetzky, P. Minssen, S. Chakrabarti,
R. Shaydulin, R. Raymond, Y. Sun, D. Herman, R. S. Andrist, G. Salton,
M. J. A. Schuetz, H. G. Katzgraber, and M. Pistoia, ``Decomposition
pipeline for large-scale portfolio optimization with applications
to near-term quantum computing,'' \emph{Phys. Rev. Research} \textbf{7},
023142 (2025).

\bibitem[21]{key-P21} S. Wang, Y. Luo, P. Zuo, L. Pan, Y. Li, and
Z. Sun, ``In-memory analog solution of compressed sensing recovery
in one step,'' \emph{Science Advances} \textbf{13}, 9 (2023).

\bibitem[22]{key-P22} M. Alser, J. Eudine, and O. Mutlu, ``Taming
large-scale genomic analyses via sparsified genomics,'' \emph{Nature
Communications} \textbf{16}, 876 (2025).

\bibitem[23]{key-P23} H. Yasuda, P. R. Buskohl, A. Gillman, T. D.
Murphey, S. Stepney, R. A. Vaia, and J. R. Raney, ``Mechanical computing,''
\emph{Nature} \textbf{598}, 39 (2021).

\bibitem[24]{key-P24} B. J. Shastri, A. N. Tait, T. F. de Lima, W.
H. P. Pernice, H. Bhaskaran, C. D. Wright, and P. R. Prucnal, ``Photonics
for artificial intelligence and neuromorphic computing,'' \emph{Nature
Photonics} \textbf{15}, 102 (2021).

\bibitem[25]{key-P25} PsiQuantum team, ``A manufacturable platform
for photonic quantum computing,'' \emph{Nature} \textbf{641}, 876
(2025).

\bibitem[26]{key-P26} J. Feldmann, N. Youngblood, M. Karpov, H. Gehring,
X. Li, M. Stappers, M. Le Gallo, X. Fu, A. Lukashchuk, A. S. Raja,
J. Liu, C. D. Wright, A. Sebastian, T. J. Kippenberg, W. H. P. Pernice,
and H. Bhaskaran, ``Parallel convolutional processing using an integrated
photonic tensor core,'' \emph{Nature} \textbf{589}, 52 (2021).

\bibitem[27]{key-P27} X. Xu, M. Tan, B. Corcoran, J. Wu, A. Boes,
T. G. Nguyen, S. T. Chu, B. E. Little, D. G. Hicks, R. Morandotti,
A. Mitchell, and D. J. Moss, ``11 TOPS photonic convolutional accelerator
for optical neural networks,'' \emph{Nature} \textbf{589}, 44 (2021).

\bibitem[28]{key-P28} H. Feng, T. Ge, X. Guo, B. Wang, Y. Zhang,
Z. Chen, S. Zhu, K. Zhang, W. Sun, C. Huang, Y. Yuan, and C. Wang,
``Integrated lithium niobate microwave photonic processing engine,''
\emph{Nature} \textbf{627}, 80 (2024).

\bibitem[29]{key-P29} D. C. Tzarouchis, B. Edwards, and N. Engheta,
\textquotedblleft Programmable wave-based analog computing machine:
a metastructure that designs metastructures,'' \emph{Nature Communications}
\textbf{16}, 908 (2025).

\bibitem[30]{key-P30} W. Bogaerts, D. Pérez, J. Capmany, D. A. B.
Miller, J. Poon, D. Englund, F. Morichetti, and A. Melloni, ``Programmable
photonic circuits,'' \emph{Nature} \textbf{586}, 207 (2020).

\bibitem[31]{key-P31} D. Pérez, I. Gasulla, P. DasMahapatra, and
J. Capmany, ``Principles, fundamentals and applications of programmable
integrated photonics,'' \emph{Advances in Optics and Photonics} \textbf{12},
709 (2020).

\bibitem[32]{key-P32} J. Capmany and D. Pérez, \emph{Programmable
Integrated Photonics}, Oxford University Press (2020).

\bibitem[33]{key-P33} Z. Zhu, A. Fardoost, F. Ghaedi Vanani, A. B.
Klein, G. Li, and S. S. Pang, ``Coherent general-purpose photonic
matrix processor,'' \emph{ACS Photonics} \textbf{11}, 1189 (2024).

\bibitem[34]{key-P34} M. Chen, Y. Wang, C. Yao, A. Wonfor, S. Yang,
R. Penty, and Q. Cheng, ``I/O-efficient iterative matrix inversion
with photonic integrated circuits,'' \emph{Nature Communications}
\textbf{15}, 5926 (2024).

\bibitem[35]{key-P35} J. Bao et al., ``Very-large-scale integrated
quantum graph photonics,'' \emph{Nature Photonics} \textbf{17}, 573
(2023).

\bibitem[36]{key-P36} Google Quantum AI and Collaborators, ``Quantum
error correction below the surface code threshold,'' \emph{Nature}
\textbf{638}, 920 (2025).

\bibitem[37]{key-P37} T. Fu, J. Zhang, R. Sun, Y. Huang, W. Xu, S.
Yang, Z. Zhu, and H. Chen, ``Optical neural networks: progress and
challenges,'' \emph{Light: Science \& Applications} \textbf{13},
263 (2024).

\bibitem[38]{key-P38} Z. Xu, B. Tang, X. Zhang, J. F. Leong, J. Pan,
S. Hooda, E. Zamburg, and A. Voon-Yew Thean, ``Reconfigurable nonlinear
photonic activation function for photonic neural network based on
non-volatile opto-resistive RAM switch,'' \emph{Light: Science \&
Applications} \textbf{11}, 288 (2022).

\bibitem[39]{key-P39} A. Macho, D. Pérez, J. Azaña, and J. Capmany,
``Analog programmable-photonic computation,'' \emph{Laser \& Photonics
Reviews} \textbf{17}, 2200360 (2023).

\bibitem[40]{key-P40} A. Macho, D. Pérez, and J. Capmany, ``Optical
implementation of 2$\times$2 universal unitary matrix transformations,''
\emph{Laser \& Photonics Reviews} \textbf{15}, 2000473 (2021).

\bibitem[41]{key-P41} J. Bass, H. Tran, W. Du, R. Soref, and S.-Q.
Yu, ``Impact of nonlinear effects in Si towards integrated microwave-photonic
applications,'' \emph{Optics Express} \textbf{29}, 30844 (2021).

\bibitem[42]{key-P42} J. A. Hartigan and M. A. Wong, ``Algorithm
AS 136: a K-Means clustering algorithm,'' \emph{Journal of the Royal
Statistical Society, Series C.} \textbf{28}, 100 (1979).

\bibitem[43]{key-P43} D. G. Luenberger and Y. Ye, \emph{Linear and
Nonlinear Programming}, Springer (2008).

\bibitem[44]{key-P44} L. Terfloth and J. Gasteiger, ``Neural networks
and genetic algorithms in drug design,'' \emph{Drug Discovery Today}
\textbf{6}, 102 (2001).

\bibitem[45]{key-P45} A. Papoulis, \emph{Probability, Random Variables,
and Stochastic Processes}, McGraw-Hill (2002).

\bibitem[46]{key-P46} J. G. Proakis and M. Salehi, \emph{Digital
Communications}, McGraw-Hill (2008).

\bibitem[47]{key-P47} M. Abramowitz and I. A. Stegun, \emph{Handbook
of Mathematical Functions}, Dover Publications (1965).

\bibitem[48]{key-P48} G. Benenti, G. Casati, and G. Strini, \emph{Principles
of Quantum Computation and Information. Volume II: Basic Tools and
Special Topics}, World Scientific (2007).

\bibitem[49]{key-P49} J. Cho and P. J. Winzer, ``Probabilistic constellation
shaping for optical fiber communications,'' \emph{J. Lightwave Technol.}
\textbf{37}, 1590 (2019).

\bibitem[50]{key-P50} M. M. Wilde, \emph{Quantum Information Theory},
Cambridge University Press (2017).
\end{thebibliography}

\begin{thebibliography}{10}
\bibitem[1]{key-S1} A. Papoulis, \emph{Probability, Random Variables,
and Stochastic Processes}, McGraw-Hill (2002).

\bibitem[2]{key-S2} A. Macho, D. Pérez, J. Azaña, and J. Capmany,
``Analog programmable-photonic computation,'' \emph{Laser \& Photonics
Reviews} \textbf{17}, 2200360 (2023).

\bibitem[3]{key-S3} M. A. Nielsen and I. L. Chuang, \emph{Quantum
Computation and Quantum Information}, Cambridge University Press (2016).

\bibitem[4]{key-S4} B. C. Hall, \emph{Lie Groups, Lie Algebras, and
Representations. An Elementary Introduction}, Springer (2003).

\bibitem[5]{key-S5} D. C. Marinescu and G. M. Marinescu, \emph{Classical
and Quantum Information}, Academic Press (2011).

\bibitem[6]{key-S6} S. M. Barnett, \emph{Quantum Information}, Oxford
University Press (2009).

\bibitem[7]{key-S7} E. Desurvire, \emph{Classical and Quantum Information
Theory: An Introduction for the Telecom Scientist}, Cambridge University
Press (2009).

\bibitem[8]{key-S8} R. Syms and J. Cozens, \emph{Optical Guided Waves
and Devices}, McGraw-Hill (1992).

\bibitem[9]{key-S9} L. B. Soldano and E. C. M. Pennings, ``Optical
multi-mode interference devices based on self-imaging: principles
and applications,'' \emph{Journal of Lightwave Technology} \textbf{13},
615 (1995).

\bibitem[10]{key-S10} H. Tan, J. Wang, W. Ke, X. Zhang, Z. Zhao,
Z. Lin, and X. Cai, ``C-Band optical 90-degree hybrid using thin
film lithium niobate,'' \emph{Optics Letters} \textbf{48}, 1946 (2023).

\bibitem[11]{key-S11} J. Capmany and D. Pérez, \emph{Programmable
Integrated Photonics}, Oxford University Press (2020).

\bibitem[12]{key-S12} J. G. Proakis and M. Salehi, \emph{Digital
Communications}, McGraw-Hill (2008).

\bibitem[13]{key-S13} E. Desurvire, \emph{Erbium-Doped Fiber Amplifiers:
Principles and Applications}, Wiley-Interscience (2002).

\bibitem[14]{key-S14} G. P. Agrawal, \emph{Fiber-Optic Communication
Systems}, Wiley (2010).

\bibitem[15]{key-S15} K. Petermann, \emph{Laser Diode Modulation
and Noise}, Kluwer Academic Publishers (1988).

\bibitem[16]{key-S16} K.-P. Ho, \emph{Phase-Modulated Optical Communications
Systems}, Springer (2005).

\bibitem[17]{key-S17} J. Bass, H. Tran, W. Du, R. Soref, and S.-Q.
Yu, \textquotedblleft Impact of nonlinear effects in Si towards integrated
microwave-photonic applications,\textquotedblright{} \emph{Optics
Express} \textbf{29}, 30844 (2021).

\bibitem[18]{key-S18} T. M. Cover and J. A. Thomas, \emph{Elements
of Information Theory}, Wiley-Interscience (2006).

\bibitem[19]{key-S19} M. Abramowitz and I. A. Stegun, \emph{Handbook
of Mathematical Functions}, Dover Publications (1965).
\end{thebibliography}
\end{document}